\documentclass[12pt]{article}
\usepackage[margin=0.75in]{geometry}

\usepackage[english]{babel}
\usepackage[utf8x]{inputenc}
\usepackage{amsmath}
\usepackage{amssymb}
\usepackage{graphicx}
\usepackage{float}
\usepackage{makecell}
\usepackage{mathtools}
\usepackage[round]{natbib}
\usepackage{multirow}
\usepackage[toc,page]{appendix}
\usepackage{xcolor}
\usepackage{url}
\usepackage{amsthm}
\usepackage{comment}
\newtheorem{theorem}{Theorem}
\newtheorem{prop}{Proposition}
\newtheorem{remark}{Remark}
\newtheorem{lemma}{Lemma}
\newtheorem*{assumption*}{Assumptions}

\DeclareMathOperator*{\argmax}{arg\,max}
\DeclareMathOperator*{\argmin}{arg\,min}

\newcommand\smallO{
  \mathchoice
    {{\scriptstyle\mathcal{O}}}
    {{\scriptstyle\mathcal{O}}}
    {{\scriptscriptstyle\mathcal{O}}}
    {\scalebox{.7}{$\scriptscriptstyle\mathcal{O}$}}
  }
  
\usepackage{algorithm}
\usepackage{algpseudocode}
\def\NoNumber#1{{\def\alglinenumber##1{}\State #1}\addtocounter{ALG@line}{-1}}

\begin{document}

  \title{ \vspace{-2cm} \bf  \huge{Robustness, model checking and \\ latent Gaussian models}  }
  \author{\Large{Rafael Cabral}\footnote{Corresponding author; Email: rafael.medeiroscabral@kaust.edu.sa}\Large{ , David Bolin, H\aa vard Rue} \\
  \small{
    Statistics Program, Computer, Electrical and Mathematical Sciences  and Engineering Division,} \\ \small{King Abdullah University of Science and Technology (KAUST),} \\ \small{Thuwal 23955-6900, Kingdom of Saudi Arabia}}
  \date{June 15, 2023}
  \maketitle

\begin{abstract}
Model checking is essential to evaluate the adequacy of statistical models and the validity of inferences drawn from them. Particularly, hierarchical models such as latent Gaussian models (LGMs) pose unique challenges as it is difficult to check assumptions about the distribution of the latent parameters. Discrepancy measures are often used to quantify the degree to which a model fit deviates from the observed data. We construct discrepancy measures by (a) defining an alternative model with relaxed assumptions and (b) deriving the discrepancy measure most sensitive to discrepancies induced by this alternative model. We also promote a workflow for model criticism that combines model checking with subsequent robustness analysis. As a result, we obtain a general recipe to check assumptions in LGMs and the impact of these assumptions on the results. We demonstrate the ideas by assessing the latent Gaussianity assumption, a crucial but often overlooked assumption in LGMs. We illustrate the methods via examples utilising Stan and provide functions for easy usage of the methods for general models fitted through R-INLA.
\end{abstract}

\noindent%

\maketitle

\section{Introduction} \label{sect:intro}

Assumptions such as linearity, normality, constant variance, or fixed smoothness are common in statistical practice. However, they may not be supported by the data $\mathbf{y}$ and can significantly impact the statistical results. Therefore, \emph{robustness analysis}, which studies the ``sensitivity of Bayesian answers to uncertain inputs'' \citep{berger1994overview}, is often employed to quantify this impact. On the other hand, \emph{model checking} procedures, such as prior or posterior predictive checks \citep{box1980sampling, gelman1996posterior}, are used to check if the model can replicate certain features in the observed data. In this paper, we explore how these seemingly different topics, \emph{robustness analysis} and \emph{model checking}, are connected in the context of checking the adequacy of assumptions in latent Gaussian models (LGMs). We also present a workflow for model criticism that integrates both topics. 

Generally, Bayesian model checking involves choosing: (a) a discrepancy measure $d(\mathbf{y})$ which measures what aspects of the model should be checked; (b) a predictive distribution $\pi(\mathbf{y}^{\text{pred}})$ to generate simulated data from the fitted model; (c) a way to compare the discrepancy measure at the observed data $d(\mathbf{y})$ with the ``reference" distribution $d(\mathbf{y}^{\text{pred}})$. The choice of discrepancy measures is usually made rather arbitrarily, based on intuitive notions, and becomes difficult when checking assumptions on the latent parameters of hierarchical models. See for example \cite{gelman1995bayesian} and \cite{sinharay2003posterior}. The latter authors checked the latent Gaussianity in a hierarchical model, but some proposed discrepancy measures did not detect model misfit, even when large misspecification was present. The discrepancy measures $d(\mathbf{y})$ should be \emph{sensitive} to the types of model misspecifications we believe may be present. First, an alternative model is seen as a potential direction of model misspecification. Then, we derive the discrepancy measure most sensitive to small discrepancies induced by it. This idea is used to construct meaningful discrepancy measures $d(\mathbf{y})$ when it is otherwise difficult. 

We consider a data vector $\mathbf{y}$ and a base model $\mathcal{M}_0$ with parameter vector $\mathbf{z}$. This base model is the special case $\eta=0$ of a more flexible model $\mathcal{M}_1$, which contains an extra flexibility parameter $\eta$ ($\mathcal{M}_0$ is nested in $\mathcal{M}_1$). The flexibility parameter $\eta$  can enter the response distribution or any of the priors, and we employ the word ``model" to encompass both the response and the prior distribution: $\pi(\mathbf{y},\mathbf{z}|\eta)$. Moreover, the parameter $\eta$ allows us to quantify deviations from the base model \citep{simpson2017penalising}, and it is linked to assumptions made in $\mathcal{M}_0$; for instance, if $\eta$ is an autocorrelation parameter then $\mathcal{M}_0$ has an independence assumption, not present in $\mathcal{M}_1$.

The goal is to assess the adequacy of model assumptions in $\mathcal{M}_0$ and the impact of these assumptions in the results. This assessment is conducted solely based on the data $\mathbf{y}$ and the $\mathcal{M}_0$ fit, i.e., $\pi(\mathbf{z}|\mathbf{y},\eta=0)$.  The aim is not that of model selection, and the model $\mathcal{M}_1$ is only used to define a plausible model perturbation. This scenario frequently arises in statistical modelling. For instance, the R-INLA software can only fit LGMs \citep{rue2007approximate}, and we would like to assess the reasonability of the latent Gaussian assumption without fitting a latent non-Gaussian model (LnGM) \citep{cabral2022fitting}, which can be prohibitively expensive. Probabilistic programming languages, such as Stan  \citep{carpenter2017stan}, provide greater flexibility in specifying models than R-INLA. Still, there are cases where it is fast and computationally feasible to perform inference on the base model $\mathcal{M}_0$, but it is strenuous or time-consuming to fit the more flexible model $\mathcal{M}_1$. The idea is that instead of spending a lot of energy implementing $\mathcal{M}_1$, we first fit $\mathcal{M}_0$ and then check if minor perturbations in the ``direction" of $\mathcal{M}_1$ do not change~the~answers~substantially. 

There is a vast literature on Bayesian model checking which is too extensive to enumerate. However, some key papers  are \cite{box1980sampling}, \cite{rubin1984bayesianly}, \cite{gelman1996posterior}, and the book \cite{gelman1995bayesian}. \cite{bayarri2007bayesian} discussed model checking in the context of hierarchical models but assumed that the discrepancy measure $d(\mathbf{y})$ is given. An important development is given in \cite{meng1994posterior}, which defined discrepancy measures as functions of both the data and model parameters $d(\mathbf{y},\mathbf{z})$. \cite{meng1994posterior} then chose the conditional likelihood ratio as a discrepancy measure, which in our setting would be \mbox{$d(\mathbf{y},\mathbf{z}) = \sup_\eta \pi(\mathbf{y}|\mathbf{z}, \eta)/\pi(\mathbf{y}|\mathbf{z}, \eta=0)$}.
However, for hierarchical models, if we are checking assumptions on the latent parameters $\mathbf{z}$, then $\pi(\mathbf{y}|\mathbf{z}, \eta) = \pi(\mathbf{y}|\mathbf{z})$, and so $d(\mathbf{y},\mathbf{z}) = 1$. Instead, we propose using the Bayes factor (BF) between $\mathcal{M}_1$ and  $\mathcal{M}_0$~given~$\eta$:
\begin{equation*}
    BF_\eta(\mathbf{y}) = \frac{\pi(\mathbf{y}|\eta)}{\pi(\mathbf{y}|\eta=0)}.
\end{equation*}
 We cannot evaluate $BF_\eta(\mathbf{y})$, because we only have access to the $\mathcal{M}_0$ fit ($\pi(\mathbf{y}|\eta)$ is unknown). However, as we will show in Section~\ref{sect:locrobmea}, under mild regularity conditions, we can perform a Taylor expansion with respect to $\eta$ around the base model to obtain
\begin{equation}\label{eq:LRexpansion}
    BF_\eta(\mathbf{y}) = 1 + s_0(\mathbf{y}) \eta + \frac{s_0(\mathbf{y})^2-I_0(\mathbf{y})}{2}\eta^2 + \smallO(\eta^2),
\end{equation}
where the sensitivity $s_0(\mathbf{y})$ can be computed solely on the data $\mathbf{y}$ and $\pi(\mathbf{z}|\mathbf{y},\eta=0)$. From now on, we will refer to $s_0(\mathbf{y})$ as the BF sensitivity, and we will motivate its use as discrepancy measure $d(\mathbf{y})$  when a more sensible choice is not available.

\begin{remark}\label{remark:1}
According to \cite{weiss1996approach} and \cite{gustafson1996local}, the results obtained from a Bayesian analysis may be called into question by a perturbation to the underlying assumptions only when all of the following criteria are satisfied:
\begin{enumerate}
    \item  The perturbation is plausible a priori.
    \item  The perturbation is supported by the observed data a posteriori.
    \item  The perturbation produces a large change in a posterior quantity of interest.
\end{enumerate}
\end{remark}
Remark \ref{remark:1} motivates a workflow for \emph{model criticism} \citep{box1979robustness}, which we demonstrate for four applications that involve both model checking and subsequent sensitivity analysis:  (a) We construct a perturbation based on the availability of a plausible alternative model $\mathcal{M}_1$; (b) Then we compute the BF sensitivity $s_0(\mathbf{y})$ of this perturbation. By contrasting it with a reference distribution $s_0(\mathbf{y}^{\text{pred}})$, predictive checks detect if $s_0(\mathbf{y})$ is unusually high, which indicates support for the perturbation; (c) Should that occur, we compute the sensitivity of posterior summaries of interest to find which results should be cast into doubt. We will revisit this workflow in Section \ref{sect:workflow_LGM}.




\sloppy  The layout of the paper is as follows. Section \ref{sect:locrobmea} shows how to compute the sensitivity of posterior summaries for small perturbations of the prior or response distribution in hierarchical models. Then the BF sensitivity $s_0(\mathbf{y})$ is shown to be the most sensitive posterior summary. Section \ref{sect:predcheck} studies model checking utilising $s_0(\mathbf{y})$ as the discrepancy measure. To assess the adequacy of the latent Gaussianity assumption in LGMs, in Section \ref{sect:sensmeasures}, we consider the case where $\mathcal{M}_0$ is an LGM and $\mathcal{M}_1$ is an LnGM. Also, as mentioned by \cite{berger2000bayesian}, ``The most important challenge for the field of Bayesian robustness is to increase its impact on statistical practice; indeed, to make it a routine component of applied work". Thus, Section \ref{sect:PPimpl} discusses the implementation of the proposed workflow in probabilistic programming languages and showcases two applications in Stan, and Section \ref{sect:inlaimpl} presents easy-to-use functionalities for R-INLA. Finally, in Section \ref{sect:discussion4}, we discuss the main results and present directions for future work. 

    

\section{Local robustness} \label{sect:locrobmea}

We study the robustness of the base model to perturbations defined in reference to the alternative model. To this end we extend the notion of perturbation of \cite{kass1989approximate} and \cite{weiss1996approach} to also include changes in the response distribution: $\pi(\mathbf{y},\mathbf{z}|\eta)/{\pi(\mathbf{y},\mathbf{z}|\eta=0)}$.  A local approach to robustness focuses on small perturbations, i.e., small values of $\eta$ \citep{berger1994overview}. Therefore, we perform the Taylor expansion:
$$\frac{\pi(\mathbf{y},\mathbf{z}|\eta)}{\pi(\mathbf{y},\mathbf{z}|\eta=0)} =  1 + p(\mathbf{y},\mathbf{z})\eta + \left(p(\mathbf{y},\mathbf{z})^2 + g(\mathbf{y},\mathbf{z}) \right)\frac{\eta^2}{2} +  \smallO(\eta^2),
$$
which requires twice differentiability of $\pi(\mathbf{y},\mathbf{z}|\eta)$ at $\eta=0$ and where,
\begin{align} \label{eq:perturbation}
    p(\mathbf{y},\mathbf{z}) = \lim_{\eta\to0} \frac{d}{d\eta}  \log \pi(\mathbf{y}, \mathbf{z}| \eta), \ \ 
    g(\mathbf{y},\mathbf{z}) = \lim_{\eta\to0} \frac{d^2}{d\eta^2}  \log \pi(\mathbf{y}, \mathbf{z}| \eta).
\end{align}
These limits are taken only from the right (if $\eta \in \mathbb{R}_0^+$) or left if needed. We shall refer to $p(\mathbf{y},\mathbf{z})$ as the ``local perturbation". These are the only derivatives that need to be computed, and they are straightforward to compute because the response and prior distribution are known analytically. Henceforward, $E\{l(\mathbf{y},\mathbf{z})|\eta\}$ and $E\{l(\mathbf{y},\mathbf{z})|\mathbf{y},\eta\}$ are used to denote the prior and posterior expectations of some function $l(\mathbf{y},\mathbf{z})$ conditioned on $\eta$, which are given by $\int l(\mathbf{y},\mathbf{z}) \pi(\mathbf{z}|\eta) d\mathbf{z}$ and~$\int l(\mathbf{y},\mathbf{z}) \pi(\mathbf{z}|\mathbf{y},\eta)d\mathbf{z}$,~respectively.

\subsection{Local robustness of the Bayes factor} \label{sect:marglik}

 For simplicity, we consider here an exponential prior on $\eta$ with rate parameter $\theta_\eta$. This prior is motivated in \cite{cabral2022controlling} for the non-Gaussianity parameter of the non-Gaussian models we will study in Section~\ref{sect:sensmeasures}. The informal but mostly used approach to Bayesian robustness is to fit the flexible model and check if our Bayesian answers change significantly. Assessing the adequacy of the base model or the necessity of using the alternative model would then involve investigating $\pi(\eta|\mathbf{y})$. We can perform a Taylor expansion of $\log \pi(\eta|\mathbf{y})$ around the base model:
\begin{equation*}
   \log \pi(\eta|\mathbf{y}) = \log \pi(\eta = 0 |\mathbf{y}) + (s_0(\mathbf{y}) -\mathbf{\theta}_\eta)\eta  - \frac{I_0(\mathbf{y})}{2} \eta^2 + \smallO(\eta^2),    
\end{equation*}
where 
\begin{equation}\label{eq:derivatives}
  s_0(\mathbf{y}) = \lim_{\eta\to0} \frac{d}{d\eta} \log \pi(\mathbf{y}|\eta), \ \ \ \ 
I_0(\mathbf{y}) = - \lim_{\eta\to0} \frac{d^2}{d\eta^2}\log \pi(\mathbf{y}|\eta),  
\end{equation}
with the limits taken only from the right (if $\eta \in \mathbb{R}_0^+$) or left if needed. The derivatives are taken over the prior predictive distribution $\pi(\mathbf{y}|\eta) = \int \pi(\mathbf{y}, \mathbf{z}|\eta) d\mathbf{z}$. The quantities $s_0(\mathbf{y})$ and $I_0(\mathbf{y})$ are the Fisher's observed score and information of $\eta$, respectively, if there are no latent parameters $\mathbf{z}$. If $s_0(\mathbf{y})<\theta_\eta$ then $\pi(\eta|\mathbf{y})$ has a local maximum at $\eta=0$, and the base model is locally robust in a (very) strict sense. For fixed parameter $\eta$, the Bayes factor (BF) between $\mathcal{M}_1$ and $\mathcal{M}_0$  is given in \eqref{eq:LRexpansion},
and $s_0(\mathbf{y})\eta$ is the increase in the BF for small values of $\eta$. 
 Computing $s_0(\mathbf{y})$ is not straightforward in hierarchical models because $\pi(\mathbf{y}|\eta)$ is seldom available in closed form. The usefulness of the following theorem is that it relates derivatives of the log evidence $\log\pi(\mathbf{y}|\eta)$ with the derivatives of $\log \pi(\mathbf{y},\mathbf{z}|\eta)$ shown in \eqref{eq:perturbation} which are readily available. The proof is given in Appendix \ref{app:proof1}.

\begin{assumption*} \label{assu:assumption}
Assumptions (A1): Let $\pi(\mathbf{y},\mathbf{z}|\eta)$ and  $\partial_\eta \pi(\mathbf{y},\mathbf{z}|\eta)$  be continuous functions of $\eta$ in a neighbourhood of \mbox{$\eta=0$}, denoted as $\zeta(\eta)$. Assume also that \mbox{$p(\mathbf{y},\mathbf{z})\pi(\mathbf{z}|\mathbf{y},\eta)$} are integrable with respect to $\mathbf{z}, \forall \eta \in \zeta(\eta)$. Assumptions (A2): Let $\partial_\eta^2 \pi(\mathbf{y},\mathbf{z}|\eta)$ be a continuous function of $\eta$ in $\zeta(\eta)$. Assume also that  \mbox{$p(\mathbf{y},\mathbf{z})^2\pi(\mathbf{z}|\mathbf{y},\eta)$}  and \mbox{$g(\mathbf{y},\mathbf{z})\pi(\mathbf{z}|\mathbf{y},\eta)$}  are integrable with respect to $\mathbf{z}, \forall \eta \in \zeta(\eta)$.
\end{assumption*}

\begin{theorem} \label{theo:scorefisher}
Under the previous assumptions, equation \eqref{eq:LRexpansion} holds and
\begin{align*}
    s_0(\mathbf{y}) = E\left.\left\{p(\mathbf{y},\mathbf{z}) \right\vert \mathbf{y}, \eta = 0 \right\} \text{and} \ 
    I_0(\mathbf{y}) = -E\left.\left\{g(\mathbf{y},\mathbf{z}) \right\vert \mathbf{y}, \eta = 0  \right\} - V\left.\left\{ p(\mathbf{y},\mathbf{z}) \right\vert \mathbf{y}, \eta = 0  \right\}.
\end{align*}
\end{theorem}

We highlight that $s_0(\mathbf{y})$ and $I_0(\mathbf{y})$ can be approximated by Monte Carlo integration from samples of \mbox{$\pi(\mathbf{z}| \mathbf{y},\eta=0)$} obtained from the base model $\mathcal{M}_0$. Further, as exemplified in Remark \ref{remark:lgm}, we only need to compute the derivative of the hierarchical component being checked, making it feasible to assess specific components of complex hierarchical models directly.

\begin{remark} \label{remark:lgm}
LGMs with joint density \mbox{$\pi(\mathbf{y},\mathbf{w}, \boldsymbol{\gamma}) = \pi(\mathbf{y}|\mathbf{w},\boldsymbol{\gamma})\pi(\mathbf{w}|\boldsymbol{\gamma})\pi(\boldsymbol{\gamma})$} are studied in Section \ref{sect:sensmeasures}, where $\mathbf{w}$ are the latent random effects and $\boldsymbol{\gamma}$ are nuisance parameters. We assess the latent Gaussian prior on $\mathbf{w}|\boldsymbol{\gamma}$ by contrasting it to a non-Gaussian prior defined by $\pi(\mathbf{w}|\boldsymbol{\gamma}, \eta)$, where $\eta$ controls the non-Gaussianity. The local perturbation in \eqref{eq:perturbation} required to compute $s_0$ then simplifies to
\begin{align*}p(\mathbf{y},\mathbf{z}) &= \lim_{\eta\to 0}\frac{d}{d\eta}\log\pi(\mathbf{y}, \mathbf{z}| \eta) = \lim_{\eta\to 0}\left\{\frac{d}{d\eta}\log\pi(\mathbf{y}|\mathbf{w},\boldsymbol{\gamma}) + \frac{d}{d\eta}\log\pi(\mathbf{w}|\boldsymbol{\gamma},\eta) + \frac{d}{d\eta} \log\pi(\boldsymbol{\gamma}) \right\} \\
&= \lim_{\eta\to 0} \frac{d}{d\eta}\log\pi(\mathbf{w}| \boldsymbol{\gamma} , \eta). \nonumber
\end{align*}
\end{remark}
We will compute this local perturbation in Section \ref{sect:sensmeasures} for a specific class of non-Gaussian models.

\subsection{Local robustness of posterior expectations} \label{sect:senspostexp}

We can also evaluate how sensitive a summary of the posterior distribution, given by $E\{l(\mathbf{y},\mathbf{z})|\mathbf{y},\eta\}$, is to small changes in $\eta$, where $l(\mathbf{y},\mathbf{z})$ is a real-valued function of the observed data $\mathbf{y}$ and parameters $\mathbf{z}$. Under mild regularity conditions given in Theorem \ref{theo:sens}, the Taylor expansion around $\eta=0$ yields:
\begin{equation}\label{eq:senslinear}
E\{l(\mathbf{y},\mathbf{z})|\mathbf{y},\eta\} = E\{l(\mathbf{y},\mathbf{z})|\mathbf{y},\eta=0\} + s_{l}\eta + \smallO(\eta),    
\end{equation}
where  $s_{l}$ is the \emph{sensitivity measure}.  If $l(\mathbf{y},\mathbf{z}) = z_1$, where $z_1$ is a regression coefficient, then $s_l$ quantifies the increase in the posterior mean of $z_1$ when we fit $\mathcal{M}_1$ instead of $\mathcal{M}_0$, for small $\eta$. Theorem \ref{theo:sens} follows from \eqref{eq:LRexpansion} and Theorem \ref{theo:scorefisher}, and it has been stated in several forms before in the robustness literature, for instance, in \cite{gustafson1996local}, and more recently in \cite{giordano2018covariances}. As before, these measures can be computed from the base model fit.



\begin{theorem} \label{theo:sens}
Assume (A1) and that $l(\mathbf{y},\mathbf{z})\pi(\mathbf{z}|\mathbf{y},\eta)$ and \mbox{$l(\mathbf{y},\mathbf{z})p(\mathbf{y},\mathbf{z})\pi(\mathbf{z}|\mathbf{y},\eta)$} are integrable with respect to $\mathbf{z}, \forall \eta \in \zeta(\eta)$. Then, \eqref{eq:senslinear} holds and $s_{l} = Cov\left.\left[ l(\mathbf{y},\mathbf{z}), \  p(\mathbf{y},\mathbf{z}) \right\vert \mathbf{y},\eta=0 \right].$

\end{theorem} 

\sloppy The proof is given in Appendix \ref{app:proof2}.  The measure $s_0(\mathbf{y})$ in \eqref{eq:LRexpansion} can be further motivated from the viewpoint of sensitivity analysis as follows. We seek the loss function $l^*(\mathbf{y},\mathbf{z})$ whose posterior expectation has the highest sensitivity $s_{l^*}$. A unique result is obtained when we consider loss functions $l^s$ with unit posterior standard deviation \mbox{($SD\left\{l(\mathbf{y},\mathbf{z})|\mathbf{y},\eta = 0\right\} =1$)}:
\begin{align*}
    l^\star(\mathbf{y},\mathbf{z}) &= \argmax_{l^s(\mathbf{y},\mathbf{z})} \left[Cov\left.\left\{ l^s(\mathbf{y},\mathbf{z}), \  p(\mathbf{y},\mathbf{z}) \right\vert \mathbf{y},\eta=0 \right\} \right] \\ &=  \argmax_{l^s(\mathbf{y},\mathbf{z})} \left[ Corr\left.\left\{ l^s(\mathbf{y},\mathbf{z}), \  p(\mathbf{y},\mathbf{z}) \right\vert \mathbf{y},\eta=0 \right\}  \right] \propto p(\mathbf{y},\mathbf{z}) + \text{const}.
\end{align*}
\sloppy The posterior expectation of the previous loss, $p(\mathbf{y},\mathbf{z})$, can then be expanded to \mbox{$E\left\{p(\mathbf{y},\mathbf{z})|\mathbf{y},\eta\right\} = s_0(\mathbf{y}) + \mathcal{O}(\eta)$}, where $s_0(\mathbf{y})$ is the approximation at the base model $\mathcal{M}_0$. The interpretation is that $s_0(\mathbf{y})$ is the posterior summary that will change the most if we fit the alternative model $\mathcal{M}_1$ (for small $\eta$) instead of $\mathcal{M}_0$. 

One challenge in the local robustness literature is calibrating the sensitivity measures \citep{sivaganesan2000global} to answer: How large is a large deviation from the base results? And does it matter? In the following section, we discuss how posterior predictive checks can inform whether a high value of $s_0(\mathbf{y})$ is unusual or if it can be attributed to the fitted model internal variability  (high values of $s_0(\mathbf{y})$ could occur even if the model assumptions are valid). Nonetheless, sensitivity analysis is useful on its own since it informs which statistical inferences are the most sensitive and to which model assumptions. Thus, it can point to model inadequacies and directions for model improvement.

\section{Model checking for LGMs}\label{sect:predcheck}

LGMs are three-stage hierarchical models containing many statistical models that are widely used \citep{rue2009approximate}. It assumed that the observed data $\mathbf{y}$ of dimension $N$ is conditionally independent given a latent Gaussian field $\mathbf{x}$ and hyperparameters $\boldsymbol{\theta}_1$, and the latent field $\mathbf{x}$ depends on a second set of hyperparameters~$\boldsymbol{\theta}_2$:
\begin{equation}\label{eq:lgm}
    \begin{array}{crl}
\mathrm{Response} & \mathbf{y} \mid \mathbf{x}, \boldsymbol{\theta}_{1} &\sim \prod_{i \in \mathcal{I}} \pi\left(y_{i} \mid x_{i}, \boldsymbol{\theta}_{1}\right) \\ 
\mathrm{Latent \ field} & \mathbf{x}|\boldsymbol{\theta}_{2} &\sim N\left( \mathbf{0}, \ \ \mathbf{Q}(\boldsymbol{\theta}_{2})\right) \\  
\mathrm{Hyperparameters} & \boldsymbol{\theta} &\sim \pi(\boldsymbol{\theta}),
\end{array}
\end{equation}
where $\boldsymbol{\theta}= (\boldsymbol{\theta}_{1}, \boldsymbol{\theta}_{2})$. LGMs are an umbrella class of models that generalize many related variants of ``additive" or ``generalized" linear models \citep{rue2017bayesian}. We 
can interpret $\{x_i, i \in \mathcal{I}\}$ in \eqref{eq:lgm} as the linear predictor $\ell_i$, which depends on a intercept and linear effects contained in $\boldsymbol{\beta}$, and random effects $\mathbf{w}_i$:
\begin{equation}\label{eq:linearpred}
\boldsymbol{\ell} = \mathbf{B}\boldsymbol{\beta} + \mathbf{A}_1\mathbf{w}_1 + \mathbf{A}_2\mathbf{w}_2 + \dotsc,
\end{equation}
where the matrix $\mathbf{B}$ is the design matrix, and $\mathbf{w}_i$ are Gaussian processes with precision matrices $\mathbf{Q}(\boldsymbol{\theta}_i)$ with dimension $n_i \times n_i$. These Gaussian processes could be, for instance, autoregressive processes to model temporal dependence, spatial models, or measurement error models. The matrices $\mathbf{A}_i$ are constant and have dimension $N \times n_i$. All components are assumed to be \emph{a priori} independent of each other, and a Gaussian prior is set on $\boldsymbol{\beta}$. Thus the joint vector $\mathbf{x} = (\boldsymbol{\ell}, \boldsymbol{\beta}, \mathbf{w}_1, \mathbf{w}_2, \dotsc)$ is also normally distributed and corresponds to the latent Gaussian field in the LGM formulation of  \eqref{eq:lgm}.

We focus on checking assumptions on a single random effect component~$\mathbf{w}$ in LGMs. Therefore, the flexibility parameter $\eta$ enters the prior $\pi(\mathbf{w}|\boldsymbol{\theta}_2,\eta)$ and the model $\mathcal{M}_1$ is defined by:
$$
\pi(\mathbf{y},\boldsymbol{\beta},\mathbf{w},\boldsymbol{\theta}|\eta) = \pi(\mathbf{y}|\boldsymbol{\beta},\mathbf{w},\boldsymbol{\theta}_1)\pi(\mathbf{w}|\boldsymbol{\theta}_2,\eta)\pi(\boldsymbol{\beta}, \boldsymbol{\theta}).
$$
Thus, the parameter vector is $\mathbf{z} = (\boldsymbol{\beta}, \mathbf{w}, \boldsymbol{\theta})$, and, as shown in Remark \ref{remark:lgm}, the local perturbation simplifies to $p(\mathbf{y},\mathbf{z}) = p( \mathbf{w},\boldsymbol{\theta}_2) = \lim_{\eta \to 0} \partial_\eta \log(\mathbf{w}|\boldsymbol{\theta}_2,\eta)$. Finally, when checking assumptions on $\mathbf{w}$, we treat the linear effects and the hyperparameters as nuisance parameters $\boldsymbol{\gamma} = (\boldsymbol{\beta}, \boldsymbol{\theta})$. In the literature, $\mathbf{w}$ is also referred to as local latent variables and $\boldsymbol{\gamma}$ as global latent variables. Section \ref{sect:sensmeasures} focuses on a non-Gaussian perturbation on $\mathbf{w}$. However, we also consider in Section \ref{sect:matern_check} perturbing the smoothness of a Gaussian Matérn model on $\mathbf{w}$.

\subsection{Why model checking?}

\cite{box1979robustness} famously said ``all models are wrong, but some are useful". Thus, as in \cite{gelman1995bayesian}, we also recognise beforehand that all models are wrong, and the goal of model checking is not to falsify the base model $\mathcal{M}_0$ but to understand in which ways it does not fit the data.  This viewpoint is elucidated further in \cite{gelman1996posterior}, where the authors present posterior predictive checks (PPCs) as the Bayesian counterpart of the classical goodness-of-fit and discuss their use in judging the fit of a single Bayesian model. 

In PPCs, the most common numerical summary of discrepancy is the probability:
\begin{equation}\label{eq:pv}
p = P(d(\mathbf{y}^{\text{pred}},\boldsymbol{\gamma}) > d(\mathbf{y},\boldsymbol{\gamma}) | \mathbf{y}, \eta = 0),
\end{equation}
where we integrate over the joint density $\pi(\mathbf{y}^{\text{pred}},\boldsymbol{\gamma} | \mathbf{y}, \eta = 0)$, $\mathbf{y}^{\text{pred}}$ is predicted data from the fitted model, and $\boldsymbol{\gamma} = (\boldsymbol{\beta}, \boldsymbol{\theta})$ are nuisance parameters. Low values indicate that the data $\mathbf{y}$ is unusual along the \mbox{$d$-dimension}. For instance, if $p = 0.01$, only 1\% of the predicted data is as extreme in the $d$-dimension as the observed data. It is important to distinguish these upper-tailed probabilities, often called \emph{Bayesian} $p$-values, with \emph{frequentist} $p$-values as defined in \cite{robins2000asymptotic}. Frequentist $p$-values are uniform under the base model. Another important consideration is that given a fixed probability of rejecting the true base model, they should have a high probability of rejecting the alternative model. These criteria are not crucial in PPCs, and it is more important that the $p$-values be interpreted as posterior probabilities of the joint model $\pi(\mathbf{y},\mathbf{y}^{\text{pred}}, \mathbf{z})$ that inform if $\pi(\mathbf{y}^{\text{pred}})$ can capture certain features in the observed data $\mathbf{y}$. PPCs have been criticised for potentially having low power and for using the data twice \citep{bayarri2000p}. Further discussions and varying perspectives on model checking can be found in \cite{bayarri2007bayesian} and \cite{evans2007comment}.

\subsection{Choice of predictive distribution}\label{sect:pred_choice}

 In Bayesian hierarchical modelling, often there is no clear line separating the prior distribution from the likelihood \citep{gelman2003bayesian, spiegelhalter2002bayesian}. When checking the prior distribution of the random effects in LGMs, it makes sense to marginalise them out and use the following density as the likelihood:
 \begin{equation} \label{eq:rep2}
 \pi(\mathbf{y}^{\text{pred}}|\boldsymbol{\beta}, \boldsymbol{\theta}, \eta=0) = \int \pi(\mathbf{y}^{\text{pred}}| \boldsymbol{\beta},\mathbf{w}, \boldsymbol{\theta}_1) \pi(\mathbf{w}|\boldsymbol{\theta}_2, \eta=0) d\mathbf{w}.
 \end{equation}

We then consider in the predictive checks of \eqref{eq:pv}, the predictive density
\begin{equation} \label{eq:rep}
\pi(\mathbf{y}^{\text{pred}},\boldsymbol{\beta}, \boldsymbol{\theta}|\mathbf{y}, \eta=0) =  \pi(\mathbf{y}^{\text{pred}}|\boldsymbol{\beta}, \boldsymbol{\theta}, \eta=0)  \pi(\boldsymbol{\beta}, \boldsymbol{\theta}|\mathbf{y}, \eta=0),
\end{equation}
where we first generate samples $\boldsymbol{\beta}^{(i)}$ and $\boldsymbol{\theta}^{(i)}$ from the posterior $\pi(\boldsymbol{\beta}, \boldsymbol{\theta}|\mathbf{y}, \eta=0)$, then sample $\mathbf{w}^{(i)}$ from the prior $\pi(\mathbf{w}|\boldsymbol{\theta}_2^{(i)}, \eta=0)$, and finally $\mathbf{y}^{\text{pred},(i)}$ is sampled from  $\pi(\mathbf{y}^{\text{pred}}|\boldsymbol{\beta}^{(i)}, \mathbf{w}^{(i)}, \boldsymbol{\theta}_1^{(i)})$.  As explained in the Appendix \ref{sect:app3}, this predictive scheme is associated with several prediction tasks, such as predicting the measurements of an unobserved individual in a longitudinal study or in a territory far away from the location of the observed data in spatial statistics. These prediction tasks involve the prior distribution of the latent random effects (and not their posterior distribution). Similar predictive densities have been used in \cite{sinharay2003posterior} and \cite{bayarri2007bayesian}. Other predictive densities can be useful, such as those considering the posterior distribution on $\mathbf{w}$, and these are discussed in Appendix \ref{sect:app3}. 

\subsection{Choice of discrepancy measure}

\cite{box1980sampling} advocated using the Fisher's score $\partial_\eta\pi(\mathbf{y}|\eta)|_{\eta=0}$ as a measure of a discrepancy for simple 1-level models in the absence of nuisance parameters. An advantage of using measures based on the Fisher's score is that for minor deviations from the base model, they yield the most powerful tests \citep{rao1948large, rao2005score}. Its use was exemplified from a Bayesian perspective to check for serial autocorrelation and the need for power transformation. We extend his ideas by using similar discrepancy measures for hierarchical models that consider the uncertainty of the nuisance parameters. It is useful when the alternative model we are entertaining can generate features in the data relevant to the statistical analysis, which the simpler model cannot generate. One example is given in the spatial application of Section \ref{sect:sensmeasures} where a non-Gaussian prior for $\mathbf{w}$ can lead to predictions with pronounced spikes, unlike the Gaussian prior. 



\begin{figure}[htp]
   \centering
   \includegraphics[width=0.32\linewidth]{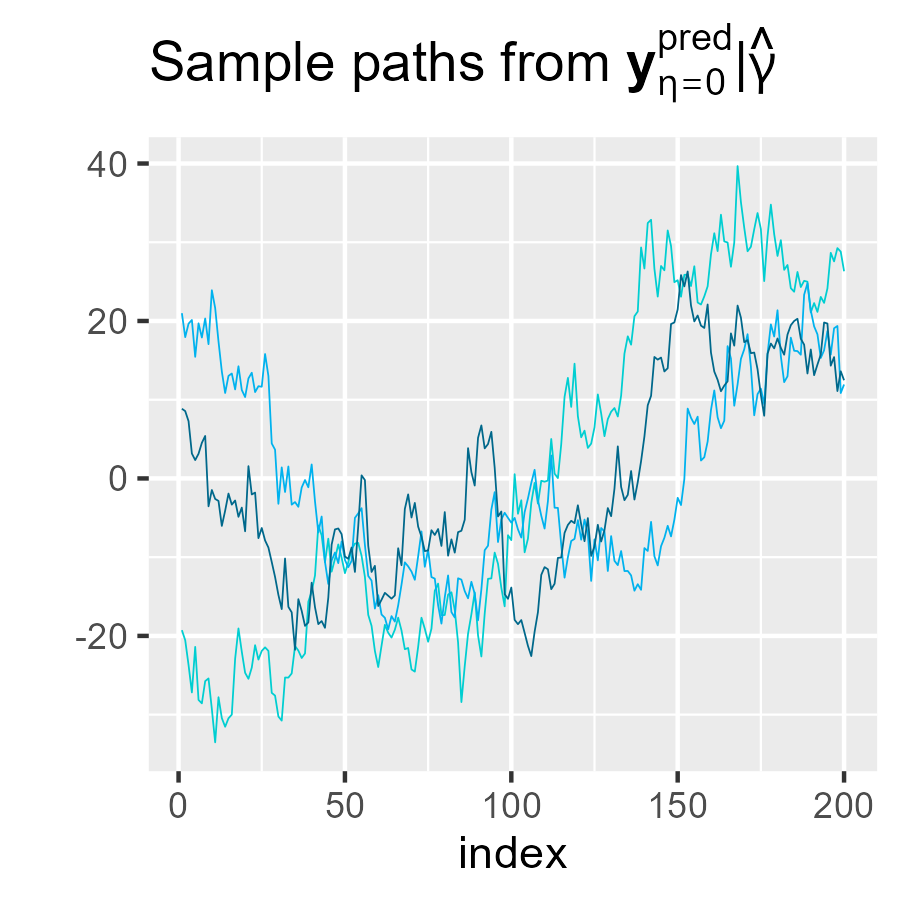}
   \includegraphics[width=0.32\linewidth]{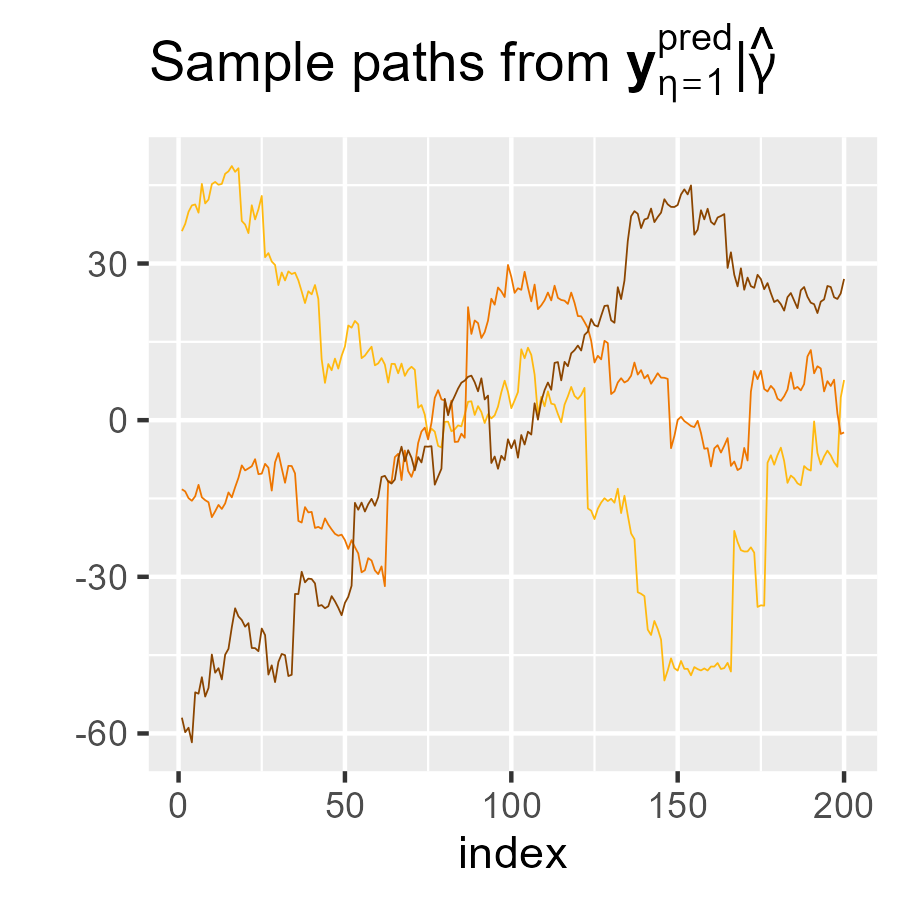}
   \includegraphics[width=0.32\linewidth]{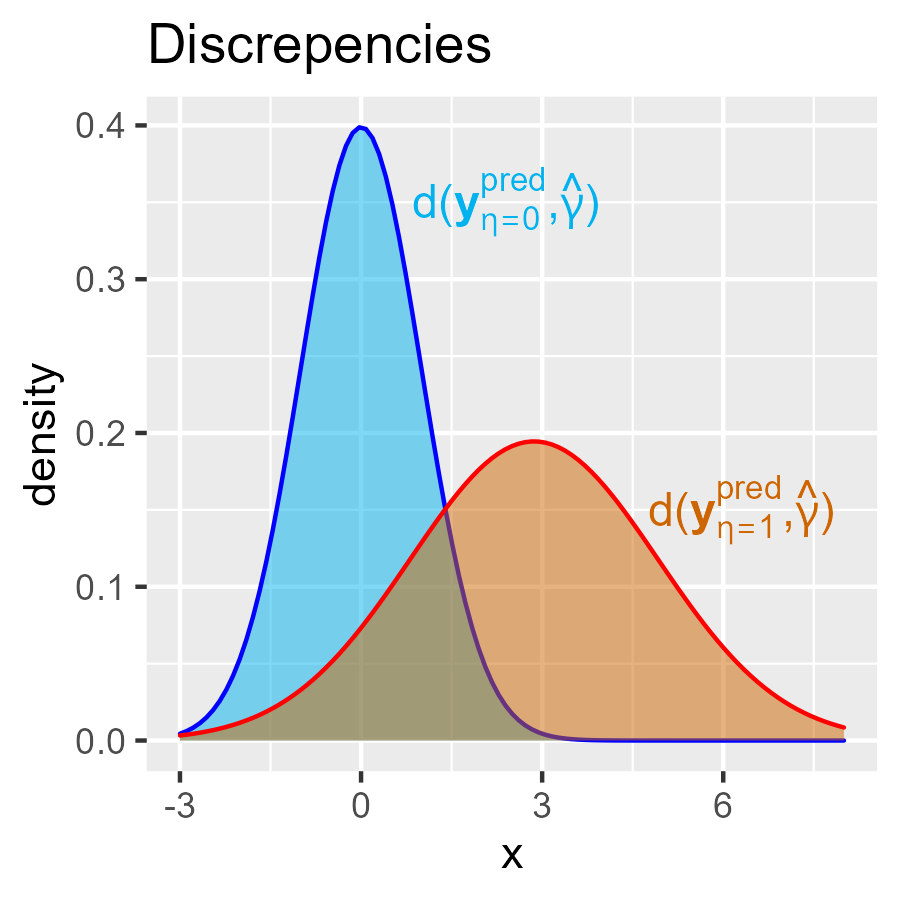}
   \caption{Sample paths generated from $\mathbf{y}^{\text{pred}}|\hat{\boldsymbol{\gamma}},\eta=0$ (left) and $\mathbf{y}^{\text{pred}}|\hat{\boldsymbol{\gamma}},\eta=1$ (center), and density plots of the resulting discrepancy distributions $d(\cdot, \cdot)$ (right). We chose $d(\cdot, \cdot)= s_0(\cdot, \cdot)$.}
  \label{fig:discrepencies} 
\end{figure}

We provide here an argument for using the BF sensitivity $s_0(\mathbf{y}, \boldsymbol{\gamma}) = E\{p(\mathbf{y},\mathbf{z})|\mathbf{y},\boldsymbol{\gamma}, \eta = 0\}$ as a discrepancy measure which directly relates to the Bayesian model check in \eqref{eq:pv} and the predictive distribution of Section~\ref{sect:pred_choice}. 
We consider the predictive distribution $ \pi(\mathbf{y}^{\text{pred}}|\boldsymbol{\gamma}, \eta)$, where, similarly to \eqref{eq:rep2}, we integrate over the prior $\pi(\mathbf{w}|\boldsymbol{\gamma},\eta)$ of the latent random effects. In Figure \ref{fig:discrepencies} we show sample paths of $\mathbf{y}^{\text{pred}}|\boldsymbol{\gamma},\eta=0$ (base model $\mathcal{M}_0$) and $\mathbf{y}^{\text{pred}}|\boldsymbol{\gamma},\eta=1$ (alternative model $\mathcal{M}_1$). The generative model is the latent random walk of order 1 (RW1) model of Section \ref{app:sim_study} for some fixed nuisance parameters $\hat{\boldsymbol{\gamma}}$. The flexibility parameter $\eta$ controls the excess kurtosis of the driving noise, and larger values lead to the appearance of sudden jumps in the sample paths.

Several criteria can be used to establish the optimality of $s_0(\cdot, \cdot)$. If the distribution of $d(\mathbf{y}^{\text{pred}}_{\eta}, \boldsymbol{\gamma})$ barely changes for increasing $\eta$, then the model checking procedure is not sensitive to model misspecification or, in other words, it fails at detecting discrepancies in the direction of the alternative model $\mathcal{M}_1$. This concern is closely linked to the model checking procedure \emph{severity} \citep{mayo1996error, mayo2006severe, gelman2013philosophy}. A model undergoes a severe check when it successfully passes a probe with a high probability of detecting an error if it is present (severity is closely related, but not identical, to power; see \cite{mayo2006severe}). 

\begin{figure}[htp]
   \centering
   \includegraphics[width=0.32\linewidth]{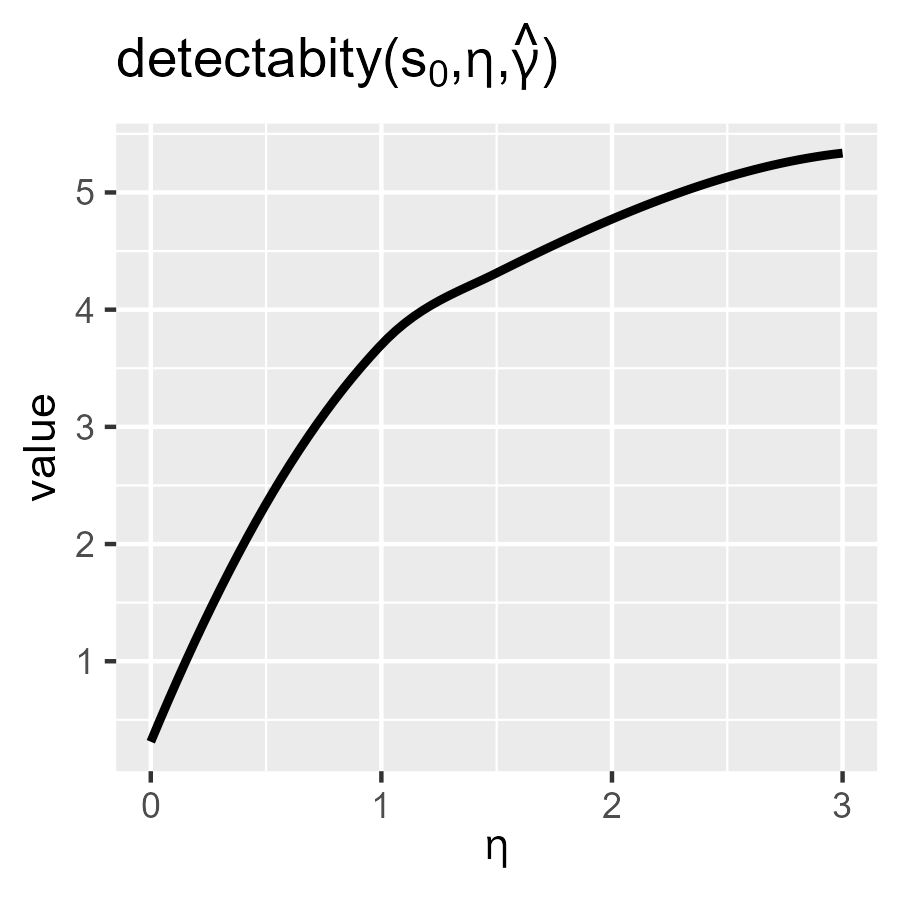}
   \includegraphics[width=0.32\linewidth]{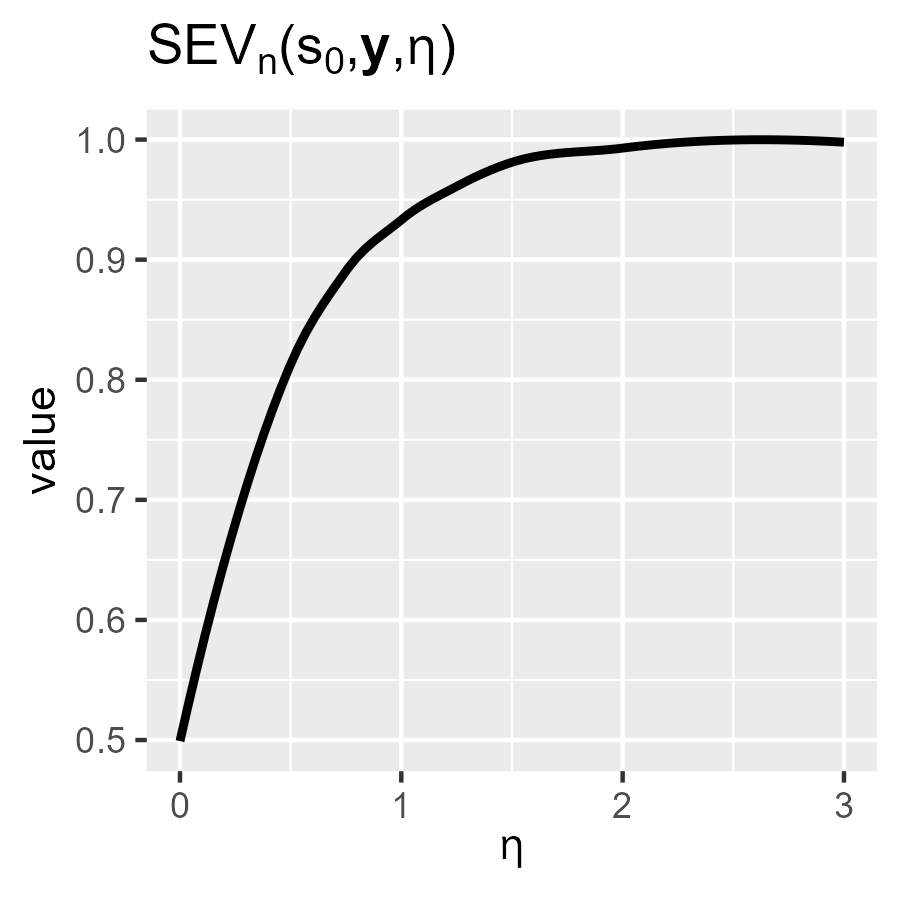}
   \caption{Detectability index $detectability(d,\eta,\boldsymbol{\gamma})$ (left), and severity $SEV_n(d, \mathbf{y}, \eta)$ (right) for varying $\eta$ and for $d_n(\mathbf{y}, \boldsymbol{\gamma})=0$.  We consider the setting of Figure \ref{fig:discrepencies} and chose $d(\cdot, \cdot)= s_0(\cdot, \cdot)$.}
  \label{fig:detectandp} 
\end{figure}

We assume, without loss of generality, that the discrepancy $d(\mathbf{y}^{\text{pred}},\boldsymbol{\gamma}, \eta=0)$ has mean 0 and variance 1. Intuitively, we want $E[d(\mathbf{y}^{\text{pred}}, \boldsymbol{\gamma})|\boldsymbol{\gamma}, \eta]$ to increase with $\eta$ and $SD[d(\mathbf{y}^{\text{pred}}_{\eta},\boldsymbol{\gamma})|\boldsymbol{\gamma}]$ to be small. This leads us to define detectability as \mbox{$
detectability(d,\eta,\boldsymbol{\gamma}) = E[d(\mathbf{y}^\text{rep},\boldsymbol{\gamma})|\boldsymbol{\gamma},\eta] \ / \ SD[d(\mathbf{y}^\text{rep},\boldsymbol{\gamma})|\boldsymbol{\gamma}, \eta]$}, shown in Figure \ref{fig:detectandp}, a quantity we aim to maximise. Lemma \ref{prop:detect} is a consequence of \eqref{eq:LRexpansion} and Theorem \ref{theo:scorefisher} and it establishes the optimality of $s_0(\cdot, \cdot)$ under the previous criteria when $\eta$ is small (proof given in Appendix \ref{app:detect}). 

 
 \begin{lemma}\label{prop:detect}
Assume (A1) and let $\pi(\mathbf{y}|\boldsymbol{\gamma},\eta)$ and $\partial_\eta \pi(\mathbf{y}|\boldsymbol{\gamma},\eta)$ be continuous functions w.r.t. $\eta$ in $\zeta(\eta)$. Also, let the functions $d(\mathbf{y}^{\text{pred}}, \boldsymbol{\gamma}) \pi(\mathbf{y}^{\text{pred}}|\boldsymbol{\gamma},\eta)$ and $d(\mathbf{y}^{\text{pred}}, \boldsymbol{\gamma}) s_0(\mathbf{y}^{\text{pred}}, \boldsymbol{\gamma}) \pi(\mathbf{y}^{\text{pred}}|\boldsymbol{\gamma},\eta)$ be integrable w.r.t. $\mathbf{y}^{\text{pred}}, \forall \eta \in \zeta(\eta)$. The slope of $detectability(d,\eta,\boldsymbol{\gamma})$  near the origin $\eta=0$ is maximised when we choose the discrepancy measure $d(\cdot, \cdot) = s_0(\cdot,\cdot)$:
\begin{align*}
 &\argmax_{d(\cdot,\cdot)} \left\{ \lim_{\eta \to 0} \frac{d}{d\eta} detectability(d,\eta,\boldsymbol{\gamma}) \right\} = a s_0(\cdot, \cdot) + b, \ a,b \in \mathbb{R}, 
\end{align*}
\end{lemma}

It is also possible to motivate $s_0(\cdot, \cdot)$ for the common summary of discrepancy in \eqref{eq:pv}, for instance, by making use of the model check's severity:
$$
SEV_n(d, \mathbf{y}, \eta) = P\left\{ d_n(\mathbf{y}^{\text{pred}}, \boldsymbol{\gamma}) > d_n(\mathbf{y}, \boldsymbol{\gamma}) | \mathbf{y}, \eta\right\},
$$
where the subscript indicates $n$ observed data. The severity can be visualized in Figure \ref{fig:detectandp}.  Theorem \ref{theo:s0best} demonstrates that the discrepancy measure $s_0(\cdot, \cdot)$ produces the model checking procedure with the highest severity for small $\eta$ and any observed data $\mathbf{y}$. The theorem follows from Lemma \ref{prop:detect} and the proof is given in Appendix \ref{app:s0best}. 
 We restrict the result to asymptotically normal discrepancy measures: \mbox{$\lim_{n\to\infty}d_n(\mathbf{y}^{\text{pred}}, \boldsymbol{\gamma}) \sim N(\cdot,\cdot)$}. While the Bayes factor would be the optimal choice for all values of $\eta$, our inability to fit the alternative model $\mathcal{M}_1$ limits us to using the BF sensitivity.

\begin{theorem}\label{theo:s0best}
Under the regularity conditions given in Appendix \ref{app:s0best}, the slope near $\eta=0$ of $SEV_n(d, \mathbf{y}, \eta)$ is maximised when $d(\cdot, \cdot) = s_0(\cdot,\cdot)$, when $n \to \infty$ and for any observed data $\mathbf{y}$:
$$
 \argmax_{d(\cdot, \cdot)}  \left[ \lim_{n \to 0} \lim_{\eta \to 0} \frac{d}{d\eta} SEV_n(d, \mathbf{y}, \eta) \right] = a s_0(\cdot, \cdot)+ b, \ a,b \in \mathbb{R}.
$$
\end{theorem}

\subsection{Approximating the reference distribution}\label{sect:ref_dist}

\sloppy  We choose the BF sensitivity $s_0$ as the discrepancy measure and consider the problem of finding the reference distribution $s_0(\mathbf{y}^{\text{pred}}, \hat{\boldsymbol{\gamma}})|\eta=0$, which is required to compute \eqref{eq:pv}. Approximating \eqref{eq:pv} by Monte Carlo requires obtaining samples from \mbox{$\pi(\mathbf{y}^{\text{pred}},\boldsymbol{\gamma} | \mathbf{y}, \eta = 0)$} of the base model fit. Then, for each sample $(\mathbf{y}^{\text{pred},(i)},\boldsymbol{\gamma}^{(i)})$ we need to compute \mbox{$s_0(\mathbf{y}^{\text{pred},(i)}, \boldsymbol{\gamma}^{(i)})= E\{p(\mathbf{w},\boldsymbol{\theta_2})| \mathbf{y}^{\text{pred},(i)}, \boldsymbol{\gamma}^{(i)}, \eta = 0\}$}, which would need to be approximated by MCMC (or INLA) if no closed-form expression is available.

A faster alternative consists in fixing the nuisance parameters at $\hat{\boldsymbol{\gamma}}$ (e.g., at their posterior mode) and considering the Gaussian approximation for the reference distribution: \mbox{$s_0(\mathbf{y}^{\text{pred}}, \hat{\boldsymbol{\gamma}})|\eta=0 \sim N[0, V\{ s_0(\mathbf{y}^{\text{pred}}, \hat{\boldsymbol{\gamma}})| \eta = 0 \} ]$}, where the variance is found analytically. In this case, \eqref{eq:pv} simplifies to
$p = \Phi[-s_0(\mathbf{y}, \hat{\boldsymbol{\gamma}})/SD\{s_0(\mathbf{y}^{\text{pred}}, \hat{\boldsymbol{\gamma}})| \eta = 0\}]$.  However, deriving the variance of the reference distribution analytically may be cumbersome for some models. As discussed in Appendix \ref{sect:refapprox}, another possibility is to approximate the variance of $s_0(\mathbf{y}^{\text{pred}}, \hat{\boldsymbol{\gamma}})|\eta=0$ by the measure $I_0(\mathbf{y}, \hat{\boldsymbol{\gamma}})$ given in Theorem \ref{theo:scorefisher} which can be computed by Monte Carlo from the $\mathcal{M}_0$ fit.


\subsection{Workflow for model criticism in LGMs}\label{sect:workflow_LGM}


Based on Remark \ref{remark:1}, we present in Algorithm \ref{alg:workflow} the workflow we use to conduct criticism on the latent random effects $\mathbf{w}$ of LGMs, which involves both model checking and subsequent sensitivity analysis. The workflow can also be applied to study the adequacy of assumptions in the response distribution or on the other latent parameters. It can also be extended to consider multiple model extensions, and an application is given in Section \ref{sect:orto}.

The first line of Algorithm \ref{alg:workflow} involves defining the local perturbation $p(\mathbf{y}, \mathbf{w},\boldsymbol{\gamma})$. This can be done analytically for specific models $\mathcal{M}_0$ and $\mathcal{M}_1$, as described in Section \ref{sect:BF_sens_compute}, or numerically, as explained in Section \ref{sect:PPimpl}, which is more natural in probabilistic programming languages as it avoids the need for model-specific derivations. The comparison of the reference distribution with the observed BF sensitivity in line 4 of Algorithm \ref{alg:workflow} can be done by computing the upper-tailed probability $p$ in \eqref{eq:pv}. However, as in \cite{gelman2003bayesian}, we favour the use of diagnostic plots such as a scatter plot of $s_0(\mathbf{y}^{\text{pred}},\boldsymbol{\gamma})$ against $s_0(\mathbf{y},\boldsymbol{\gamma})$ (see Figure \ref{fig:crime_result}) or the density plot of $s_0(\mathbf{y}^{\text{pred}},\hat{\boldsymbol{\gamma}})$  compared to $s_0(\mathbf{y},\hat{\boldsymbol{\gamma}})$ (see Figure \ref{fig:ortho_score}). Model misfit is revealed if the observed BF sensitivity is in the tails of reference distribution, which prompts a sensitivity analysis to investigate the most sensitive answers.

\begin{algorithm}[h]
\caption{Workflow for model criticism in LGMs}\label{alg:workflow}
\begin{algorithmic}[1]

\Require  \Statex $\pi(\mathbf{w},\boldsymbol{\gamma}|\mathbf{y},\eta = 0)$ \Comment{$\mathcal{M}_0$ fit (samples or a deterministic result as in INLA)}
          \Statex $\log \pi(\mathbf{y},\mathbf{w},\boldsymbol{\gamma}|\eta)$ \Comment{$\mathcal{M}_1$ log density}
          \Statex $l_i(\mathbf{y}, \mathbf{w},\boldsymbol{\gamma}), \ \ i = 1, \dots, m$ \Comment{loss functions of interest}
         \Statex 

\State Define $p(\mathbf{y}, \mathbf{w},\boldsymbol{\gamma}) \gets \lim_{\eta\to0} \partial_\eta \log \pi(\mathbf{y}, \mathbf{w}, \boldsymbol{\gamma} |\eta)$ \Comment{Local perturbation}
\Statex 
\NoNumber{\emph{// Model checking}}
\State Define $\pi(\mathbf{y}^{\text{pred}},\boldsymbol{\gamma}|\mathbf{y}, \eta=0)$ \Comment{Replicated data from the fitted $\mathcal{M}_0$ model} 
\State Define $s_0(\cdot, \boldsymbol{\gamma}) = E[p(\cdot, \mathbf{w},\boldsymbol{\gamma}) | \cdot]$ \Comment{BF sensitivity} 

\State Compare $s_0(\mathbf{y}^{\text{pred}},\boldsymbol{\gamma})$ with $s_0(\mathbf{y},\boldsymbol{\gamma})$ under the distribution $\pi(\mathbf{y}^{\text{pred}},\boldsymbol{\gamma}|\mathbf{y}, \eta = 0)$ 

\Statex 
\NoNumber{\emph{// Sensitivity analysis}}
\If{model check reveals important misfit}
\State Compute $s_{l_i} = Cov\left.\left[ l_i(\mathbf{y},\mathbf{w},\boldsymbol{\gamma}), \  p(\mathbf{y}, \mathbf{w},\boldsymbol{\gamma}) \right\vert \mathbf{y},\eta=0 \right]$ \Comment{Sensitivity measures}
\State Rank $s_{l_i}$ \Comment{Determine most sensitive answers; See Figure \ref{fig:pressd2} (right) and Table \ref{table:crime_mean}}
\EndIf

\end{algorithmic}
\end{algorithm}

Model checking and sensitivity analysis go hand-in-hand. The former informs if the fitted model fails to capture important features in the data. However, the model may still be useful for certain purposes even if it does not fit some aspects of the data \citep{gelman2005multiple}, and sensitivity analysis can inform if the perturbation to the assumptions has a large impact on posterior inferences of interest. For example, if one is interested in estimating the intercept in a large data problem, then an inadequate Gaussian assumption on the random effects can be of little practical importance.

\section{Robustness and model checking for latent non-Gaussianity} \label{sect:sensmeasures}

The latent random effects $\mathbf{w}$ are used to model spatial and temporal dependence in LGMs, and the Gaussian distribution is the canonical choice of distribution for $\mathbf{w}$. Checking this Gaussian prior is of particular interest, since, as shown in Appendix \ref{sect:app3}, if the Gaussian prior is a bad \emph{generative prior} \citep{gelman2017prior}, in that it fails to replicate certain non-Gaussian features that may be found in the data, such as sudden jumps or spikes, then the posterior distribution will also fail to generate those features in the future predictions (or at unobserved locations in spatial statistics).

The Gaussian prior assumption imposes constraints on the tail behaviour of the posterior distribution. For instance, the posterior distribution satisfies $\pi(\mathbf{w}|\mathbf{y}) \leq C \pi(\mathbf{w})$, where $C$ is a constant and $\mathbf{y}$ is the observed data \citep{chiuchiolo2022extended}. Hence, the posterior marginals of $\mathbf{w}$ are constrained to have tails that behave like Gaussian tails or are lighter. However, there are many applied problems where leptokurtic behaviour is present \citep{bolin2014spatial,wallin2015geostatistical} and in these problems, an LnGM that assumes a leptokurtic prior on $\mathbf{w}$ can lead to improved predictions~\citep{cabral2022controlling}. In this section, we perturb the Gaussian prior assumption on $\mathbf{w}$ and demonstrate how to carry out the workflow in Section \ref{sect:workflow_LGM}. We start by defining the alternative model $\mathcal{M}_1$.


\subsection{Latent non-Gaussian models (LnGMs)} \label{sect:lngm}


 We focus on LGMs with a single random effects component $\mathbf{w}$ with mean $\mathbf{0}$ and precision matrix $\mathbf{Q}(\boldsymbol{\theta}_2) = \mathbf{D}(\boldsymbol{\theta}_2)^T\text{diag}(\mathbf{h})^{-1}\mathbf{D}(\boldsymbol{\theta}_2)$, where $\mathbf{h}$ is a known vector. We consider LnGMs \citep{cabral2022fitting} as flexible extensions of LGMs with the following prior model on the latent random effect: $\mathbf{D}(\boldsymbol{\theta}_2)\mathbf{w} = \mathbf{\Lambda}(\eta)$, where $\mathbf{\Lambda}(\eta)$ is a vector of independent random variables such that $\mathbf{\Lambda}(0) \sim N\{0, \text{diag}(\mathbf{h})\}$ results in the simpler Gaussian model, and larger values of $\eta$ are related to increasing levels of leptokurtosis. The multivariate transformation method then yields 
\begin{equation}\label{eq:multtrans}
    \pi(\mathbf{w}|\boldsymbol{\theta}_2, \eta)= |\mathbf{D}(\boldsymbol{\theta}_2)|\pi_{\mathbf{\Lambda}}\{\mathbf{D}(\boldsymbol{\theta}_2)\mathbf{w} | \eta\} = |\mathbf{D(\boldsymbol{\theta}_2)}|\prod_{i=1}^n\pi_{\Lambda_i}[\{\mathbf{D(\boldsymbol{\theta}_2)}\mathbf{w}\}_i | \eta],
\end{equation}
where $\pi_{\Lambda_i}(w|\eta)$ is the density function of a univariate non-Gaussian distribution, e.g., the symmetric normal inverse-Gaussian (NIG) and generalized asymmetric Laplace (GAL) distributions in \cite{cabral2022controlling}. Their densities are:
$$\pi_{\Lambda_i}^{\text{NIG}}(w|\eta) \propto \frac{K_1\left(\sqrt{\eta  w^2+h_i^2}/\eta\right)}{\sqrt{\left(h_i^2+\eta  w^2\right)}}, \ \ \  \pi_{\Lambda_i}^{\text{GAL}}(w|\eta) \propto |w| ^{\eta  h_i-\frac{1}{2}} K_{\eta  h_i-\frac{1}{2}}\left(\sqrt{2 \eta } |w| \right),$$
where $K_a(x)$ is the modified Bessel function of the second kind of order $a$. The vector $\mathbf{h}$ is necessary when approximating continuous models. For example, it contains the distances between observed time points in stochastic processes, or the area of the basis functions in the SPDE approach (see Section \ref{sect:pressure} and \cite{lindgren2011explicit}). For models defined in discrete space, $\mathbf{h}$ is a vector of ones. This model extension for the random effects can be used for time series models, models for graphical and areal data, and Matérn processes, which are useful in several applications, such as geostatistics and spatial point processes \citep{cabral2022fitting}.


\subsection{BF sensitivity measure} \label{sect:BF_sens_compute}
We consider the previous non-Gaussian model extension for the latent random effect $\mathbf{w}$ in LGMs. To compute the sensitivity measures in Theorems \ref{theo:scorefisher} and \ref{theo:sens}, we only need to calculate the local perturbation in \eqref{eq:perturbation}, which from \eqref{eq:multtrans} is given by
\begin{align}\label{eq:ngpert}
    p(\mathbf{w},\boldsymbol{\theta}_2) = \sum_i p_i[\{\mathbf{D}(\boldsymbol{\theta}_2)\mathbf{w}\}_i], \  \ \  p_i(w) = \lim_{\eta \to 0} \frac{d}{d\eta} \log \pi_{\Lambda_i}(w|\eta) = \frac{(w^2 - 3 h_i)^2 - 6 h_i^2}{8 h_i^3},  
\end{align}
when the alternative model considers random effects driven by either NIG or GAL noise. We note that the BF sensitivity $s_0(\mathbf{y})$ in \eqref{eq:LRexpansion} is then given by $s_0(\mathbf{y}) = \sum_{i=1}^n d_i(\mathbf{y})$, where 
\begin{align}\label{eq:scorebayes}
d_i(\mathbf{y}) = E\left.\left[\frac{(\{\mathbf{D}(\boldsymbol{\theta}_2)\mathbf{w}\}_i^2-3h_i)^2-6h_i^2}{8h_i^3} \right\vert \mathbf{y}, \eta=0 \right].
\end{align} 
An informal check of plotting each element $d_i$ can be useful, as it illustrates the increase in the Bayes factor when the noise element $\Lambda_i(0)$ is made slightly non-Gaussian (see Section~\ref{sect:marglik}). For spatial and temporal models, each index $i$ is associated with a particular time point or location in space, so such plots can indicate the most sensitive points in time or space (see Figure~\ref{fig:pressd1}). This observation highlights the potential benefits of exploring the sensitivity of the Bayes factor to deviations from the Gaussian assumption.

\subsection{LGMs with Gaussian response}\label{sect:LGMnormal}

The examples and applications in this paper are all part of the general model with  Gaussian response and a Gaussian prior on the latent random effects $\mathbf{w}$:
\begin{equation} \label{eq:model}
\mathbf{y}|\boldsymbol{\beta}, \mathbf{w}, \tau_{\boldsymbol{\epsilon}}  \sim N(\mathbf{B}\boldsymbol{\beta} + \mathbf{A}\mathbf{w}, \tau_{\boldsymbol{\epsilon}}^{-1}\mathbf{I}), \ \ \ \mathbf{w}|\boldsymbol{\theta}_2\sim  N[\mathbf{0}, \{\mathbf{D}(\boldsymbol{\theta}_2)^T\text{diag}(\mathbf{h})^{-1}\mathbf{D}(\boldsymbol{\theta}_2)\}^{-1} ].
\end{equation}

We derive next the BF sensitivity $s_0(\mathbf{y},\boldsymbol{\gamma})$ and the mean and variance of the reference distribution $s_0(\mathbf{y}^{\text{pred}},\boldsymbol{\gamma})|\eta=0$ for the previous model. We consider fixed nuisance parameters $\boldsymbol{\gamma}=(\boldsymbol{\beta}, \tau_\epsilon, \boldsymbol{\theta}_2)$, although we can also integrate the results over the posterior distribution of $\boldsymbol{\gamma}$. The results make the model checking of Section \ref{sect:predcheck} both straightforward and computationally efficient. In the case of LGMs with a non-normal response, an asymptotic approximation of the reference distribution is given in Appendix \ref{sect:refapprox}. 

For the LGM in \eqref{eq:model}, the latent random effects  $\mathbf{w}$ have a full conditional with precision matrix \mbox{$\mathbf{Q}_{\mathbf{w}} = \tau_{\boldsymbol{\epsilon}} \mathbf{A}^T\mathbf{A}  + \mathbf{D}^T\text{diag}(\mathbf{h})^{-1}\mathbf{D}$}, where  $\mathbf{D} = \mathbf{D}(\boldsymbol{\theta}_2)$ and $\mathbf{Q}_{\mathbf{w}} = \mathbf{Q}_{\mathbf{w}}(\tau_{\boldsymbol{\epsilon}}, \boldsymbol{\theta}_2)$. The distribution is:
\begin{equation}\label{eq:posterior_w}
    \mathbf{w}|\mathbf{y}, \boldsymbol{\beta}, \tau_{\boldsymbol{\epsilon}}, \boldsymbol{\theta}_2 \sim N\left\{ \boldsymbol{\mu}_\mathbf{w} = \tau_{\boldsymbol{\epsilon}}  \mathbf{Q}_{\mathbf{w}}^{-1} \mathbf{A}^T(\mathbf{y}-\mathbf{B}\boldsymbol{\beta}), \mathbf{Q}_{\mathbf{w}}^{-1} \right\}.
\end{equation}
The sensitivity measures $d_i$ are
\begin{equation}\label{eq:sensitivityconditional}
d_i(\mathbf{y}, \boldsymbol{\gamma}) =  \frac{ b_i^4 + 3\Gamma_{ii}^2 - 6 b_i^2\Gamma_{ii}}{8 h_i^3}, \ \ \mathbf{b} = \mathbf{D} \boldsymbol{\mu}_\mathbf{w}, \ \ \mathbf{\Gamma} = -\mathbf{D} \mathbf{Q}_{\mathbf{w}}^{-1} \mathbf{D}^T + \text{diag}(\mathbf{h}),
\end{equation}
and finally $s_0(\mathbf{y}, \boldsymbol{\gamma}) = \sum_i d_i(\mathbf{y}, \boldsymbol{\gamma})$. To perform the model checking of Section \ref{sect:predcheck}, we also need to compare $s_0(\mathbf{y}, \boldsymbol{\gamma})$ with the reference distribution $s_0(\mathbf{y}^{\text{pred}},\boldsymbol{\gamma})|\eta=0$. The reference distribution can be approximated to a Gaussian distribution with mean and variance given in Proposition \ref{prop:variance_approx} (proof given in Appendix \ref{app:variance_approx}). An example of this comparison is shown in Figure \ref{fig:ortho_score}.

\begin{prop}\label{prop:variance_approx}
The mean and variance of the reference distribution for replicated data from \eqref{eq:model} with fixed $\boldsymbol{\gamma}$ is
$E\{s_0(\mathbf{y}^{\text{pred}},\boldsymbol{\gamma})|\eta=0\} = 0$ and  $V\{s_0(\mathbf{y}^{\text{pred}},\boldsymbol{\gamma})|\eta=0\} =  \sum_{i,j} 3 \Gamma_{ij}^4/(8 h_i^3 h_j^3).$
\end{prop}

\subsection{Example}\label{sect:pressure}

\begin{figure}[htp]
    \centering
    \includegraphics[width=0.49\linewidth, height = 0.4\linewidth]{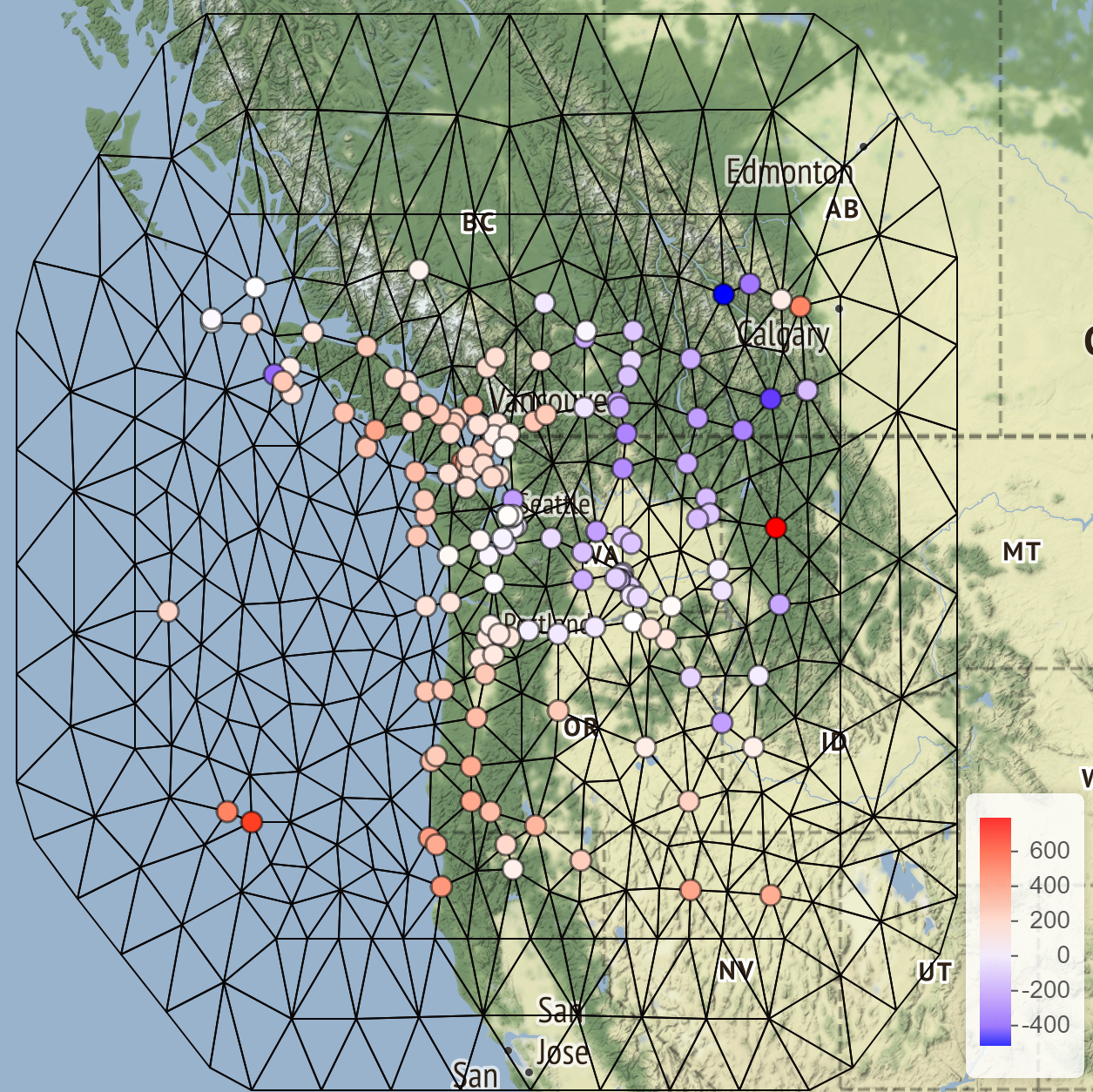}      \includegraphics[width=0.49\linewidth, height = 0.4\linewidth]{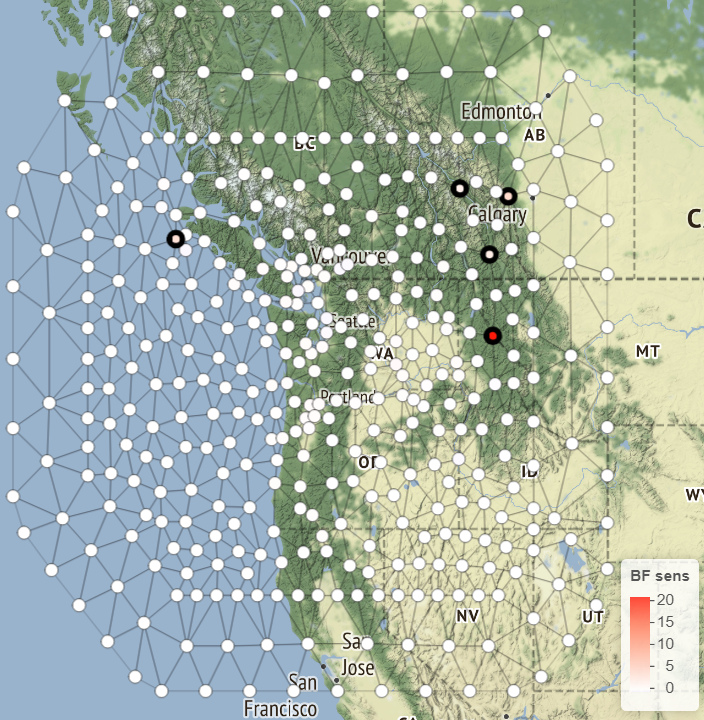}
  \caption{Measurements of pressure with the FEM mesh on the background (left). Spatial plot of $d_i$'s revealing the location of the most sensitive nodes (right).}
  \label{fig:pressd1}
\end{figure}

Figure \ref{fig:pressd1} shows pressure measurements at 157 different locations in the North American Pacific Northwest, where the sample mean was subtracted from the data. \cite{bolin2020multivariate} considered an LnGM for the pressure data which appeared to have some localized spikes and short-range variations that were not well captured by an LGM. This dataset was further analysed in \cite{cabral2022controlling}, where a leave-one-out cross-validation study gave substantial preference to the LnGM. 

We fitted the model $\mathbf{y} \sim N(\mathbf{A}\mathbf{w}, \sigma_\epsilon^2\mathbf{I})$, where the random effects $\mathbf{w}$ approximate a Gaussian Matérn random field obtained by the stochastic partial differential equation  approach \citep{lindgren2011explicit}.  It begins by expressing the underlying continuous process $X(\mathbf{s})$ as a weighted sum of basis functions, $X(\mathbf{s}) = \sum_{i=1}^{n}w_i\psi_i(\mathbf{s})$ where $\psi_i(\mathbf{s})$ are basis functions associated with the mesh in Figure~\ref{fig:pressd1}. The weights $\mathbf{w}=(w_1,\dotsc,w_n)$ follow the system \mbox{$\mathbf{D}\mathbf{w}=\boldsymbol{\Lambda}(0)$}, for a particular matrix $\mathbf{D}$ shown in \cite{lindgren2011explicit}. Finally, the matrix $\mathbf{A}$ is the projector matrix with elements $A_{ij}=\psi_i(\mathbf{s}_j)$, where $\mathbf{s}_j$ is the~location~of~the $j$th~observation. 

\begin{figure}[htp]
   \centering
   \includegraphics[width=0.5\linewidth]{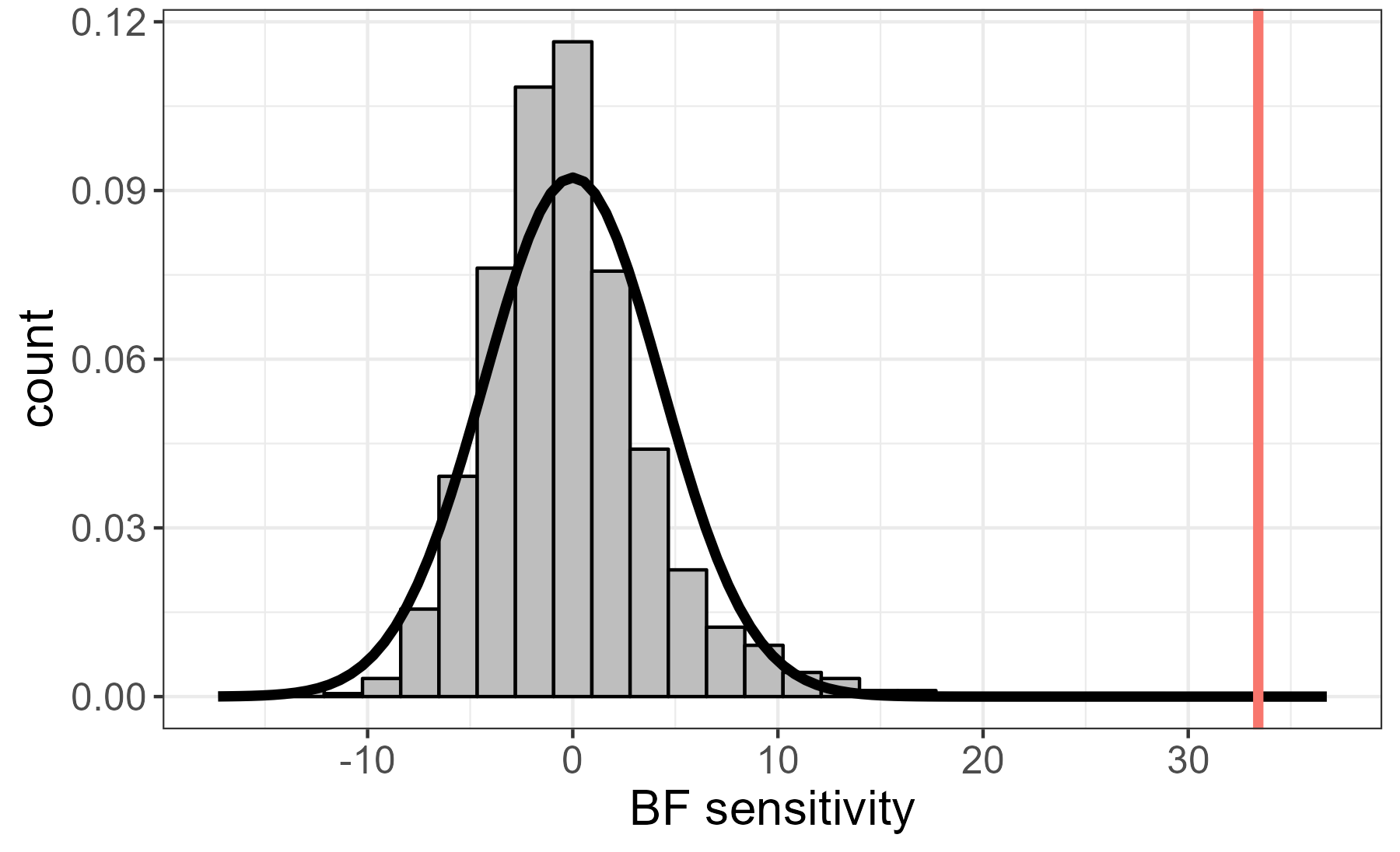}
   \includegraphics[width=0.46\linewidth]{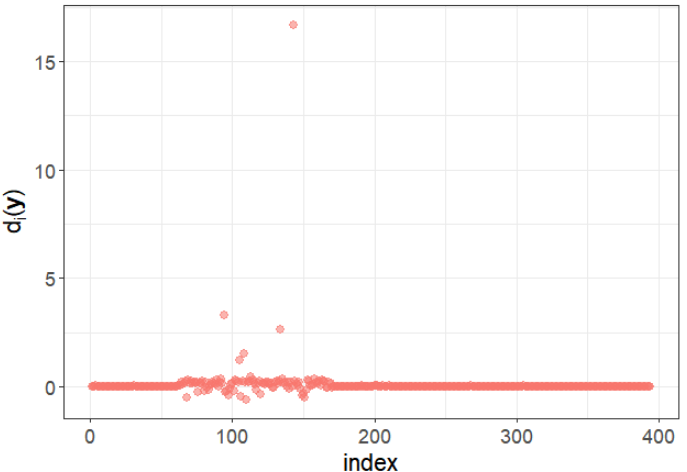}
   \caption{The left plot shows samples from the reference distribution (histogram), its Gaussian approximation (black line) and observed BF sensitivity (red line) for the geostatistical data problem. The right plot shows $d_i$, the contribution of each spatial index to the BF sensitivity.} 
  \label{fig:press_score} 
\end{figure}

The previous LGM was fitted in R-INLA (with the hyperparameters fixed at the posterior mode), and the sensitivity measures $d_i$ in \eqref{eq:scorebayes} are shown in Figures~\ref{fig:pressd1} (right) and \ref{fig:press_score} (right). We observe five locations that stand out for leading to a larger increase in the Bayes factor when we apply a non-Gaussian perturbation to the associated driving noise elements. In contrast, for most spatial indices, the same non-Gaussian perturbation has minimal impact on the Bayes factor. These locations match closely with the locations with non-Gaussian departures in the latent field found after fitting the LnGM. In Appendix B, we fitted this LnGM, which considers the prior \mbox{$\mathbf{D}\mathbf{w}=\boldsymbol{\Lambda}(\eta)$}, to the pressure data, and then generated similar diagnostics as the one in Figure \ref{fig:pressd1} (right). However, The LnGM fit took about 2.5 hours in Stan (with 1000 total iterations), while the LGM fit in R-INLA and the diagnostics shown in Figures \ref{fig:pressd1} and \ref{fig:press_score} took just a few seconds.


We simulated the reference distribution $s_0(\mathbf{y}^{\text{pred}},\hat{\boldsymbol{\gamma}})|\hat{\boldsymbol{\gamma}},\eta=0$ by Monte Carlo and the samples are shown in Figure~\ref{fig:press_score} (left), where $\hat{\boldsymbol{\gamma}}$ is the posterior mode of the nuisance parameters. We also show the Gaussian approximation of the reference distribution, based on Proposition \ref{prop:variance_approx}. The BF sensitivity of the observed data is considerably larger than the samples of the reference distribution, indicating that the fitted LGM fails to replicate the pronounced spikes observed in the data. 

\begin{figure}[htp]
    \centering
    \includegraphics[width=0.49\linewidth,  height = 0.4\linewidth]{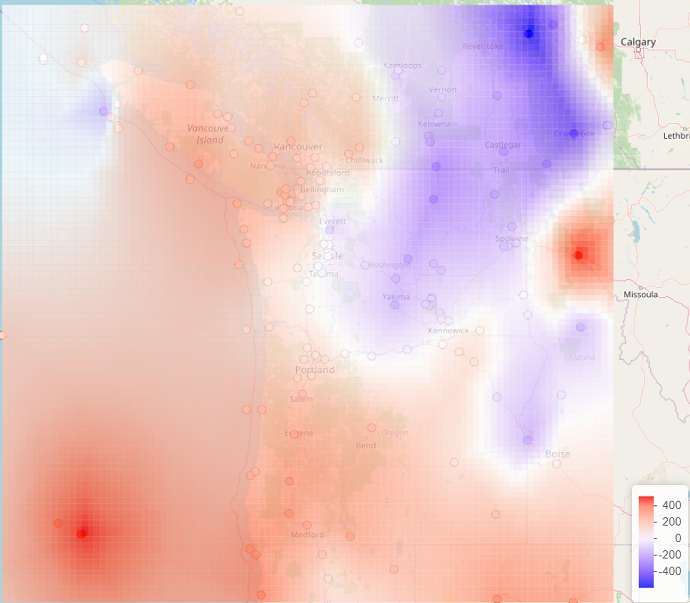}      \includegraphics[width=0.49\linewidth,  height = 0.4\linewidth]{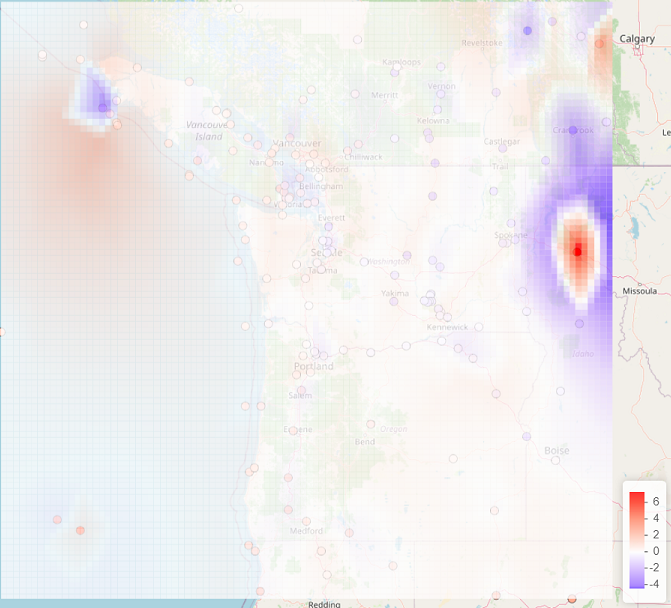}
  \caption{Spatial predictions (left) of pressure and their scaled sensitivity (right).}
  \label{fig:pressd2} 
\end{figure}

After detecting model misfit, as outlined in Algorithm \ref{alg:workflow}, we proceed to calculate the sensitivity of posterior summaries of interest using Theorem \ref{theo:sens} and equation \eqref{eq:ngpert}. This application focused on spatial predictions, which are the posterior mean of the spatial random effects at unobserved locations. These predictions are shown in Figure~\ref{fig:pressd2}, which displays a grid of $100\times100$ locations encompassing the studied region. We also show in Figure~\ref{fig:pressd2} the sensitivity of those predictions for each unobserved location. We scaled the sensitivity measures by dividing them with the standard deviation of each prediction. The plot informs us about the difference in predictions between the LGM and LnGM (without requiring fitting the LnGM). We observe that a small non-Gaussian perturbation to the random effects' prior will have a larger impact on the spatial regions near the spikes of the observed data. The results suggest fitting an LnGM to improve the predictions near those regions. Details about how the spatial predictions and their sensitivity were computed in R-INLA are in Appendix \ref{sect:geostatapp}. 


\subsection{Simulation study} \label{app:sim_study}


To evaluate the effectiveness of the model checking procedure of Section \ref{sect:predcheck} in identifying latent non-Gaussianity, we generate simulated data from a latent random walk of order 1 (RW1) model that is driven by NIG noise. The model for the data is $y_i = \sigma_{\mathbf{w}}w_i + \sigma_{\boldsymbol{\epsilon}} \epsilon_i$, where $w_{i}-w_{i-1} = \Lambda(\eta), \ i=2,\dots,N$, and $\Lambda(\eta)$ is standardized NIG noise with parameter $\eta$. The Gaussian model is obtained when $\eta = 0$ (see Section \ref{sect:lngm}).

\begin{figure}[!htp]
   \centering
   \begin{tabular}{ccc}
  & $N=200$ & $N=1000$ \\  
   $\sigma_\mathbf{w}=3$ &   \raisebox{-.5\height}{\includegraphics[width=0.43\linewidth , height = 3.1cm]{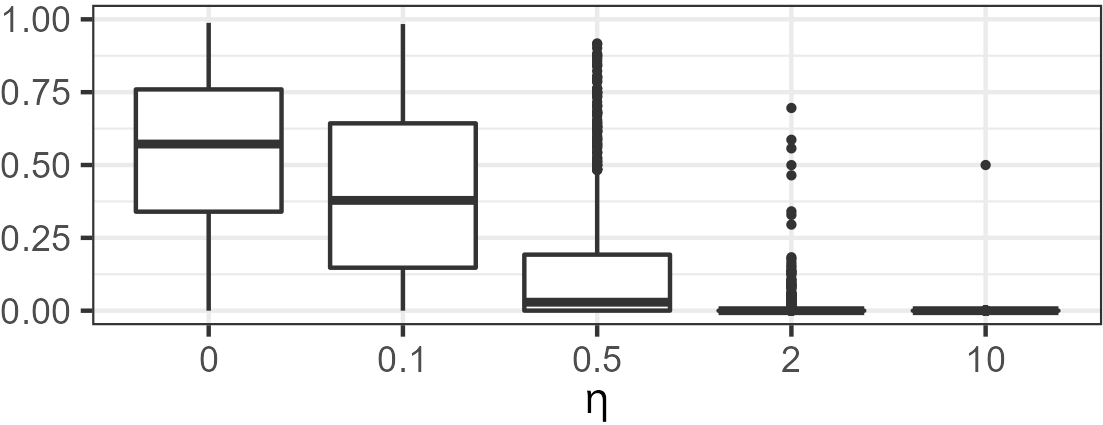}} &  \raisebox{-.5\height}{\includegraphics[width=0.43\linewidth , height = 3.1cm]{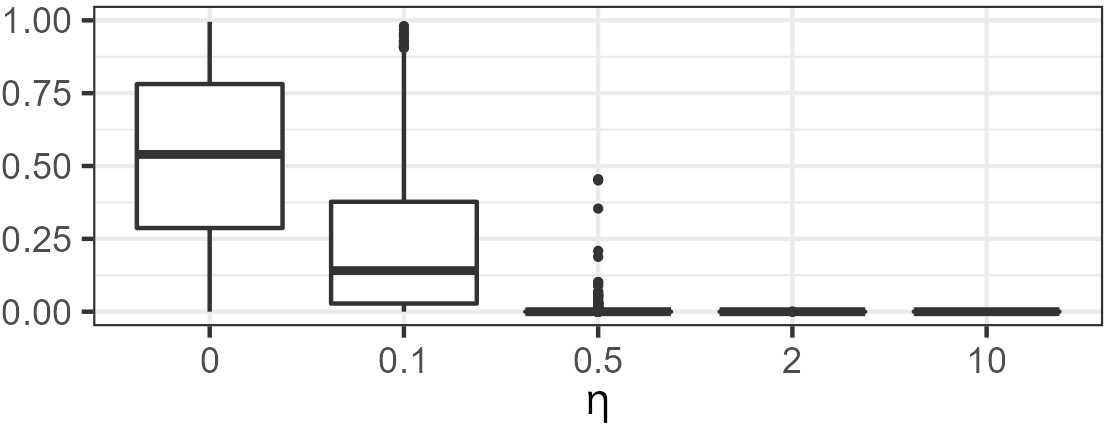}}
\\
 $\sigma_\mathbf{w}=1$ & \raisebox{-.5\height}{\includegraphics[width=0.43\linewidth , height = 3.1cm]{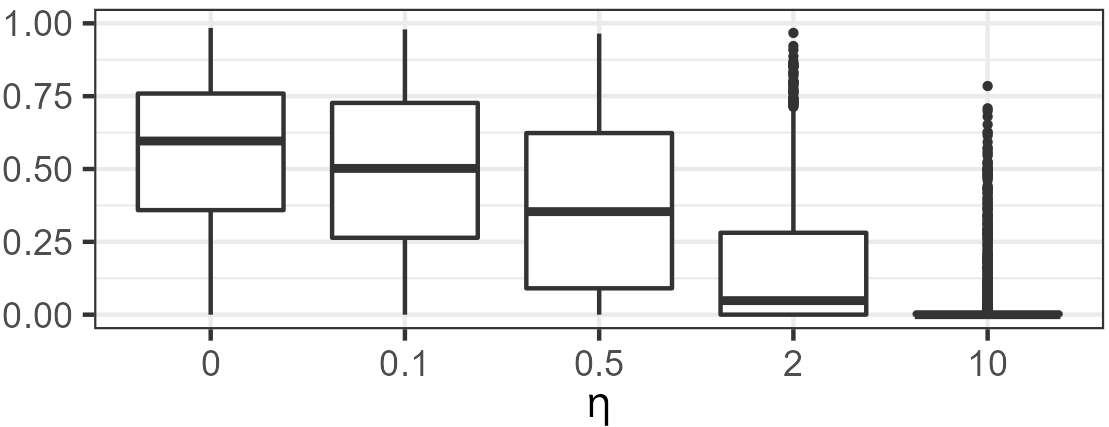}} &   
      \raisebox{-.5\height}{\includegraphics[width=0.43\linewidth , height = 3.1cm]{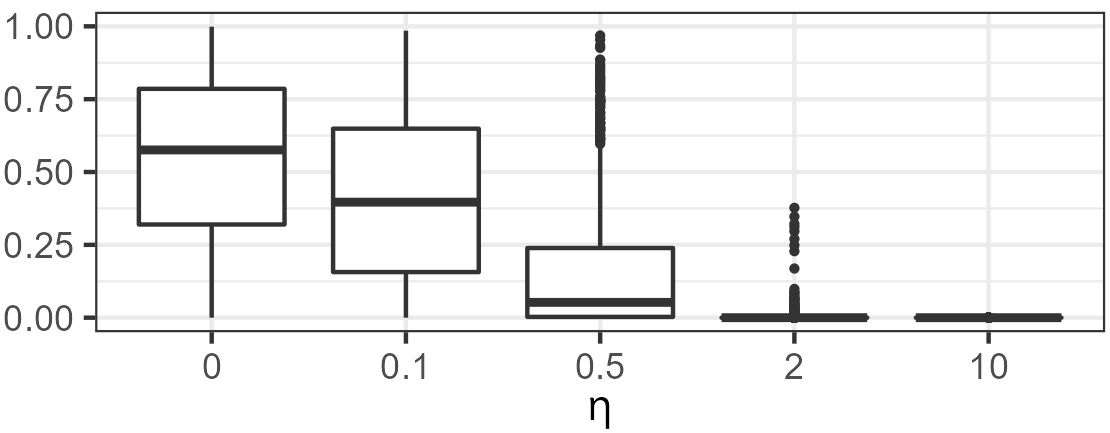}}
\\    
  $\sigma_\mathbf{w}=\frac{1}{3}$  & \raisebox{-.5\height}{\includegraphics[width=0.43\linewidth , height = 3.1cm]{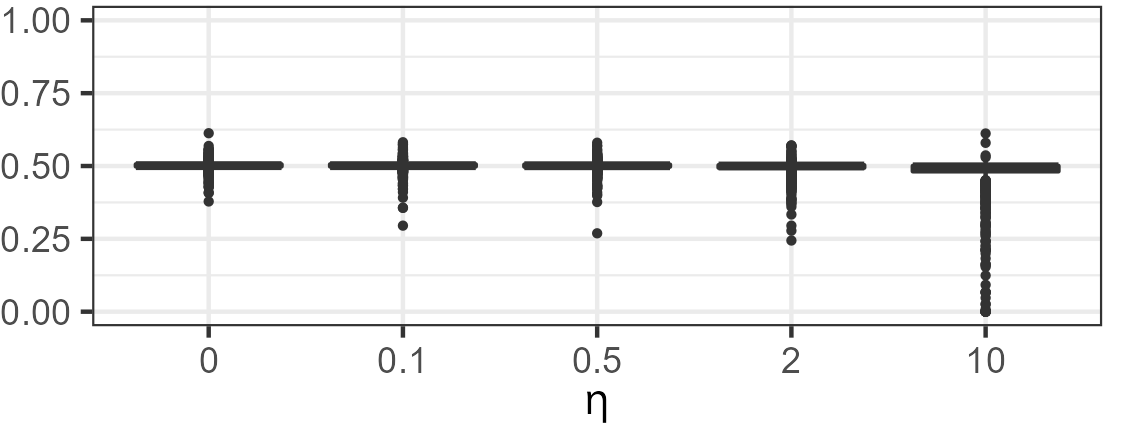}} &    	    \raisebox{-.5\height}{\includegraphics[width=0.43\linewidth , height = 3.1cm]{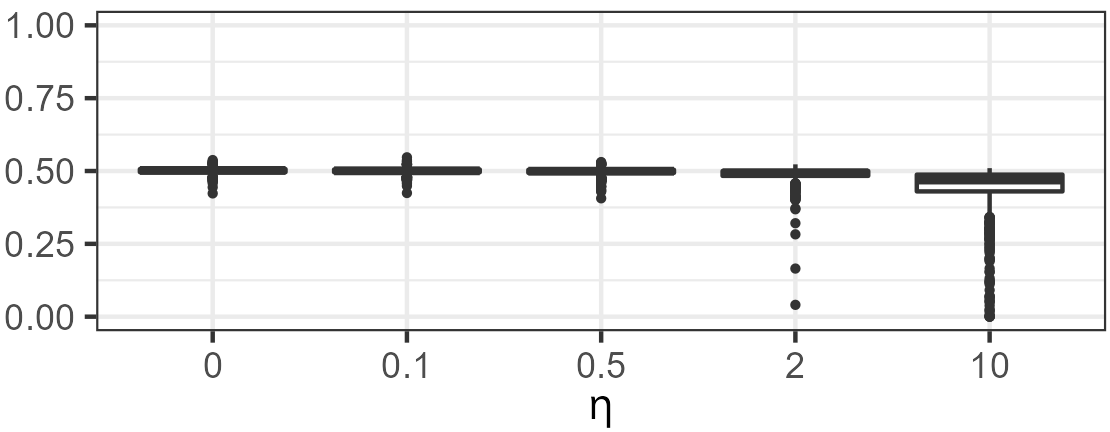}}
   \end{tabular}
   \caption{Box plots of the upper-tailed probabilities $p$ for 1000 simulated latent RW1 datasets driven by NIG noise with dimension $N$, and parameters $\eta$, $\sigma_{\boldsymbol{\epsilon}}=1$ and $\sigma_\mathbf{w}$.} 
  \label{fig:severity} 
\end{figure}

For the simulated data we considered the parameters $\eta \in (0, 0.1, 0.5, 2, 10)$, $\sigma_{\boldsymbol{\epsilon}} = 1$,  \mbox{$\sigma_\mathbf{w} \in (1/3, 1,3)$}, and dimension $N \in (200, 1000)$. We simulated 1000 datasets for each parameter configuration. We fitted each simulated dataset to an LGM (that assumes a latent Gaussian RW1 model) in R-INLA with the empirical Bayes approximation, meaning that the hyperparameters are fixed at their posterior mode $\hat{\boldsymbol{\gamma}} = (\hat{\sigma}_{\mathbf{w}},\hat{\sigma}_{\boldsymbol{\epsilon}})$. For each fit, we then computed the upper-tailed probability in \eqref{eq:pv} with the \verb|ngvb| package (see Section \ref{sect:inlaimpl}) which uses the approximation $p \approx \Phi[-s_0(\mathbf{y},\hat{\boldsymbol{\gamma}})/SD\{s_0(\mathbf{y}^{\text{pred}},\hat{\boldsymbol{\gamma}})|\eta=0\} ]$. The results are shown in Figure~\ref{fig:severity}. As expected, discrepancies are more easily detected when we increase the non-Gaussianity ($\eta$) and scale ($\sigma_\mathbf{w}$) of the latent process and the dimension $N$ of the simulated data because the upper-tailed probabilities $p$ are closer to 0.


When $\sigma_\mathbf{w}=1/3$, the latent signal is barely detectable, and the model checks do not detect discrepancy. This behaviour is also observed in
\cite{sinharay2003posterior}. Indeed, when we consider $\sigma_{\boldsymbol{\epsilon}} \to \infty$, we have \mbox{$\pi(\mathbf{w}|\mathbf{y}, \sigma_{\mathbf{w}})\sim N\{ \mathbf{0}, \mathbf{D}(\sigma_{\mathbf{w}})^{-1}\mathbf{D}(\sigma_{\mathbf{w}})^{-T}\}$}, which implies that $s_0(\mathbf{y},\hat{\boldsymbol{\gamma}})=0$. Thus the Bayes factor $BF_{\eta}$ in \eqref{eq:LRexpansion} is locally flat, giving no preference to the non-Gaussian latent prior. However, this behaviour is desired because, in this limit, the latent random effects are not detectable and play no role in the fitted model.


\section{Implementation in probabilistic programming languages} \label{sect:PPimpl}


In the probabilistic programming (PP) paradigm models are specified usually through the logarithm of the joint distribution, $\log \pi(\mathbf{y},\mathbf{z},\eta)$, and then inference is performed automatically.  Popular PP languagues are Stan \citep{carpenter2017stan}, Tensorflow Probability \citep{dillon2017tensorflow} and PyMC3 \citep{salvatier2016probabilistic}, among others. Robustness studies can be easily incorporated in this programming paradigm, because we only additionally require posterior samples of the local perturbation \mbox{$p(\mathbf{y},\mathbf{z}) = \lim_{\eta\to0} \partial_\eta \log \pi(\mathbf{y},\mathbf{z}, \eta)$}. This perturbation could be computed via the automatic differentiation algorithms of PP languages or through
\begin{equation*}
    p(\mathbf{y},\mathbf{z}) \approx  (\log \pi(\mathbf{y}, \mathbf{z}, \eta = \epsilon) -  \log \pi(\mathbf{y}, \mathbf{z}, \eta = 0))/\epsilon,
\end{equation*}
for a small $\epsilon$. Then the sensitivity measures of Theorem \ref{theo:sens} are obtained by computing the posterior covariance between $l(\mathbf{y},\mathbf{z})$ and $p(\mathbf{y},\mathbf{z})$. To perform model checking, we also need to specify the predictive density and now the joint model is $\pi(\mathbf{y}, \mathbf{y}^{\text{pred}}, \mathbf{z}, \eta)$.

\subsection{Application 1: Checking a Matérn model with fixed smoothness} \label{sect:matern_check}

The Matérn covariance function is widespread in Gaussian processes applications to time series, spatial statistics and beyond. Its smoothness parameter $\nu$ provides complete control over the mean-square differentiability of the process. However, calculating derivatives of this covariance function with respect to $\nu$ requires computing derivatives of the modified Bessel function of the second kind, which can be costly. As a result, PP languages and software packages often require fixing $\nu$ instead of estimating it.

\begin{figure}[htp]
   \centering
   \includegraphics[width=0.49\linewidth]{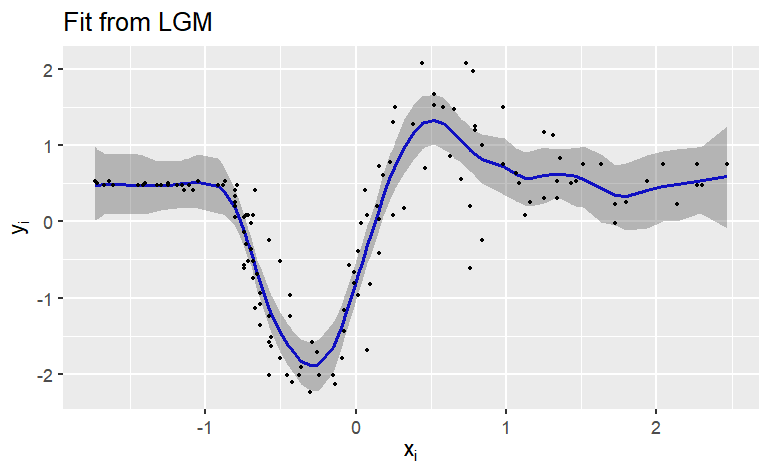}
      \includegraphics[width=0.49\linewidth]{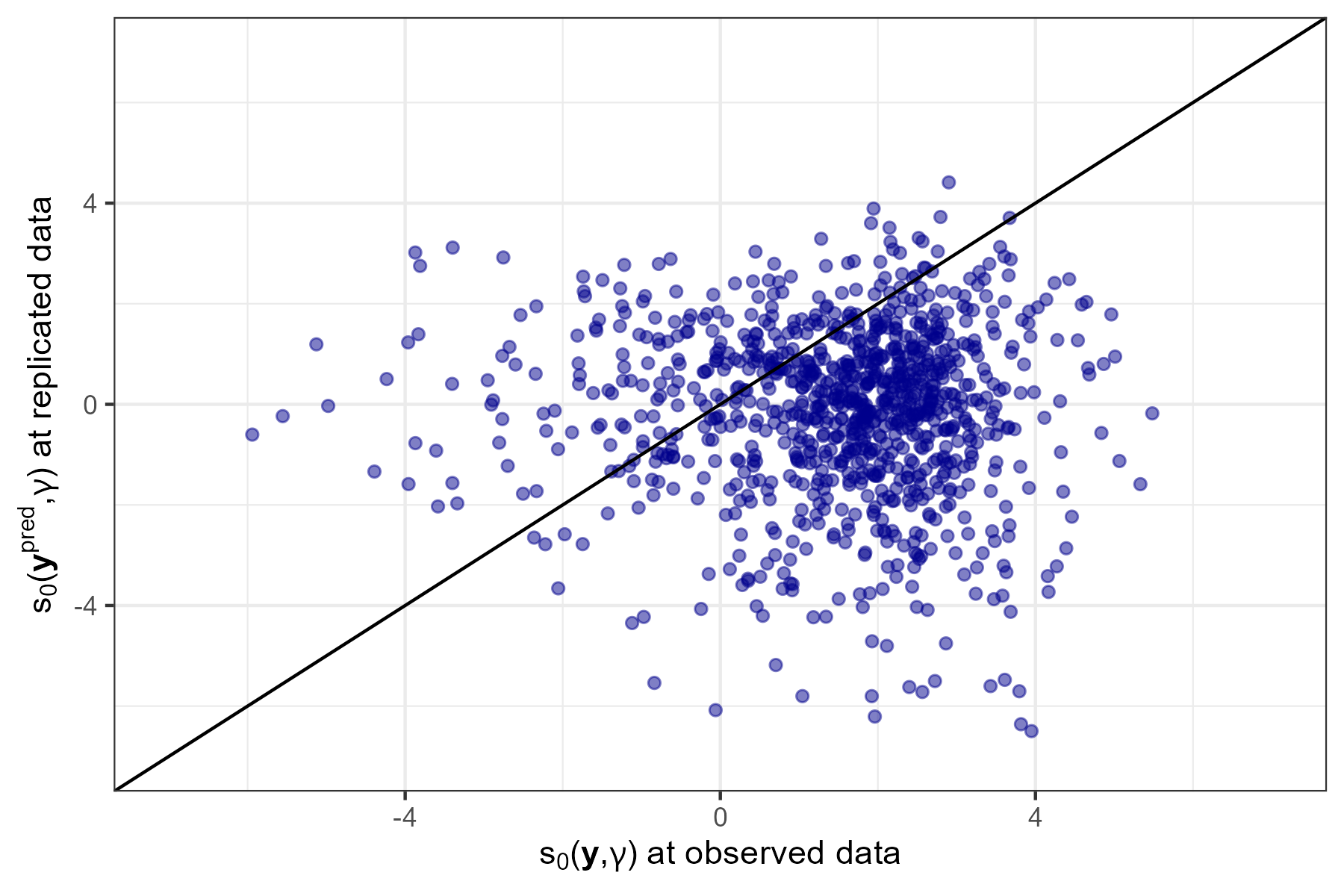}
   \caption{The left plot shows the data as black circle markers, the posterior mean of the random effects (blue line) and 97.5\% credible intervals (shaded area). The right plot shows $d(\mathbf{y}, \boldsymbol{\gamma})$ at the observed and replicated data, for each hyperparameter sample from the posterior distribution.} 
  \label{fig:matern} 
\end{figure}

The data shown in Figure \ref{fig:matern} are standardised measurements of head acceleration per millisecond in a simulated motorcycle accident, and it is available in the \verb|MASS| R-package. We consider noisy observations $\mathbf{y}|\mathbf{w}, \sigma_{\boldsymbol{\epsilon}} \sim N(\mathbf{w}, \sigma_{\boldsymbol{\epsilon}}^2 \mathbf{I})$, where $\mathbf{w}$ is a Gaussian process with Matérn covariance function with distribution $\mathbf{w}|\sigma_\mathbf{w}, \rho, \nu \sim N \{\mathbf{0}, \mathbf{\Sigma}(\sigma_\mathbf{w}, \rho, \nu)\}$, conditioned on the scale, range and smoothness parameters, respectively. We can marginalise out the random effects $\mathbf{w}$ and declare the following model in Stan:
\begin{equation}\label{eq:fullmatern}
\pi(\mathbf{y}|\sigma_{\boldsymbol{\epsilon}}, \sigma_\mathbf{w}, \rho, \nu) \sim N \{\mathbf{0}, \sigma_{\boldsymbol{\epsilon}}^2 \mathbf{I} + \mathbf{\Sigma}(\sigma_\mathbf{w}, \rho, \nu)\},    
\end{equation}
and then samples from the random effects $\mathbf{w}$ can be generated through \eqref{eq:posterior_w}. Stan's built-in Matérn covariance functions include the cases $\nu = ( 1/2, 3/2, 5/2, \infty )$. We fix $\nu = 3/2$ and compute:
\begin{equation}\label{eq:pertmatern}
    p(\mathbf{y}, \boldsymbol{\gamma}) \approx  (\log \pi(\mathbf{y}| \boldsymbol{\gamma},  \nu = 3/2 + \epsilon) -  \log \pi(\mathbf{y}|\boldsymbol{\gamma}, \nu = 3/2))/\epsilon,
\end{equation}
at each iteration, where  $\boldsymbol{\gamma} = (\sigma_{\boldsymbol{\epsilon}}, \sigma_\mathbf{w}, \rho)$ contains the hyperparameters. Finally, for model checking, we define replicates $\mathbf{y}^{\text{pred}}$ that have the same distribution as \eqref{eq:fullmatern} for $\nu=3/2$, and we also compute $p(\mathbf{y}^{\text{pred}}, \boldsymbol{\gamma})$ based on \eqref{eq:pertmatern}. Since we do not need to integrate out the random effects in \eqref{eq:pertmatern}, we can use $p(\mathbf{y}, \boldsymbol{\gamma})$ directly as a discrepancy measure $d(\mathbf{y}, \boldsymbol{\gamma})$. We show in Figure \ref{fig:matern} a  scatter plot of $d(\mathbf{y}^{\text{pred}}, \boldsymbol{\gamma})$ and $d(\mathbf{y}, \boldsymbol{\gamma})$, where $(\mathbf{y}^{\text{pred}}, \boldsymbol{\gamma})$ is draw from the posterior distribution. The fraction of points above the diagonal gives us the upper-tail probability $P\{ d(\mathbf{y}^{\text{pred}}, \boldsymbol{\gamma}) > d(\mathbf{y}, \boldsymbol{\gamma})|\mathbf{y}, \nu = 3/2\}$  which is about 0.24, and so the fitted model can generate replicated data with similar levels of smoothness as the observed data as quantified by the discrepancy measure $d$. We repeated the model cheeking procedure for $\nu=1/2$ and $\nu=5/2$. In the first case, $p=0.08$, and so the Matérn model with $\nu=1/2$ may not be sufficiently smooth, while for $\nu=5/2$, no large misfit was detected.

\subsection{Application 2: Checking latent non-Gaussianity in Stan}

Sampling algorithms like Stan's Hamiltonian Monte Carlo can struggle to explore the parameter space of hierarchical models because of the problematic geometry induced by the posterior distribution \citep{margossian2020hamiltonian, betancourt2015hamiltonian}. For LGMs the problematic geometry is due to the interaction between the latent Gaussian variable $\mathbf{x}$ in \eqref{eq:lgm} and the hyperparameters $\boldsymbol{\theta}_2$. For LGMs with a Gaussian response an easy remedy is to remove explicit hierarchical correlations by marginalising out $\mathbf{w}$ as in \eqref{eq:fullmatern}. This (base) model can be easily fitted, however, the previous marginalisation is no longer possible if we want to consider non-Gaussian distributions for the random effects. 

\begin{figure}[htp]
   \centering
   \includegraphics[width=0.49\linewidth,height = 0.35\linewidth]{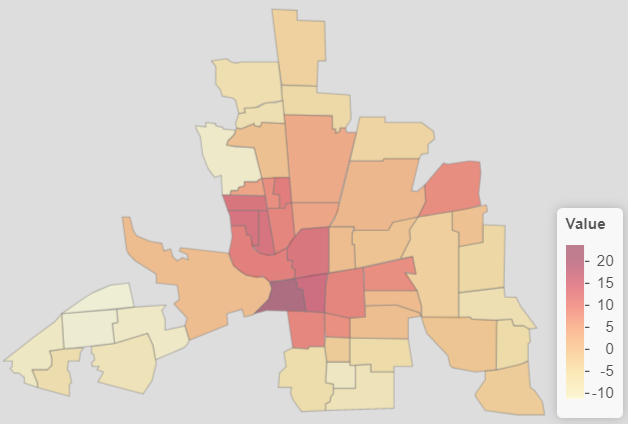}
   \includegraphics[width=0.49\linewidth,height = 0.35\linewidth]{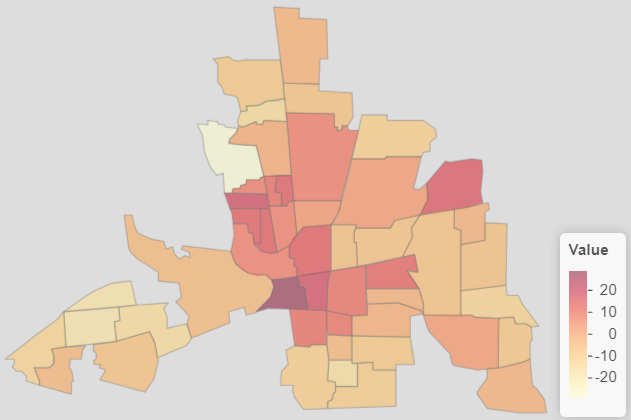}
    \caption{Crime rates in Columbus, Ohio, per thousand households (left), and posterior mean of the spatial effects $\mathbf{w}$ (right).}
  \label{fig:crime_data}   
\end{figure}

We study areal data, consisting of the number of residential burglaries and vehicle thefts per thousand households ($y_i$) in 49 counties of Columbus, Ohio, in 1980. This dataset, illustrated in Figure~~\ref{fig:crime_data}, is found in the \verb|spdep| package in R. We observe several sharp variations in the crime rate of neighbouring counties, which, with a deeper examination of the data, do not seem fully explained by the available covariates. These sharp variations suggest that we could benefit from using a non-Gaussian model to account for the spatial effects. \cite{walder2020bayesian} analysed this dataset using a non-Gaussian model to account for the spatial dependency. We consider the same set of covariates as the previous authors and fit the following model:
\begin{equation} \label{eq:columbus}
y_{i}= \beta_0 + \beta_1 \mathrm{HV}_i + \beta_2 \mathrm{HI}_i +  \sigma_{\mathbf{w}}w_i + \sigma_{\epsilon}\epsilon_i, 
\end{equation}
where $\mathrm{HV}_i$ and $\mathrm{HI}_i$ are the average household value and household income for county $i$, $\mathbf{w}$ accounts for structured spatial effects, whereas $\epsilon_i \overset{i.i.d}{\sim} N(0,1)$.  We consider a simultaneous autoregressive (SAR) model \citep{besag1974spatial,wall2004close,ver2018relationship} for the spatially structured effects $\mathbf{w}$. The Gaussian SAR model is defined by $\mathbf{D}_{SAR}(\sigma_{\mathbf{w}}, \rho)\mathbf{w} = \mathbf{\Lambda}(0)$ with $\mathbf{D}_{SAR}(\sigma_{\mathbf{w}}, \rho)=(\mathbf{I}-\rho\mathbf{W})/\sigma_{\mathbf{w}}$ where $\mathbf{W}$ is the row standardized adjacency matrix, $-1<\rho<1$ is the autocorrelation parameter, $\sigma_{\mathbf{w}}$ is the scale parameter, and $\mathbf{\Lambda}(0)$ is a vector of i.i.d. standard Gaussian noise. 

\begin{figure}[htp]
   \centering
      \includegraphics[width=0.49\linewidth]{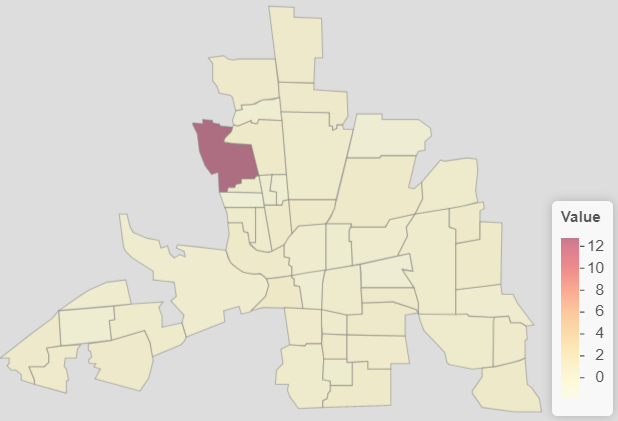}
      \includegraphics[width=0.49\linewidth]{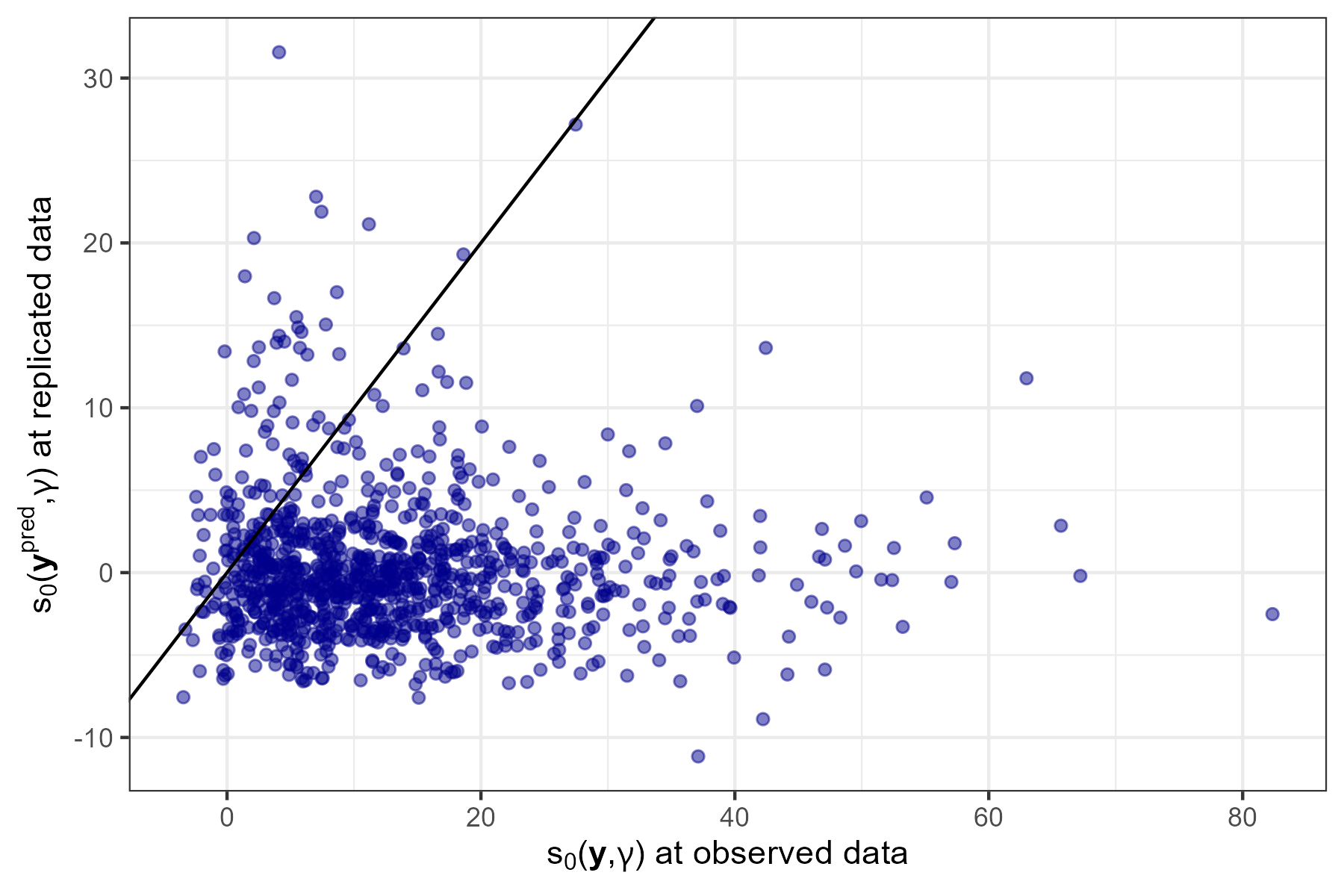}
    \caption{BF sensitivity for each county $d_i$ (left) and sensitivity measure $s_0$ at the observed and replicated data, for each sample from the posterior distribution (right).}
  \label{fig:crime_result}   
\end{figure}

We fit this LGM  by declaring $\pi(\mathbf{y}|\boldsymbol{\gamma}) \sim N\{ \beta_0 + \beta_1 \mathrm{HV} + \beta_2 \mathrm{HI}, \sigma_{\boldsymbol{\epsilon}} \mathbf{I} +  (\mathbf{D}_{SAR}^T \mathbf{D}_{SAR})^{-1}\}$ in Stan. An equivalent LnGM considers $\mathbf{D}_{SAR}(\sigma_{\mathbf{w}}, \rho)\mathbf{w} = \mathbf{\Lambda}(\eta)$, where $\mathbf{\Lambda}(\eta)$ is a vector of i.i.d. standardized NIG noise \citep{cabral2022controlling}, but it cannot be as easily fitted in Stan. Therefore, at each iteration of the LGM fit we compute the sensitivity measures  $s_0(\mathbf{y},\boldsymbol{\gamma})$ and $s_0(\mathbf{y}^{\text{pred}},\boldsymbol{\gamma})$, from Section \ref{sect:LGMnormal}, and we show the scatter plot in Figure \ref{fig:crime_result}.  The result suggest that predictive replicates have some difficulty in replicating features in the observed data since $P\{s_0(\mathbf{y}^{\text{pred}},\boldsymbol{\gamma}) > s_0(\mathbf{y},\boldsymbol{\gamma}) | \mathbf{y}, \eta = 0\} \approx 0.05$. Figure \ref{fig:crime_result} also shows the sensitivity measures $d_i$ in \eqref{eq:scorebayes}, where we observe that only one county is detected as an ``outlying" county. Following Algorithm \ref{alg:workflow}, since the Gaussian prior assumption seems inadequate, we compute the sensitivity of the posterior mean of the intercept and linear effects (divided by the standard deviation) and show the results in Table \ref{table:crime_mean}. The sign of the sensitivity measures suggest an increase of the posterior mean of $\beta_1$, and a decrease in the posterior mean of $\beta_2$, when considering an LnGM.
An LnGM was fitted to this data in \cite{cabral2022fitting} and the results corroborate the previous conclusions. 


\begin{table}[htp]
\centering
\begin{tabular}{cccc}
\hline
  \textbf{Parameter} & $\beta_0$     & $\beta_1$  & $\beta_2$    \\ \hline
 Mean (sd) & 60.10 (6.33) & -0.30 (0.09) & -0.94 (0.37) \\ 
  Scaled Sensitivity & -0.93 & 7.43 & -2.75 \\ \hline
\end{tabular}
\caption{Mean and standard deviation of the linear effects (second row), and sensitivity of the linear effects scaled by the standard deviation (third row).}
\label{table:crime_mean}
\end{table}

\section{Implementation in R-INLA} \label{sect:inlaimpl}

To facilitate model checking and the computation of the sensitivity measures of Section \ref{sect:sensmeasures}, we built the function \verb|ng.check|. This function was added to the \verb|ngvb| package, which is available at \url{github.com/rafaelcabral96/ngvb}. The package also contains functions that fit LnGMs using INLA and variational Bayes approximations \citep{cabral2022fitting}. After fitting an LGM with R-INLA we can call the \verb|ng.check| function using the \verb|inla| object as the input. We consider the linear predictor in \eqref{eq:linearpred} which contains several random effects $\mathbf{w}_i$ that are traditionally Gaussian processes and to which we entertain the possibility of being driven by NIG or GAL noise with flexibility parameter $\eta_i$. For each random effects, the \verb|ng.check| function produces diagnostic plots (see Figure~\ref{fig:ortho_score}) and computes the following:
\begin{enumerate}
    \item  Measures $d_i(\mathbf{y})$ of Section \ref{sect:LGMnormal}, and $s_0(\mathbf{y})$ and $I_0(\mathbf{y})$ of Theorem \ref{theo:scorefisher}.
    \item  Upper-tailed probability $p$ in \eqref{eq:pv} for LGMs with Gaussian response.
    \item  Sensitivity of the random effects, $\partial_{\eta_i} E(w_{i,j}|\mathbf{y},\eta_i)\rvert_{\eta_i=0}$, and likewise, the sensitivity of the intercept, linear effects, and linear predictor.
\end{enumerate}

These sensitivity measures are computed based on the Gaussian mixture approximation in R-INLA \citep{rue2009approximate}, 
\begin{equation}\label{eq:Gaussian_mixture}
    \pi(\mathbf{x}, \boldsymbol{\theta} \mid \mathbf{y}) \propto \sum_{k=1}^K \pi_{\mathrm{G}}(\mathbf{x} \mid \mathbf{y}, \boldsymbol{\theta}_k ) \pi(\boldsymbol{\theta}_k \mid \mathbf{y})  \Delta_k.
\end{equation}
The previous approximation considers only a set of $K$ hyperparameter values $\boldsymbol{\theta}_k$ with associated integration weights $\Delta_k$ which are obtained by placing a regular grid or using a central composite design centred on the posterior mode. More details are included in Appendix \ref{app:INLA}.
\subsection{Application 3: Model with random intercept and slope} \label{sect:orto}

\begin{figure}[htp]
   \centering
   \includegraphics[width=0.7\linewidth]{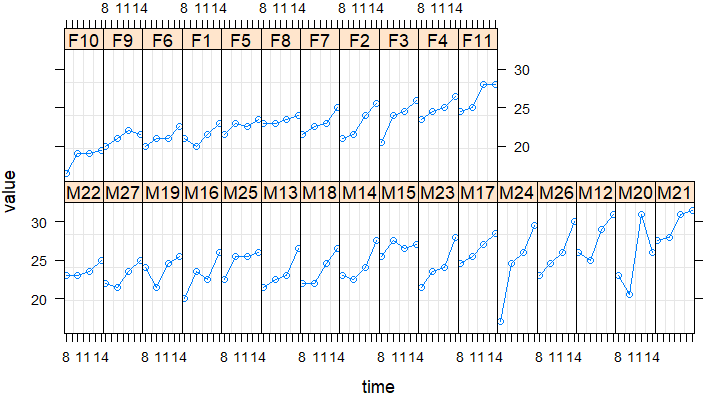}
   \caption{Trellis display of the distance in millimetres between the pituitary and pterygomaxillary fissure for girls (first row) and boys (second row).} 
  \label{fig:ortho_data} 
\end{figure}

 The data comes from an orthodontic study reported by \cite{potthoff1964generalized}. The response variable, shown in  Figure~~\ref{fig:ortho_data}, is the distance in millimetres between the pituitary and pterygomaxillary fissure, measured for 11 girls and 16 boys at ages 8, 10, 12, and~14. The data suggest that the intercept and slope vary with the subject. Thus, \cite{pinheiro2001efficient} proposed the following linear mixed-effects model to describe the response growth with age:
\begin{equation}\label{eq:app1}
y_{i j}=\beta_0+\beta_1 I_i(F) +\left(\beta_2+\beta_3 I_i(F)\right) t_j + \sigma_0 b_{0 i}+ \sigma_1 b_{1 i} t_j+\epsilon_{i j},
\end{equation}
\sloppy where $y_{i j}$ denotes the response for the $i$th subject at age $t_j$, $i=1, \ldots, 27$ and \mbox{$j=1, \ldots, 4$};  $\beta_0$ and $\beta_2$ denote the intercept and slope linear effects, respectively, for boys; and $\beta_1$ and $\beta_4$ denote the difference in intercept and slope fixed effects, respectively, between girls and boys. In addition, $I_i(F)$ denotes an indicator variable for females, and $\mathbf{b}_i=\left(b_{0 i}, b_{1 i}\right)$ is the random effects vector for the $i$th subject. Finally,  $\epsilon_{i j}\sim N(0,\sigma_{\boldsymbol{\epsilon}}^2)$ is~the~within-subject~error.

\begin{figure}[htp]
   \centering
   \includegraphics[width=0.49\linewidth]{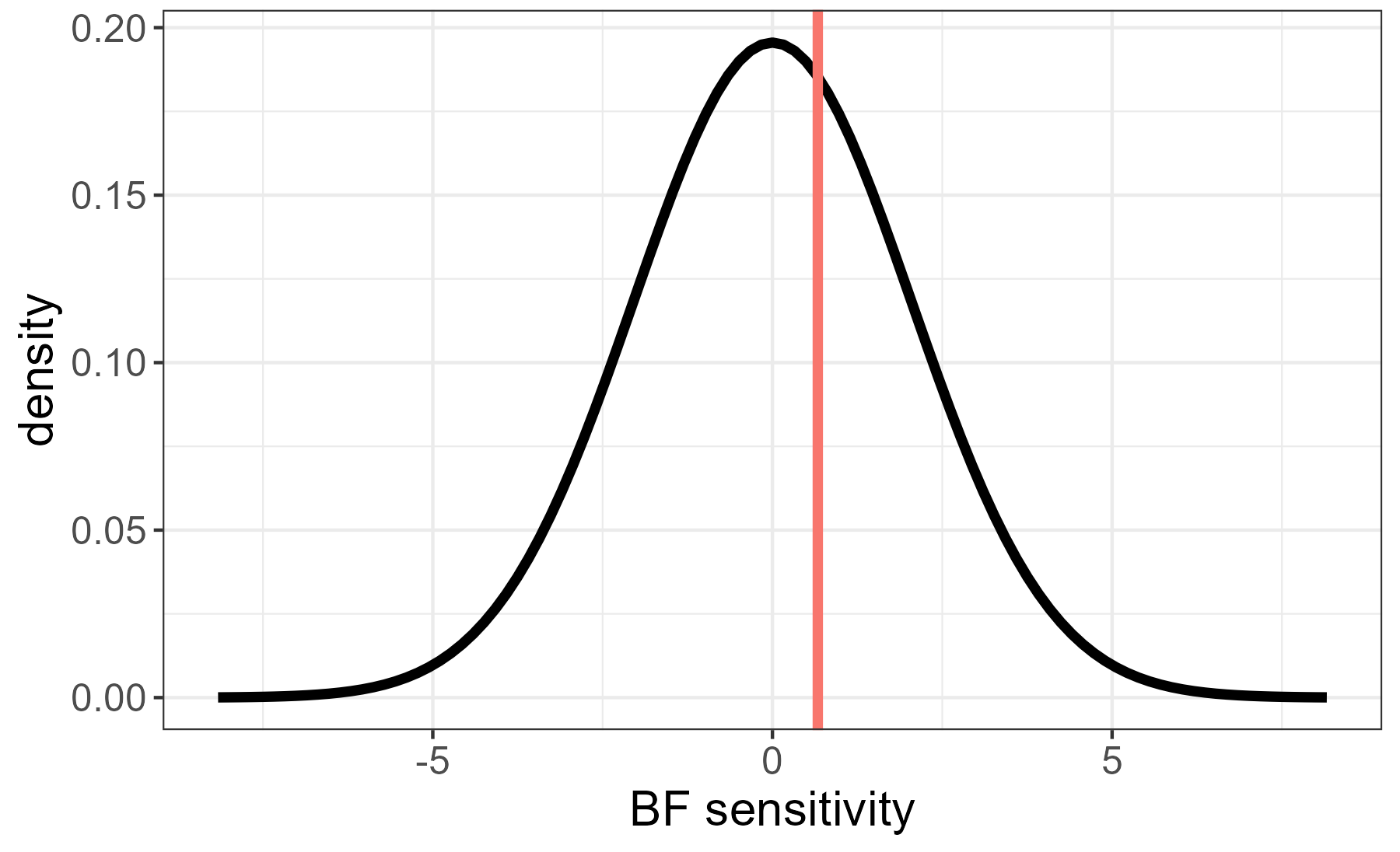}
    \includegraphics[width=0.49\linewidth]{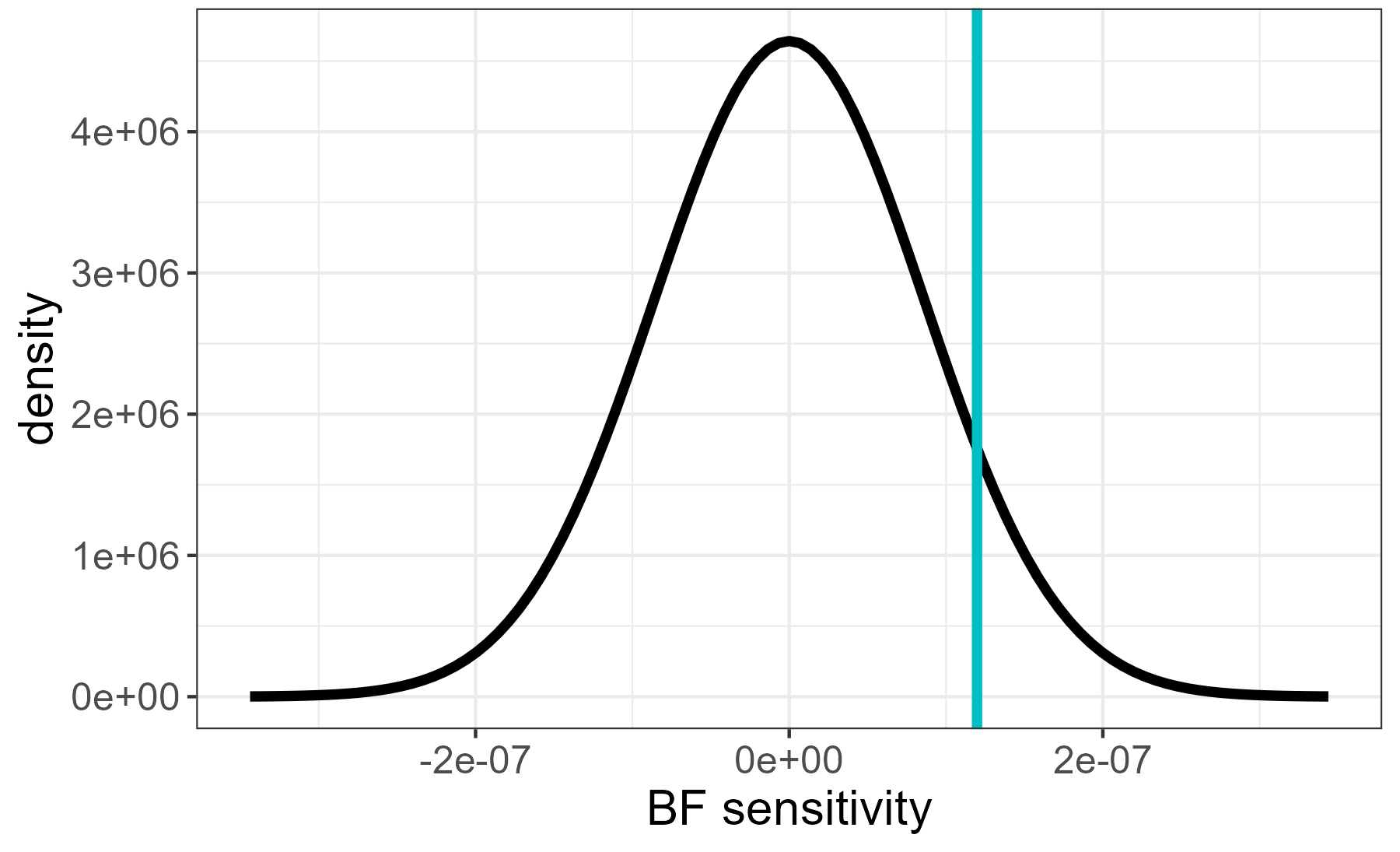}
   \caption{Gaussian approximation of the reference distribution based on Proposition \ref{prop:variance_approx} (black curve) for the random intercept component (left) and random slope component (right).} 
  \label{fig:ortho_score} 
\end{figure}

We consider Gaussian priors for the latent random intercepts \mbox{$b_{0 i} \overset{i.i.d}{\sim} \mathrm{N}(0,1)$} and random slopes $b_{1 i} \overset{i.i.d}{\sim} \mathrm{N}(0,1)$. The resulting model is an LGM that can be fitted with R-INLA, and we check the previous latent Gaussianity assumption. We show the observed and reference distribution for $s_0$ in Figure~\ref{fig:ortho_score}, where no large departure from latent Gaussianity are detected. The BF sensitivities were 0.67 and $1.20\cdot10^{-7}$ for the random intercept and random slope components, so the Bayes factor is more sensitive to the first component than the second. These results are consistent with the LnGM fit in \cite{cabral2022fitting} which indicated only a slight departure from latent Gaussianity for the random intercept component and no significant departure for the random slope component.

The LGM fit in R-INLA and the diagnostic plots were generated with the following code, where the data is stored in \verb|Orthodont| which is included in the \verb|ngvb| package.

\begin{verbatim}
formula <- value ~ 1 + Female + time + tF  + f(subject, model = "iid")
            + f(subject2, time, model = "iid")
LGM     <- inla(formula, data = Orthodont,
                control.compute = list(config = TRUE))
check   <- ng.check(fit = LGM) 
\end{verbatim}

\section{Discussion} \label{sect:discussion4}

Prior modelling should not be overlooked, as features of the prior distribution can carry over into the posterior and affect the model's overall performance. Therefore, 
predictive checks and sensitivity analysis that incur little additional computational cost were developed to check the latent Gaussianity assumption in LGMs. The idea is that the analyst will first fit a simpler LGM and assess if it can replicate important features in the data and then perturb it slightly in the direction of the LnGM to evaluate the impact on the results. This follows from the ideas of scientific learning, iterative model building, and continuous model expansion of \cite{box1980sampling}, \cite{rubin1984bayesianly}, and \cite{gelman1996posterior}. Although it is possible to perform model checking without an explicit alternative model \citep{evans2007comment},  it may be challenging to elicit intuitive discrepancy measures when checking the latent components of a complex hierarchical model. In these cases, if a more realistic model for the latent components is plausible, a statistically motivated choice is the Bayes factor sensitivity. On the other hand, from a robustness analysis perspective, posterior predictive checks can help in the interpretation of the sensitivity measures, such as the Bayes factor sensitivity, by indicating if the observed sensitivities are unusually high according to the fitted model. Finally, model checking and robustness analysis are also related through the workflow for model criticism we present in Section \ref{sect:workflow_LGM}.



 

An interesting observation in the time series and spatial statistics applications we studied is that significant deviations from Gaussianity, when present, usually occurred only for a few time points or locations. For example, in the pressure data example,  Figure~\ref{fig:pressd2} shows that deviations from Gaussianity are more prevalent for 5 spatial locations. Consequently, rather than robustifying the entire latent process by fitting an LnGM, it is more efficient to robustify only those specific locations. This can be done, for instance, by conditioning the random effects' driving noises on mixture variables that control their variance in order to add more flexibility to the model (see \cite{cabral2022fitting}). By doing so, the latent process can account for non-Gaussian behaviour in specific locations while preserving the Gaussian assumption in most areas.


Future work includes investigating the use of these tools to check other common assumptions, such as independence, stationarity, or symmetry, both at the response distribution or at any latent layer of hierarchical models. We gave primacy to the sensitivity of the Bayes factor. However, there are other measures for model assessment and selection (see \cite{gelman2014understanding} and \cite{vehtari2012survey}), such as the expected log predictive density, $E\{\log\pi(\mathbf{y}|\mathbf{z})| \mathbf{y},\eta\}$, and the widely applicable information criterion (WAIC). Most of these measures involve a posterior expectation and, therefore, can be assessed for sensitivity using Theorem \ref{theo:sens}.

\bibliographystyle{abbrvnat}
\bibliography{main}

\newpage
\begin{appendices}

The appendices provide supporting material for the main article. Firstly, Appendix \ref{sect:localsensproof} includes the proof of the theorems, lemma and proposition. Then, Appendix \ref{sect:geostatapp} examines different ways of simulating data from the fitted model and explains the reasons behind choosing the particular sampling scheme outlined in Section \ref{sect:pred_choice}. Furthermore, in Appendix \ref{sect:refapprox}, we discuss the asymptotic approximation of the reference distribution, which is useful when checking LGMs with non-Gaussian responses. Finally, Appendix \ref{app:INLA} details how the various sensitivity measures are computed in R-INLA.

\section{Proofs}\label{sect:localsensproof}

The formulas for $s_0$ and $I_0$ in Theorem \ref{theo:scorefisher} can be proven by applying the interchange of derivative with integral, similar to the proofs of Fisher's and Louis' identities \citep{louis1982finding} used in maximum-likelihood estimation problems involving hidden variables or unobserved data. Similarly, the formula for $s_l$ in Theorem \ref{theo:scorefisher} has been proven in \cite{perez2006mcmc} using the same derivative-integral exchange. We provide here more concise proofs based on Taylor expansions.

\subsection{Proof of Theorem \ref{theo:scorefisher}}\label{app:proof1}
The regularity conditions  of the theorem are necessary for the Taylor expansions of $BF_\eta(\mathbf{y})$ with a remainder that behaves like $\smallO(\eta^2)$. We start by expanding the Bayes factor  of equation \eqref{eq:LRexpansion}:
\begin{align*}
    BF_{\eta}(\mathbf{y}) &= \frac{\pi(\mathbf{y}|\eta)}{\pi(\mathbf{y}|\eta=0)} = \frac{\int \pi(\mathbf{y},\mathbf{z}|\eta)d\mathbf{z}}{\pi(\mathbf{y}|\eta=0)} \\ &\overset{(a)}{=} \frac{\int \pi(\mathbf{y},\mathbf{z}|\eta=0)(1+p(\mathbf{y},\mathbf{z})\eta + (p(\mathbf{y},\mathbf{z})^2 + g(\mathbf{y},\mathbf{z}))\eta^2/2 + \smallO(\eta^2))d\mathbf{z}}{\pi(\mathbf{y}|\eta=0)} \\
    &= \int \pi(\mathbf{z}|\mathbf{y},\eta=0)(1+p(\mathbf{y},\mathbf{z})\eta + (p(\mathbf{y},\mathbf{z})^2 + g(\mathbf{y},\mathbf{z}))\eta^2/2 + \smallO(\eta^2))d\mathbf{z} \\
    &= 1 + E\{ p(\mathbf{y},\mathbf{z}) | \mathbf{y}, \eta = 0\}\eta + (E\{ p(\mathbf{y},\mathbf{z})^2 | \mathbf{y}, \eta = 0\} + E\{ g(\mathbf{y},\mathbf{z}) | \mathbf{y}, \eta = 0\})\eta^2/2 + \smallO(\eta^2).
\end{align*} 
where step (a) follows from \eqref{eq:perturbation}. Comparing the previous expansion with equation \eqref{eq:LRexpansion}, $$BF_{\eta}(\mathbf{y}) = 1 + s_0(\mathbf{y}) \eta + (s_0(\mathbf{y})^2-I_0(\mathbf{y}))\eta^2/2 + \smallO(\eta^2),$$ yields the result of the theorem. \qed

\subsection{Proof of Theorem \ref{theo:sens}}\label{app:proof2}

The regularity conditions of the theorem are necessary for the Taylor expansions of $\pi(\mathbf{z}|\mathbf{y},\eta)$ and $E\{ l(\mathbf{y},\mathbf{z}) | \mathbf{y}, \eta\}$ with a remainder that behaves like $\smallO(\eta)$. We start by doing a Taylor expansion of the posterior density:
\begin{align*}
\pi(\mathbf{z}|\mathbf{y},\eta) &= \frac{\pi(\mathbf{y},\mathbf{z}|\eta)}{\pi(\mathbf{y}|\eta)} \overset{(a)}{=} \frac{\pi(\mathbf{y},\mathbf{z}|\eta=0)(1 + p(\mathbf{y},\mathbf{z})\eta  + \smallO(\eta) )}{\pi(\mathbf{y}|\eta=0)(1 + E\{p(\mathbf{y},\mathbf{z})|\mathbf{y},\eta=0\} \eta  + \smallO(\eta))} \\ & \overset{(b)}{=} \pi(\mathbf{z}|\mathbf{y},\eta=0)(1+ (p(\mathbf{y},\mathbf{z})-E\{p(\mathbf{y},\mathbf{z})|\mathbf{y},\eta=0\})\eta + \smallO(\eta)),   
\end{align*}
where in step (a), the Taylor expansion in the numerator follows from \eqref{eq:perturbation} and the Taylor expansion in the denominator follows from Theorem \ref{theo:scorefisher}. Step (b) results from another Taylor expansion to the fraction.  Now, the posterior expectation of $l(\mathbf{y},\mathbf{z})$ is given by
\begin{align*}
    E\{ l(\mathbf{y},\mathbf{z}) | \mathbf{y}, \eta\} &= \int l(\mathbf{y},\mathbf{z}) \pi(\mathbf{z}|\mathbf{y},\eta)d\mathbf{z}  = \int l(\mathbf{y},\mathbf{z})  \pi(\mathbf{z}|\mathbf{y},\eta=0) d\mathbf{z} \ + \\ &+ \left(\int l(\mathbf{y},\mathbf{z})  (p(\mathbf{y},\mathbf{z})-E\{p(\mathbf{y},\mathbf{z})|\mathbf{y},\eta=0\})\pi(\mathbf{z}|\mathbf{y},\eta=0)  d\mathbf{z}  \right)\eta + \smallO(\eta) \\
    &= E\{ l(\mathbf{y},\mathbf{z}) | \mathbf{y}, \eta = 0\} + Cov\left.\left\{l(\mathbf{y}, \mathbf{z}), \ p(\mathbf{y},\mathbf{z}) \ \right\vert \ \mathbf{y},\eta=0 \right\} \eta + \smallO(\eta).
\end{align*} \qed

\subsection{Proof of Lemma \ref{prop:detect}}\label{app:detect}

We consider the predictive distribution $ \pi(\mathbf{y}^{\text{pred}}|\boldsymbol{\gamma}, \eta)$: $$\pi(\mathbf{y}^{\text{pred}}|\boldsymbol{\gamma}, \eta) = \int \pi(\mathbf{y}^{\text{pred}}|\mathbf{w},\boldsymbol{\gamma}) \pi(\mathbf{w}|\boldsymbol{\gamma},\eta) d\mathbf{w},$$ where we integrate over the prior $\pi(\mathbf{w}|\boldsymbol{\gamma},\eta)$ and the nuisance parameters $\gamma$ are fixed.  We also define $E(d,\boldsymbol{\gamma},\eta) = E\{d(\mathbf{y}^{\text{pred}},\boldsymbol{\gamma})|\boldsymbol{\gamma}, \eta\}$ and $SD(d,\boldsymbol{\gamma},\eta) = [V\{d(\mathbf{y}^{\text{pred}},\boldsymbol{\gamma})|\boldsymbol{\gamma},\eta\}]^{1/2}$ to be the mean and variance of the discrepancy measure $d$ under the simulated data. The steps of the proof are given next. As the previous proofs, it relies on regularity conditions that permit derivative-integral and limit-integral exchange.

\begin{align*}
 &\argmax_{d(\cdot, \cdot)} \left\{ \lim_{\eta \to 0} \frac{d}{d\eta} \frac{E(d,\boldsymbol{\gamma},\eta)}{SD(d,\boldsymbol{\gamma},\eta)} \right\} =\argmax_{d(\cdot, \cdot)} \left\{ \frac{1}{SD[d,  \boldsymbol{\gamma}, 0]} \lim_{\eta\to0}\frac{d}{d\eta} E(d, \boldsymbol{\gamma}, \eta) \right\} \\
 &=\argmax_{d(\cdot, \cdot)} \left[  \frac{1}{SD(d,  \boldsymbol{\gamma}, 0)} \int d(\mathbf{y}^{\text{pred}}, \boldsymbol{\gamma}) \left\{\lim_{\eta \to 0} \partial_\eta \pi(\mathbf{y}^{\text{pred}}|\boldsymbol{\gamma},\eta)\right\}d\mathbf{y}^{\text{pred}} \right]  \\
 &\overset{(a)}{=}\argmax_{d(\cdot, \cdot)} \left\{ \frac{1}{ SD(d,  \boldsymbol{\gamma}, 0)} \int d(\mathbf{y}^{\text{pred}}, \boldsymbol{\gamma}) s_0(\mathbf{y}^{\text{pred}}, \boldsymbol{\gamma}) \pi(\mathbf{y}^{\text{pred}}|\boldsymbol{\gamma},\eta=0) d\mathbf{y}^{\text{pred}} \right\} \\
 &=\argmax_{d(\cdot, \cdot)} \left[ \frac{Cov\left\{d(\mathbf{y}^{\text{pred}},\boldsymbol{\gamma}),  s_0(\mathbf{y}^{\text{pred}}, \boldsymbol{\gamma})|\boldsymbol{\gamma},\eta=0\right\}}{SD\left\{d(\mathbf{y}^{\text{pred}},\boldsymbol{\gamma})|\boldsymbol{\gamma}, \eta = 0\right\}}  \right] \\
 &=\argmax_{d(\cdot, \cdot)} \left[ Corr\left\{d(\mathbf{y}^{\text{pred}},\boldsymbol{\gamma}),  s_0(\mathbf{y}^{\text{pred}}, \boldsymbol{\gamma})|\boldsymbol{\gamma},\eta=0\right\} \ SD\left\{ s_0(\mathbf{y}^{\text{pred}}, \boldsymbol{\gamma})|\boldsymbol{\gamma}, \eta = 0\right\} \right] \\
  &=\argmax_{d(\cdot, \cdot)} \left[  Corr\left\{d(\mathbf{y}^{\text{pred}},\boldsymbol{\gamma}),  s_0(\mathbf{y}^{\text{pred}}, \boldsymbol{\gamma})|\boldsymbol{\gamma},\eta=0\right\} \right]  = a s_0(\cdot, \cdot) + b, \ a,b \in \mathbb{R}, 
\end{align*}
where $Corr$ stands for correlation. Step (a) is based on \eqref{eq:LRexpansion} and Theorem \ref{theo:scorefisher}: $$\pi(\mathbf{y}^{\text{pred}}|\boldsymbol{\gamma},\eta) = \pi(\mathbf{y}^{\text{pred}}|\boldsymbol{\gamma},\eta=0)( 1 + s_0(\mathbf{y}^{\text{pred}},\boldsymbol{\gamma})\eta + \smallO(\eta) ).$$
\qed

\subsection{Proof of Theorem \ref{theo:s0best}}\label{app:s0best}


We utilise here the same notation of Appendix \ref{app:detect} and for fixed $\boldsymbol{\gamma}$ consider $$\lim_{\eta\to\infty}d_n(\mathbf{y}^{\text{pred}}, \boldsymbol{\gamma}) = d(\mathbf{y}^{\text{pred}}, \boldsymbol{\gamma})\sim N\{E(d,\boldsymbol{\gamma},\eta),V(d,\boldsymbol{\gamma},\eta)\}.$$ 

\textbf{Assumptions:}  For step (a), we assume that $SEV_n(d, \mathbf{y}, \eta)$ is integrable w.r.t. $\eta$ and converges uniformly to a continuous function, and that $d_n(\mathbf{y},\boldsymbol{\gamma})$ is asymptotically normal. Let $\alpha(d,\eta,\boldsymbol{\gamma}) = \Phi\left\{ (d(\mathbf{y},\boldsymbol{\gamma}) - E(d,\boldsymbol{\gamma},\eta))/SD(d,\boldsymbol{\gamma},\eta)\right\}$.  For step (b) we need to assume that $\alpha(d,\eta,\boldsymbol{\gamma})$ and $\partial_\eta \alpha(d,\eta,\boldsymbol{\gamma})$ are continuous functions w.r.t. $\eta$ in $\zeta(\eta)$, and that the functions $\alpha(d,\eta,\boldsymbol{\gamma})\pi(\boldsymbol{\gamma}|\mathbf{y},\eta)$ and  $\partial_\eta \alpha(d,\eta,\boldsymbol{\gamma})\pi(\boldsymbol{\gamma}|\mathbf{y},\eta)$ are integrable w.r.t. $\boldsymbol{\gamma}$, $\forall \eta \in \zeta(\eta)$. Lastly, for step (c) we need the same assumptions as Lemma \ref{prop:detect}.

We then have that:
$$
\lim_{n \to \infty}  \lim_{\eta \to 0} \frac{d}{d\eta} SEV_n(d, \mathbf{y}, \eta) \overset{(a)}{=} \lim_{\eta \to 0} \frac{d}{d\eta}  \lim_{n \to \infty}  SEV_n(d, \mathbf{y}, \eta) = \lim_{\eta \to 0} \frac{d}{d\eta}   SEV(d, \mathbf{y}, \eta),
$$
where $SEV(d, \mathbf{y}, \eta) = P\left\{ d(\mathbf{y}^{\text{pred}}, \boldsymbol{\gamma}) > d(\mathbf{y}, \boldsymbol{\gamma}) | \mathbf{y}, \eta\right\}$. The regularity assumptions on $SEV_n(d, \mathbf{y}, \eta)$ mentioned at the end of the proof allow exchanging limit with derivative in step (a). Then, by leveraging on the normality of $d$ we have:
\begin{align*}
SEV(d, \mathbf{y}, \eta) &= P\left\{ d(\mathbf{y}^{\text{pred}}, \boldsymbol{\gamma}) >  d(\mathbf{y}, \boldsymbol{\gamma}) | \mathbf{y}, \eta=0\right\} =  1 - \int \Phi\left\{ \frac{d(\mathbf{y}, \boldsymbol{\gamma}) - E(d,\boldsymbol{\gamma},\eta)}{SD(d,\boldsymbol{\gamma},\eta)}\right\} \pi(\boldsymbol{\gamma}|\mathbf{y},\eta) d\boldsymbol{\gamma}.
\end{align*}
Finally,
\begin{align*}
 \argmax_{d(\cdot, \cdot)} \left[ \lim_{\eta \to 0} \frac{d}{d\eta} SEV(d, \mathbf{y}, \eta) \right] &=  \argmin_{d(\cdot, \cdot)} \Bigg[  \lim_{\eta \to 0} \frac{d}{d\eta} \int \Phi\left\{ \frac{d(\mathbf{y}, \boldsymbol{\gamma}) - E(d,\boldsymbol{\gamma},\eta)}{SD(d,\boldsymbol{\gamma},\eta)}\right\} \pi(\boldsymbol{\gamma}|\mathbf{y},\eta) d\boldsymbol{\gamma} \Bigg] \\
  &\overset{(b)}{=} \argmin_{d(\cdot, \cdot)} \Bigg[ \lim_{\eta \to 0} \frac{d}{d\eta}\Phi\left\{ \frac{d(\mathbf{y}, \boldsymbol{\gamma}) - E(d,\boldsymbol{\gamma},\eta)}{SD(d,\boldsymbol{\gamma},\eta)}\right\} \Bigg] \\
   &= \argmax_{d(\cdot, \cdot)} \Bigg\{ \frac{1}{ SD[d,  \boldsymbol{\gamma}, 0]} \lim_{\eta\to0}\frac{d}{d\eta} E(d, \boldsymbol{\gamma}, \eta) \Bigg\} \\
  &\overset{(c)}{=} a s_0(\cdot, \cdot) + b, \ a,b \in \mathbb{R},
\end{align*}

Step (b) follows by exchanging the limit and derivative with integral and because $\pi(\boldsymbol{\gamma}|\mathbf{y},\eta)$ is positive and does not depend on $d$. Step (c) follows from Lemma \ref{prop:detect}.

\qed

 
\subsection{Proof of Proposition \ref{prop:variance_approx}}\label{app:variance_approx}

Here we consider the nuisance parameters $\boldsymbol{\gamma}$, which are $\boldsymbol{\beta}$ and $\boldsymbol{\theta} = (\tau_\epsilon, \boldsymbol{\theta}_2)$ to be fixed. The replicated data $\mathbf{y}^{\text{rep}}_0$ from the LGM in \eqref{eq:model} follows the distribution

\begin{equation*}\label{eq:replicated_density}
\pi(\mathbf{y}^{\text{rep}}_0) \sim N(\mathbf{B}\boldsymbol{\beta} , \ \  \tau_\epsilon^{-1} \mathbf{I} + \mathbf{A}(\mathbf{D}^T\text{diag}(\mathbf{h})^{-1}\mathbf{D})^{-1} \mathbf{A}^T).
\end{equation*}

Now, as in \eqref{eq:sensitivityconditional}, the BF sensitivity is  $s_0(\mathbf{y}^{\text{rep}}_0, \boldsymbol{\gamma}) = \sum_i d_i(\mathbf{y}^{\text{rep}}_0, \boldsymbol{\gamma})$, and the contribution from index $i$ is

\begin{equation*}\label{eq:sens_proof}
d_i(\mathbf{y}^{\text{rep}}_0, \boldsymbol{\gamma}) =  \frac{ b_i^4(\mathbf{y}^{\text{rep}}_0) + 3\Gamma_{ii}^2 - 6 b_i^2(\mathbf{y}^{\text{rep}}_0)\Gamma_{ii}}{8 h_i^3},
\end{equation*}
where,
$$ \mathbf{b}(\mathbf{y}^{\text{rep}}_0)  = \mathbf{D} \tau_{\boldsymbol{\epsilon}}  \mathbf{Q}_{\mathbf{w}}^{-1} \mathbf{A}^T(\mathbf{y}^{\text{rep}}_0-\mathbf{B}\boldsymbol{\beta}), \ \ \ 
\boldsymbol{\Gamma} = -\mathbf{D} \mathbf{Q}_{\mathbf{w}}^{-1} \mathbf{D}^T + \text{diag}(\mathbf{h}), \ \ \ \mathbf{Q}_{\mathbf{w}} = \tau_{\boldsymbol{\epsilon}} \mathbf{A}^T\mathbf{A}  + \mathbf{D}^T\text{diag}(\mathbf{h})^{-1}\mathbf{D}.
$$

The distribution of $\mathbf{b}(\mathbf{y}^{\text{rep}}_0)$ can be simplified to $N(\mathbf{0},\boldsymbol{\Gamma})$. Then, it follows from the central mixed moments' formula for the multivariate normal distribution \citep{guiard1986general} that $E[d_i(\mathbf{y}^{\text{rep}}_0, \boldsymbol{\gamma})]=0$, and $Cov(d_i(\mathbf{y}^{\text{rep}}_0, \boldsymbol{\gamma}), d_j(\mathbf{y}^{\text{rep}}_0, \boldsymbol{\gamma})) = (3\Gamma_{ij}^4)/(8 h_i^3 h_j^3)$. Therefore, the mean and variance of the reference distribution are $
E[s_0(\mathbf{y}^{\text{rep}}_0, \boldsymbol{\gamma})]=0$, and $
Var[s_0(\mathbf{y}^{\text{rep}}_0, \boldsymbol{\gamma})] = \sum_{i,j} Cov(d_i(\mathbf{y}^{\text{rep}}_0, \boldsymbol{\gamma}), d_j(\mathbf{y}^{\text{rep}}_0, \boldsymbol{\gamma})) =  \sum_{i,j} (3\Gamma_{ij}^4)/(8 h_i^3 h_j^3)$.

\qed

\section{Details about the pressure data application}\label{sect:geostatapp}

\subsection{Fitting the latent non-Gaussian model}

\begin{figure}[htp]
   \centering
   \includegraphics[width=0.49\linewidth]{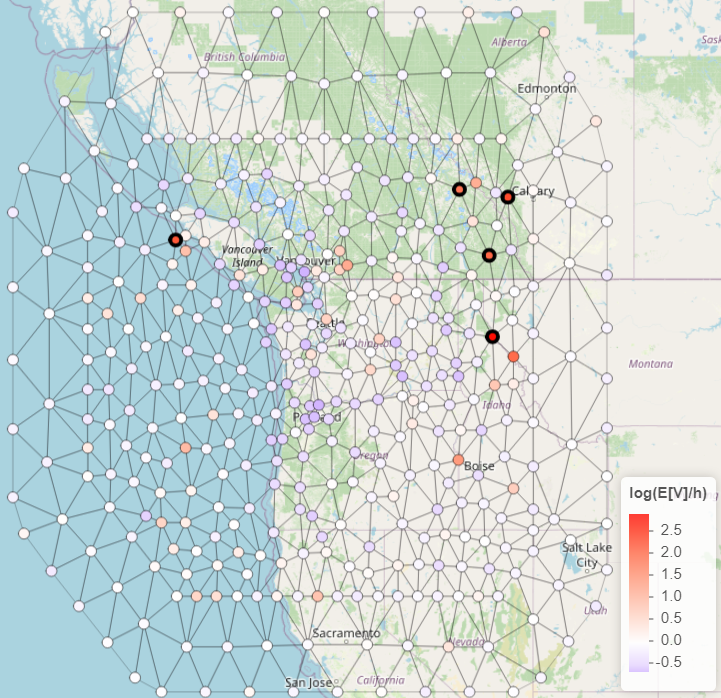}
   \caption{Plot of the posterior means of $\log(V_i/h_i)$ for each spatial index. The 5 nodes where significant departures from Gaussianity are detected are illustrated as circles with thicker contours.} 
  \label{fig:Vmean} 
\end{figure}

We fitted the latent non-Gaussian model (LnGM) to the pressure data using the Stan implementation in \cite{cabral2022controlling}. Details about how to fit this model in Stan can be found in \url{rafaelcabral96.github.io/nigstan/}. The LnGM considers $\mathbf{y} \sim N(\mathbf{A}\mathbf{w}, \sigma_\epsilon^2\mathbf{I})$, where the random effects $\mathbf{w}$ approximate a Matérn random field driven by NIG noise, and are defined by $\mathbf{D}\mathbf{w} = \boldsymbol{\Lambda}(\eta)$, for matrices $\mathbf{A}$ and $\mathbf{D}$ specified in \cite{cabral2022controlling}. A useful alternative representation is 
    \begin{align*}
\mathbf{w}|\mathbf{V} &\sim N\left(\mathbf{0}, (\mathbf{D}^{T}\text{diag}(\mathbf{V})\mathbf{D})^{-1} \right), \\
V_{i}|\eta &\sim \text{IGaussian}(h_i,\eta^{-1}h_i^2)   \ \ i=1,\dotsc, N,
\end{align*}
where now the latent variables $\mathbf{w}$ are normally distributed conditioned on a set of mixing variables $\mathbf{V}$ that follow the inverse-Gaussian distribution. These mixing variables add extra flexibility to the initial Gaussian model.
For the Gaussian base model, we have \mbox{$V_i=h_i$}, so values of $V_i$ much larger or smaller than $h_i$ indicate more flexibility or rigidity of the latent field at node $i$. We show in Figure \ref{fig:Vmean} the posterior means of $V_i/h_i$, for each spatial node. We highlighted the nodes where the 90\% posterior credible intervals of $V_i$ do not include the value $h_i$ which indicates a significant departure from Gaussianity. The resulting plot highlights the same 5 nodes as the ones shown in Figure \ref{fig:pressd1}, which shows the BF sensitivity for each spatial node. The advantage of the BF sensitivity plot is that it identifies the locations where the latent Gaussian assumptions lack flexibility solely from the output of the LGM fit in R-INLA, i.e., without fitting the more expensive LnGM.

\subsection{Sensitivity of spatial prediction}

After fitting the LGM in R-INLA, we can obtain the posterior distribution of the weights $\mathbf{w}$, which correspond to the latent spatial field $X(\mathbf{s})$ evaluated at the mesh nodes. The triangulation mesh of the finite element method is shown in Figure~\ref{fig:pressd2}. The random field $X(\mathbf{s})$ evaluated at the unobserved locations  $\mathbf{s}_1, \mathbf{s}_2, \dotsc$ of the $100\times100$ grid composes the vector $\mathbf{x}_P= \{X(\mathbf{s}_1), X(\mathbf{s}_2), \dotsc\}$ and it is given by the linear combination $\mathbf{x}_P= \mathbf{A}_P\mathbf{w}$, where $\mathbf{A}_P$ is the projector matrix with elements $\psi_i(\mathbf{s}_j)$. This matrix is computed with the command \verb|inla.spde.make.A| available in R-INLA. Then, the posterior mean of the predictions is $\mathbf{A}_P E(\mathbf{w}|\mathbf{y})$, where $E(\mathbf{w}|\mathbf{y})$ is the posterior mean of the random effects obtained from R-INLA. The sensitivity of the predictions illustrated in Figure~\ref{fig:pressd2} is derived from Theorem \ref{theo:sens}. Specifically, we have $
s_{P}  =   \mathbf{A}_P \lim_{\eta \to 0}\partial_\eta E(\mathbf{w}|\mathbf{y},\eta)$,
where the term on the right pertains to the sensitivity of the random effects $\mathbf{w}$ and is obtained from the function \verb|ng.check|  (see Section \ref{sect:inlaimpl}).

\section{Predictive distributions in model checking} \label{sect:app3}

There are several ways to simulate data from the fitted model (see \cite{robins2000asymptotic}). The number of choices increase when dealing with hierarchical models, because we might choose, for instance, to simulate the second level variables from their prior distribution, and the third level variables from their posterior distribution or vice-versa. Before discussing the usage of different predictive distributions, we present an illustrative example that sheds light on some of the challenges of performing predictive checking on the latent levels of hierarchical models.

\subsection{Illustrative example} \label{sect:illust}

\begin{figure}[htp]
   \centering
   \includegraphics[width=0.49\linewidth]{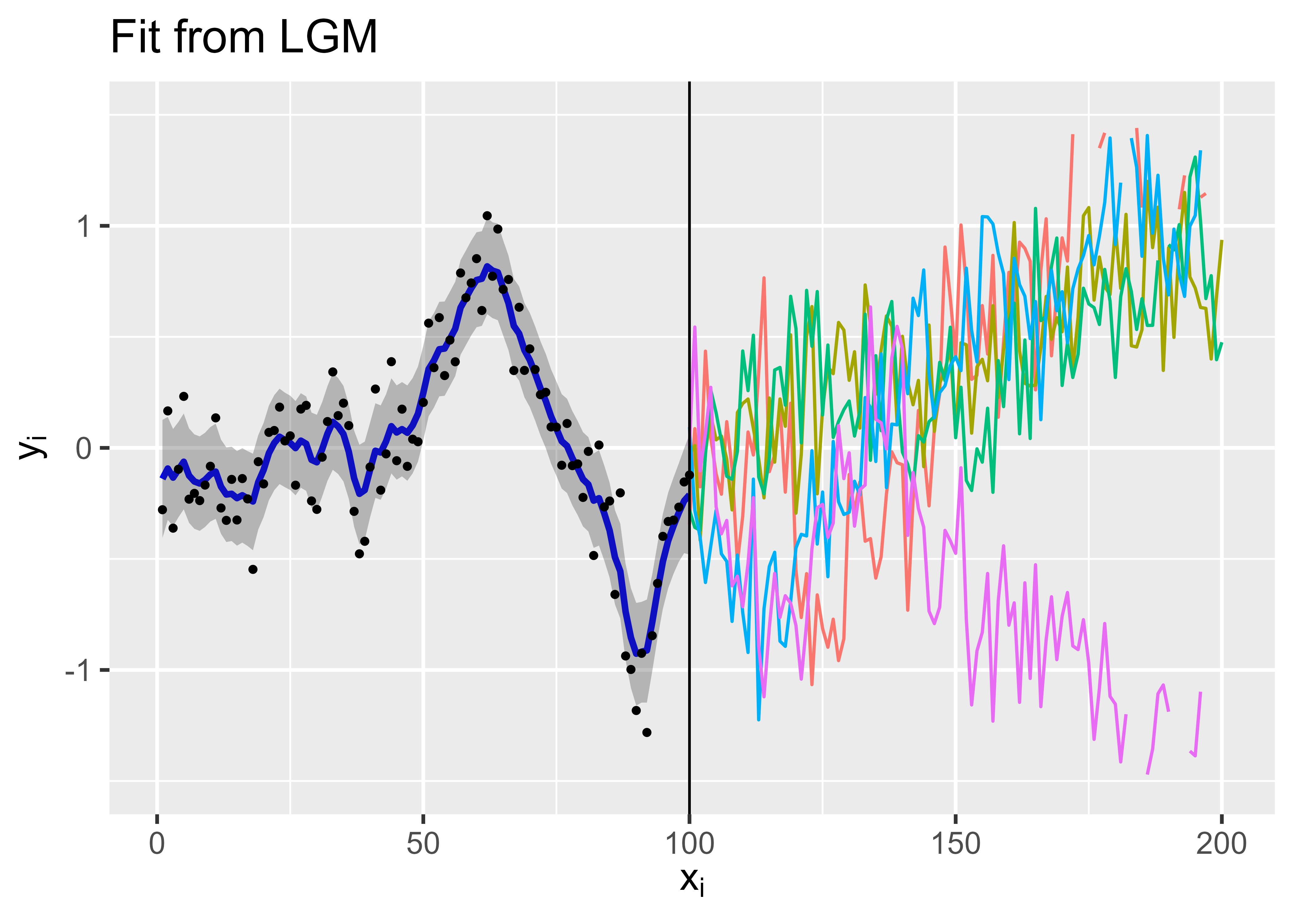}
   \includegraphics[width=0.49\linewidth]{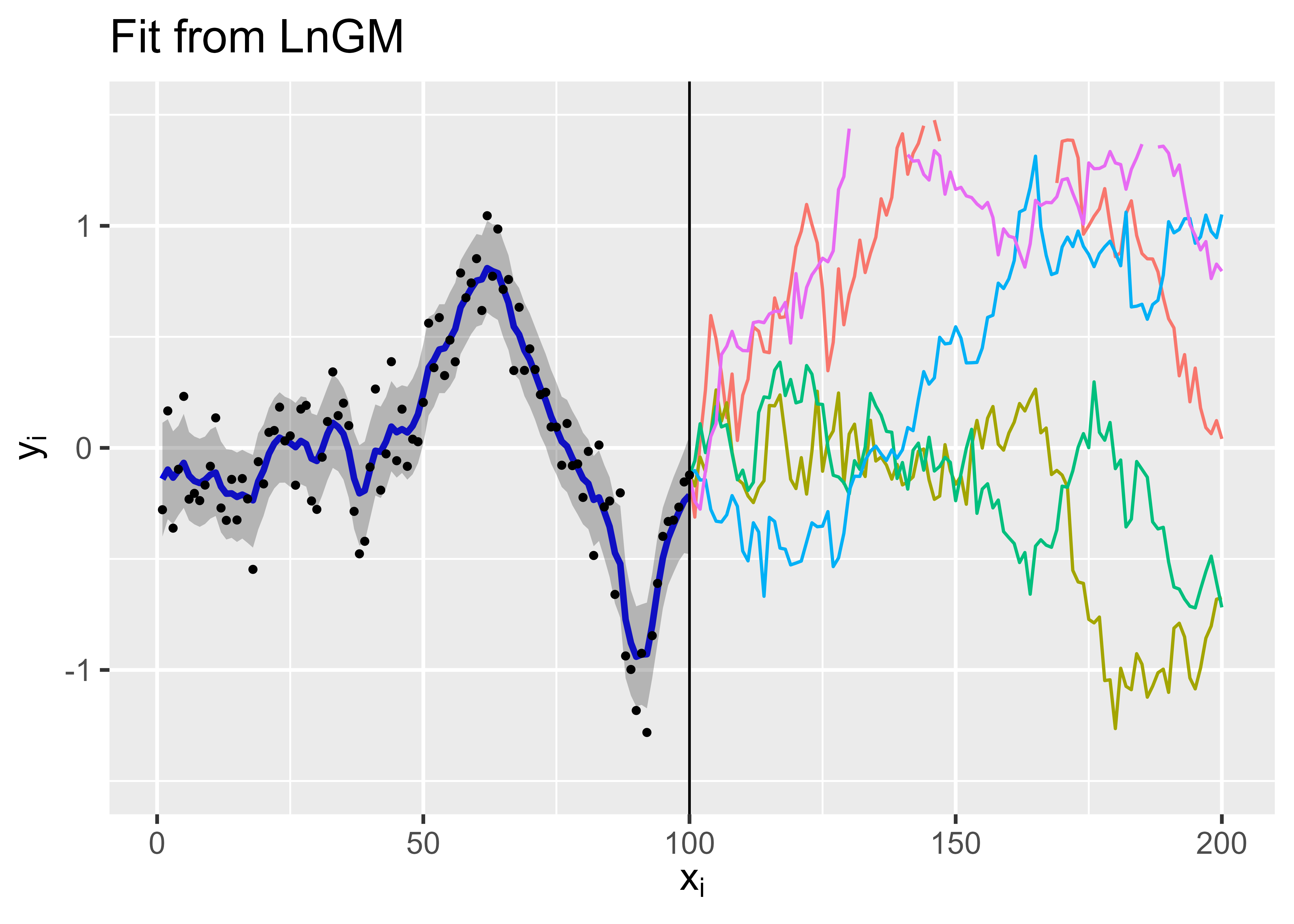} 
   \includegraphics[width=0.49\linewidth]{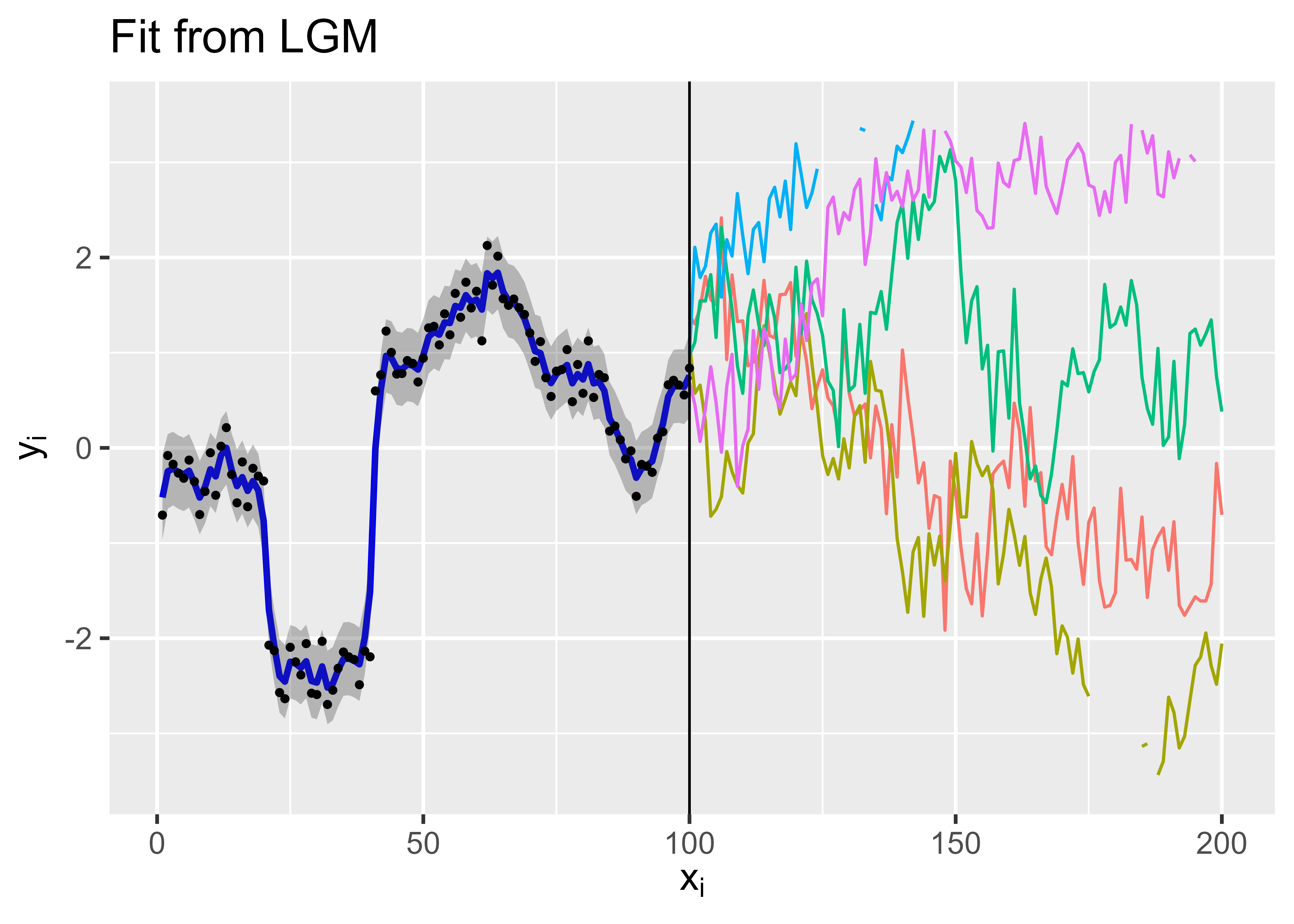}
   \includegraphics[width=0.49\linewidth]{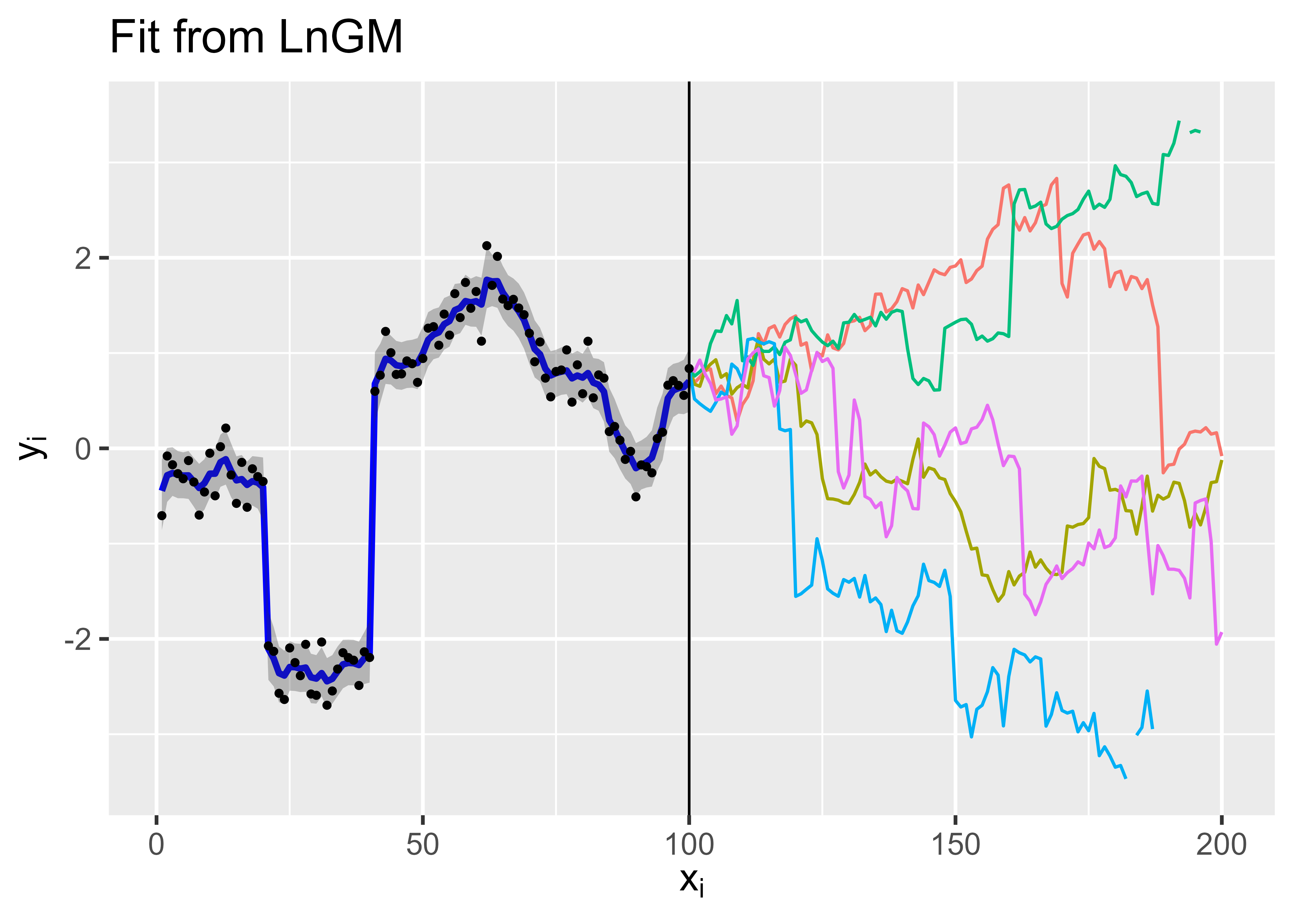} 
   \caption{The data is shown as black circle markers for the original time series (top) and the modified time series (bottom). The posterior mean of the latent field (blue line) and 97.5\% credible intervals (shaded area) are shown in the interval $x\in [1,100]$, and five prediction replicates are shown in the interval $x\in [101,200]$.} 
  \label{fig:datafit} 
\end{figure}

We consider the problem of smoothing time series data containing sudden jumps. Similarly to \cite{wang2018bayesian}, the data were simulated from the following model:
$$
y_i = sin^3(2\pi (w_i/10000)^3) + \sigma_{\boldsymbol{\epsilon}} \epsilon_i, \ \  \epsilon_i\sim N(0,1),
$$
where $w_i\in [0,100]$ are equally spaced with distance 1 and $\sigma_{\boldsymbol{\epsilon}}^2=0.04$.  We added a jump of size -2 at location 20 and another one of size 3 at location 40 and subtracted the simulated processes by the sample mean. The original and modified time series with two added jumps are shown in Figure~ \ref{fig:datafit}. We considered the following model:
\begin{equation}\label{eq:illusmodel}
y_i = w_i + \sigma_{\boldsymbol{\epsilon}} \epsilon_i,    
\end{equation} 
where $\mathbf{w}=(w_1,\dotsc,w_{100})$ follows a random walk of order 1 (RW1) prior. The dependency matrix $\mathbf{D}$ that defines this RW1 component is
$$
\mathbf{D}(\sigma_{\mathbf{w}}) = \frac{1}{\sigma_{\mathbf{w}}}\begin{pmatrix}
-1 & 1  &   &  & \\
 & -1 & 1 &    &\\
 &  &   \ddots & \ddots & \\
 &  &     & -1 & 1
\end{pmatrix}, 
$$
where $\mathbf{D}(\sigma_{\mathbf{w}})\mathbf{w} = \mathbf{\Lambda}(0)$, with $\mathbf{\Lambda}(0)$ being an i.i.d. standard Gaussian noise vector.


\sloppy The two sudden jumps can be seen as process outliers in that, for a Gaussian RW1 model, the likelihood of those sudden jumps occurring is very low. To account for these events, we also considered an equivalent non-Gaussian RW1 component driven by NIG noise, where \mbox{$w_{i+1}-w_{i} = \Lambda_i, \ i=1,\dotsc, 99$}. Large events in the noise process $\Lambda_i$ will lead to large jumps in the process $\mathbf{w}$. We chose the priors \mbox{$\tau_{\boldsymbol{\epsilon}} = 1/\sigma_{\boldsymbol{\epsilon}}^2 \sim \text{Gam}(1,0.5)$}, \mbox{$\tau_{\mathbf{w}} = 1/\sigma_{\mathbf{w}}^2 \sim \text{Gam}(1,0.005)$} (using the shape and rate reparameterization), 
and \mbox{$\eta \sim \text{Exp}(1)$}. The smoothed process $\pi(\mathbf{w}|\mathbf{y})$ and prediction replicates drawn from $\pi(\mathbf{y}_{101:200}|\mathbf{y})$ for the LGM, and LnGM fits are shown in Figure~\ref{fig:datafit}.


\begin{figure}[htp]
   \centering
   \includegraphics[width=0.49\linewidth,height=0.4\linewidth]{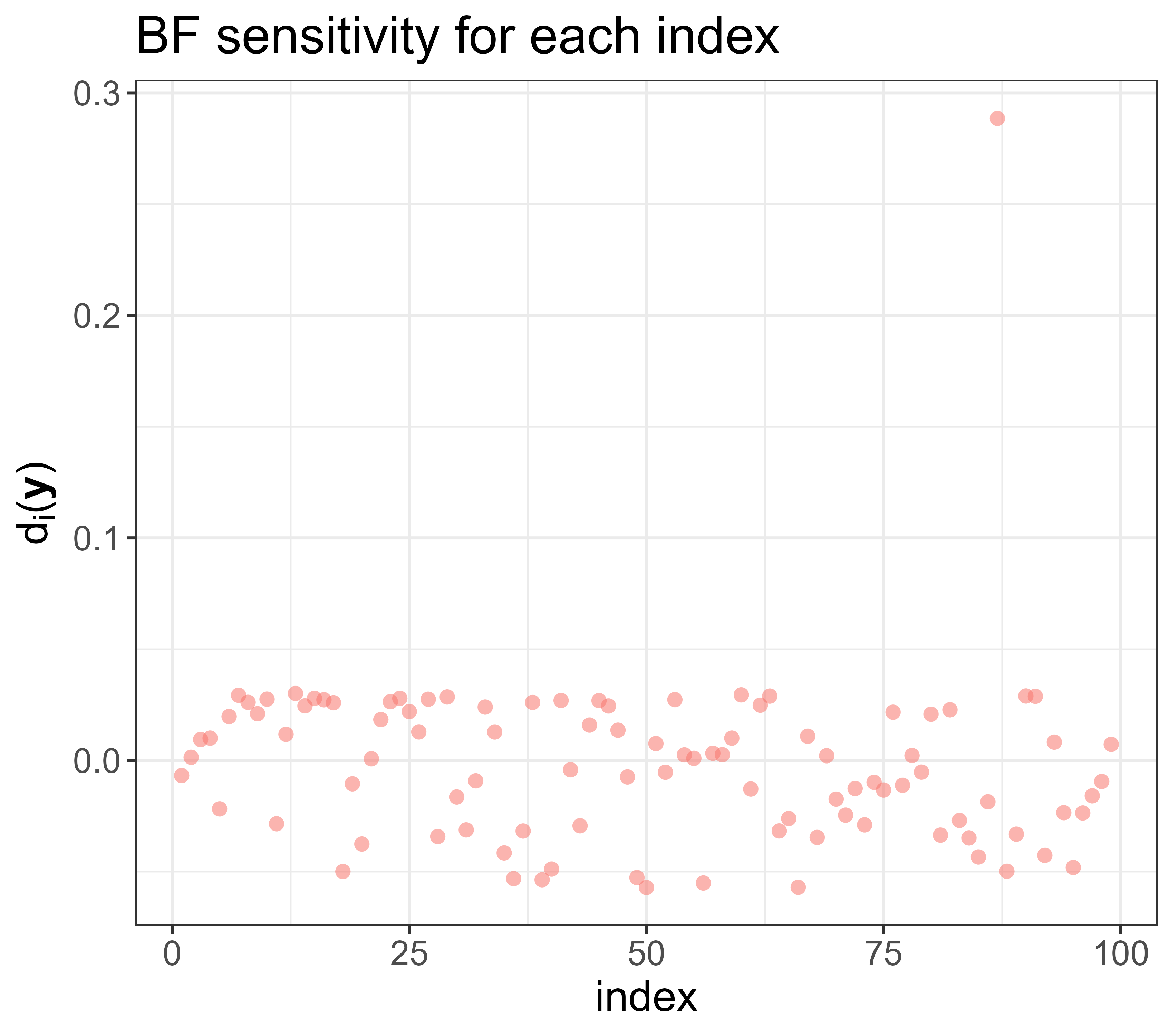}
   \includegraphics[width=0.49\linewidth,height=0.4\linewidth]{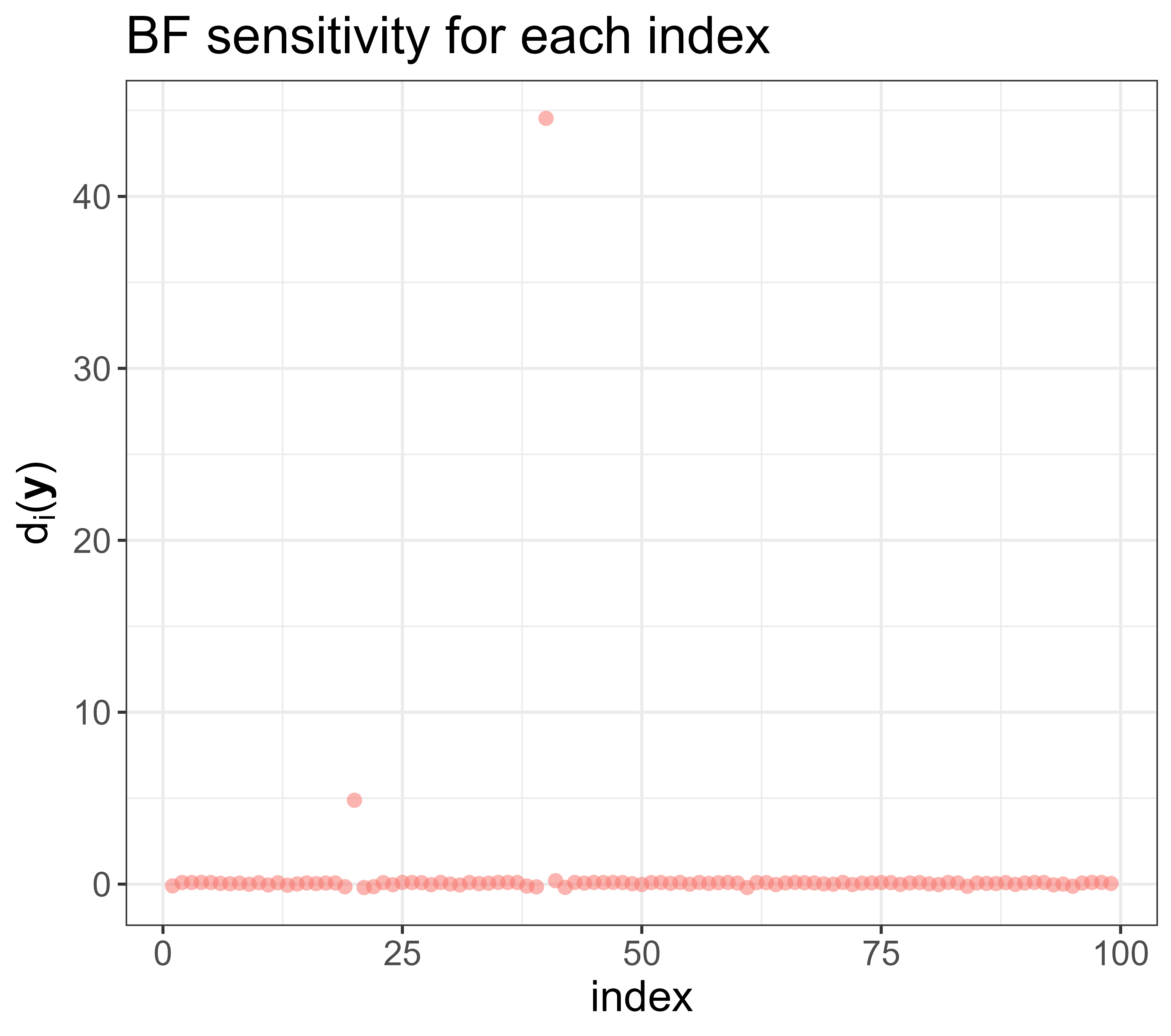}
   \caption{Marginal likelihood sensitivity measures $d_i$ for the LGM fit of the original time series (left) and time series with jumps (right).} 
  \label{fig:scorefit1} 
\end{figure}

The sensitivity measures $d_i(\mathbf{y})$ computed from the LGM fit are shown in Figure~\ref{fig:scorefit1}, where the locations 20 and 40 of the two sudden jumps clearly stand out for the modified time series. We make a few observations regarding the LGM fit:



\begin{enumerate}
    \item  The two jumps were captured by posterior distribution of the LGM's latent field $\pi(\mathbf{w}|\mathbf{y})$ although at the cost of a $45\%$ increase in its marginal scale, compared with the original time series with no jumps. On the other hand, the marginal scale of the LnGM only increased by $8\%$.
    \item  Each term of the score $s_0(\mathbf{y}) = \sum_i d_i(\mathbf{y})$ in \eqref{eq:scorebayes} can be simplified to
    $$d_i(\mathbf{y}) = E\left.\left[\frac{(\{\mathbf{D}(\sigma_{\mathbf{w}})\mathbf{w}\}_i^2-3)^2-6}{8} \right\vert \mathbf{y}, \eta=0 \right],$$
    where $\{\mathbf{D}(\sigma_{\mathbf{w}})\mathbf{w}\}_i = (w_{i+1}-w_i)/\sigma_{\mathbf{w}}$, and so $d_i(\mathbf{y})$ will be large when there is a sudden jump at location $i$ of the smoothed process $\pi(\mathbf{w}|\mathbf{y})$, which is visible in Figure~\ref{fig:scorefit1}.
    \item  Unlike the pressure example of Section \ref{sect:pressure}, the LGM does not oversmooth the data since the two sudden jumps are present in the posterior mean of the latent field. Thus, one may think there is no benefit in considering an LnGM beyond having less uncertain predictions. However, the future replicates of the LnGM fit exhibit jumps for the modified time series, while the future replicates of the LGM fit do not. 
\end{enumerate}

The last point exemplifies that even if no under fitting can be detected at the observed locations, the LGM may fail to generate important path features in future predictions. To detect predictive hindrance of this kind, we can consider replicates drawn from the  predictive density given in Section \ref{sect:pred_choice}, which we will further explore here.


\subsection{Predictive distributions}

We consider here several predictive distributions for LGMs. We use the notation $\mathbf{y}^{\text{obs}}$ to refer to the observed data and we we define three types of replicates: $\mathbf{y}^{\text{prior}}$, $\mathbf{y}^{\text{mixed}}$, and $\mathbf{y}^{\text{post}}$, which correspond to prior, mixed, and posterior predictive replicates, respectively.

\subsubsection{Prior predictive distribution}

\begin{figure}[htp]
   \centering
   \includegraphics[width=0.49\linewidth]{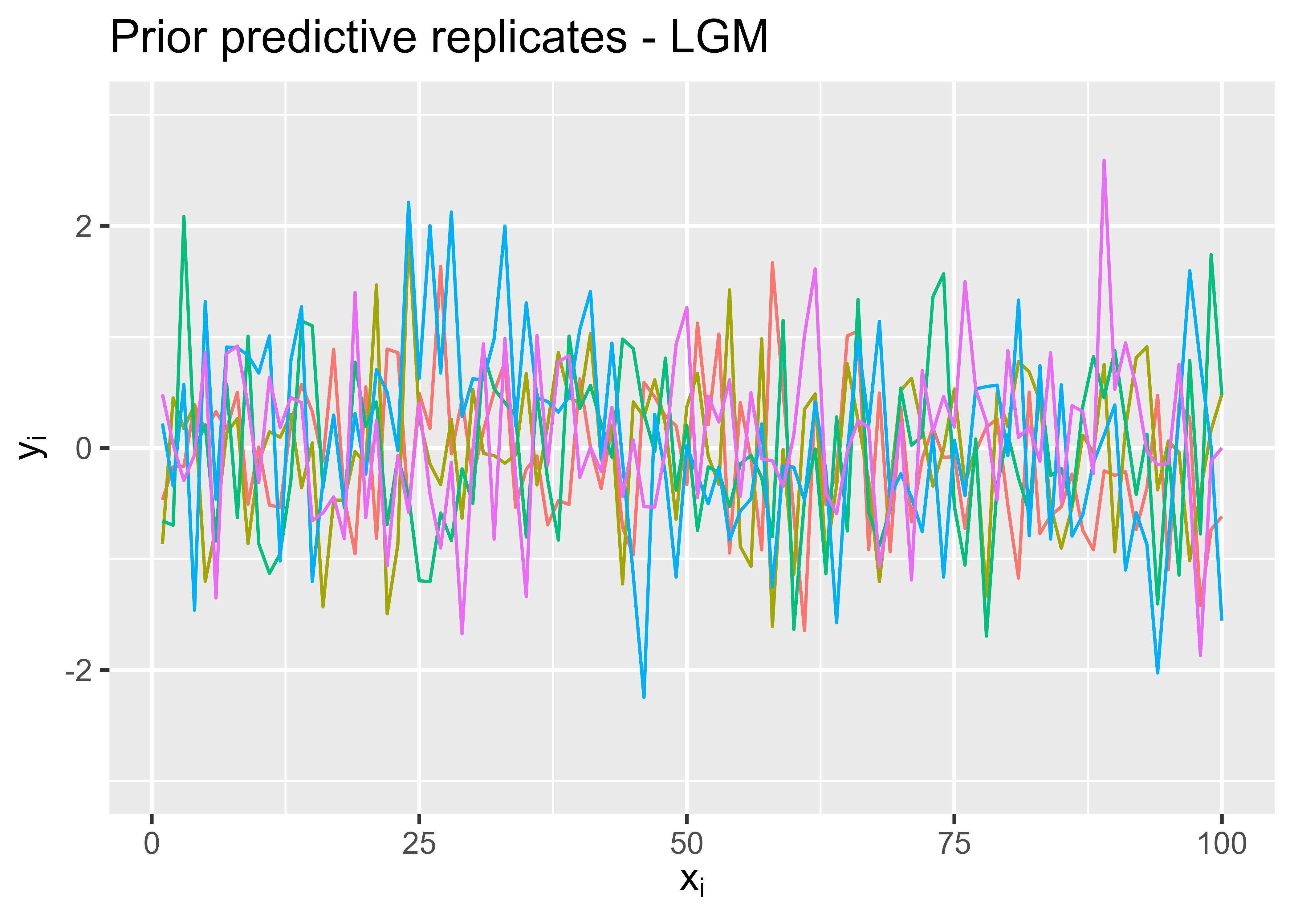}
      \includegraphics[width=0.49\linewidth]
      {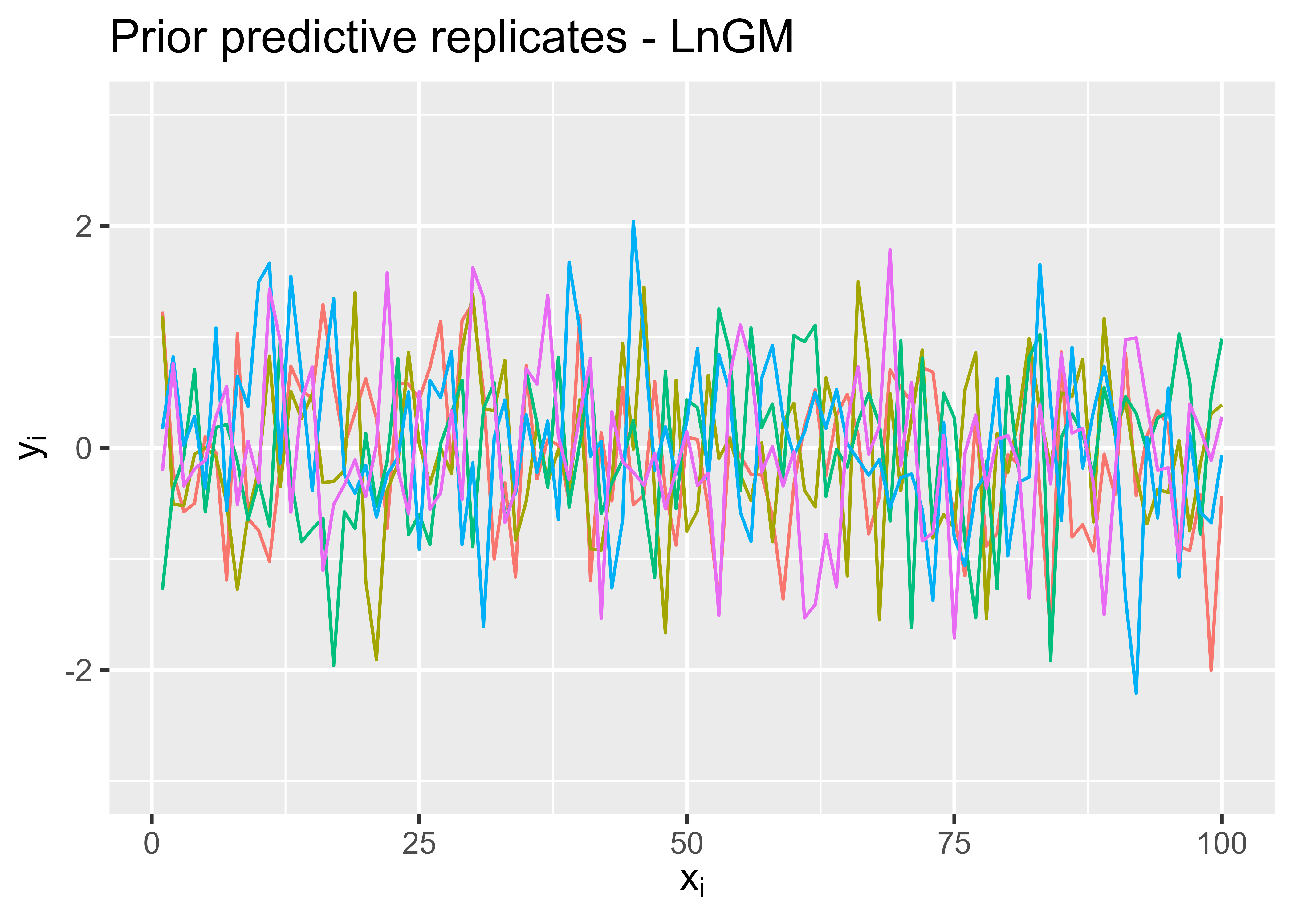}
   \caption{Prior predictive replicates from the LGM (left) and LnGM (right) fit, for the data with two sudden jumps.} 
  \label{fig:priorrep} 
\end{figure}

\cite{box1980sampling} proposed checking Bayesian models by contrasting the observed data with the prior predictive distribution, which for LGMs with only one random effects $\mathbf{w}$ takes the form $$\pi(\mathbf{y}^{\text{prior}}|\eta=0) = \int \pi(\mathbf{y}^{\text{prior}}| \boldsymbol{\beta},\mathbf{w}, \boldsymbol{\theta}_1) \pi(\boldsymbol{\beta}, \mathbf{w}, \boldsymbol{\theta}|\eta=0) d\boldsymbol{\beta}d\mathbf{w} d\boldsymbol{\theta}.$$
In the present context this predictive distribution might not be desirable, since it does not allow distinguishing an inadequate prior for the latent random effects $\mathbf{w}$ with an inadequate prior for $\boldsymbol{\beta}$ and $\boldsymbol{\theta}$. We often choose very diffuse priors for the linear effects $\boldsymbol{\beta}$ and hyperparameters $\boldsymbol{\theta}$ or shrinkage priors meant to penalize model complexity \citep{simpson2017penalising}. Although they can lead to reasonable posterior distributions, these priors are often not generative. A prior is not generative if it does not lead to replicates consistent with our understanding of the problem \citep{gelman2017prior}. Figure \ref{fig:priorrep} shows five prior predictive replicates for the model of Section  \ref{sect:illust}. The marginal standard deviation of the prior predictive process is considerably higher than the fitted process in Figure~\ref{fig:datafit}. At the same time, the replicates appear to be comparatively less smooth than the data. 
 

\begin{figure}[htp]
   \centering
   \includegraphics[width=0.49\linewidth]{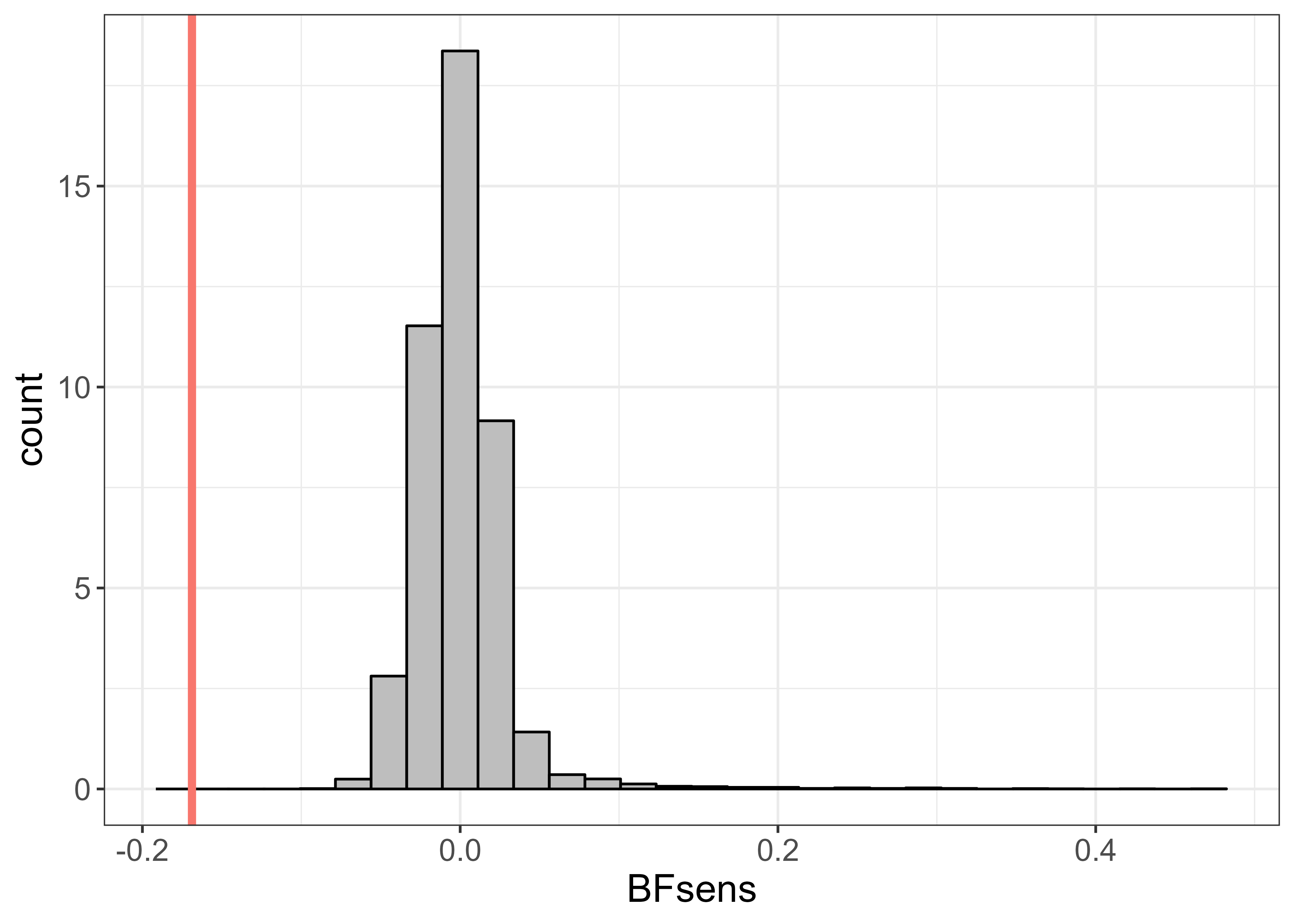}
      \includegraphics[width=0.49\linewidth]
      {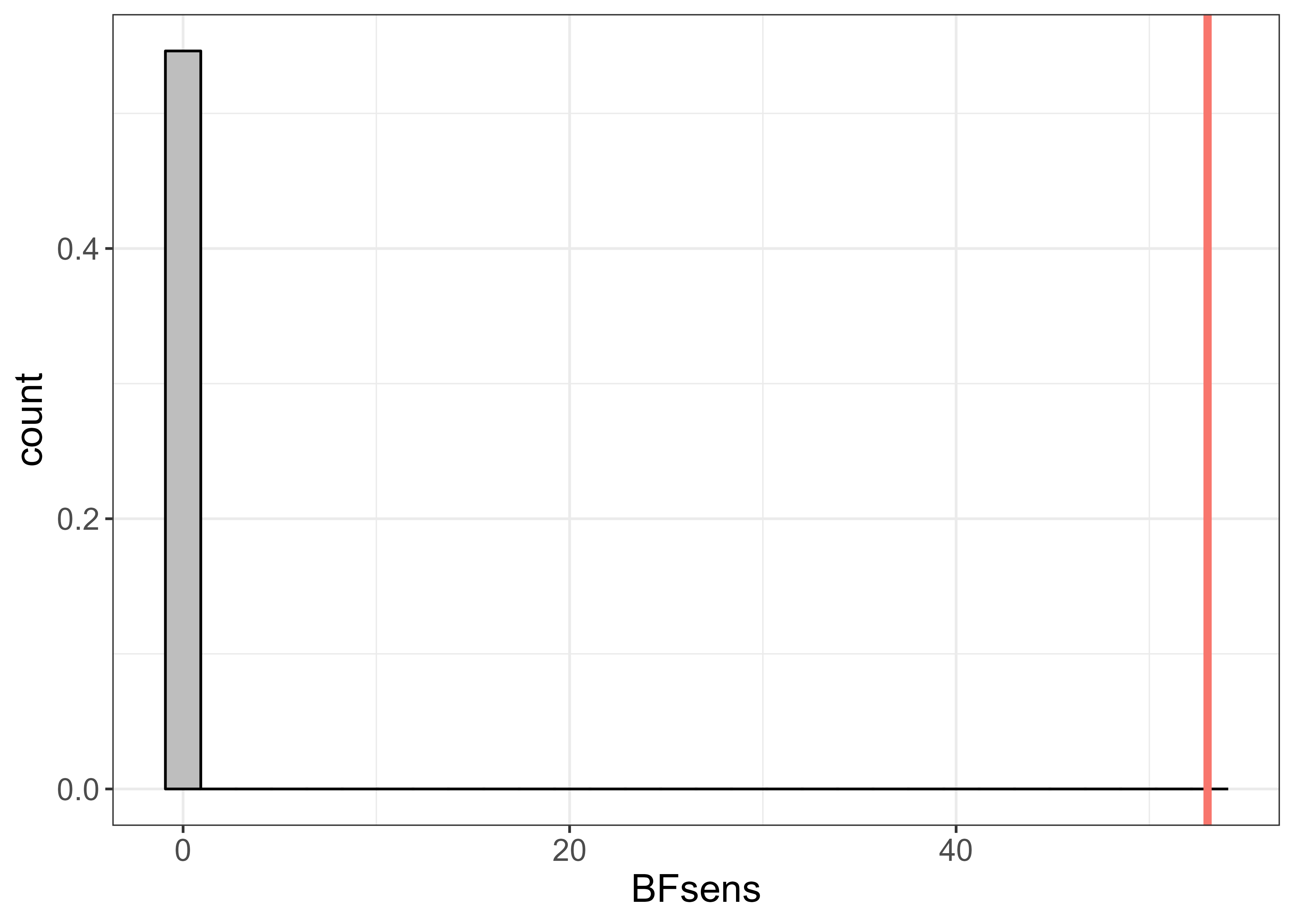}
   \caption{Reference distribution $s_0(\mathbf{y}_0^\text{prior}, \hat{\boldsymbol{\gamma}})$ obtained for the original time series with no jumps (left) and the time series with the added jumps (right) using prior predictive replicates. The red vertical line indicates the BF sensitivity at the observed data.} 
  \label{fig:priorpredictive} 
\end{figure}

We simulated the reference distribution of the BF sensitivity $s_0(\mathbf{y}_0^\text{prior},\hat{\boldsymbol{\gamma}})$, for the time series with and without the sudden jumps, where $\hat{\boldsymbol{\gamma}} = (\hat{\sigma}_{\mathbf{w}}, \hat{\sigma}_{\boldsymbol{\epsilon}})$ is the prior mode of the nuisance parameters. The posterior distribution of $\eta$ is close to 0 for the original time series with no jumps, suggesting adequacy of the LGM. However, as shown in Figure~\ref{fig:priorpredictive}, in the $s_0$-dimension, the observed data is highly unlikely given this prior predictive distribution since the prior distribution of the nuisance parameters is not representative of the data.

\vspace{0.5cm}

\subsubsection{Posterior predictive density}

\begin{figure}[htp]
   \centering
   \includegraphics[width=0.49\linewidth]{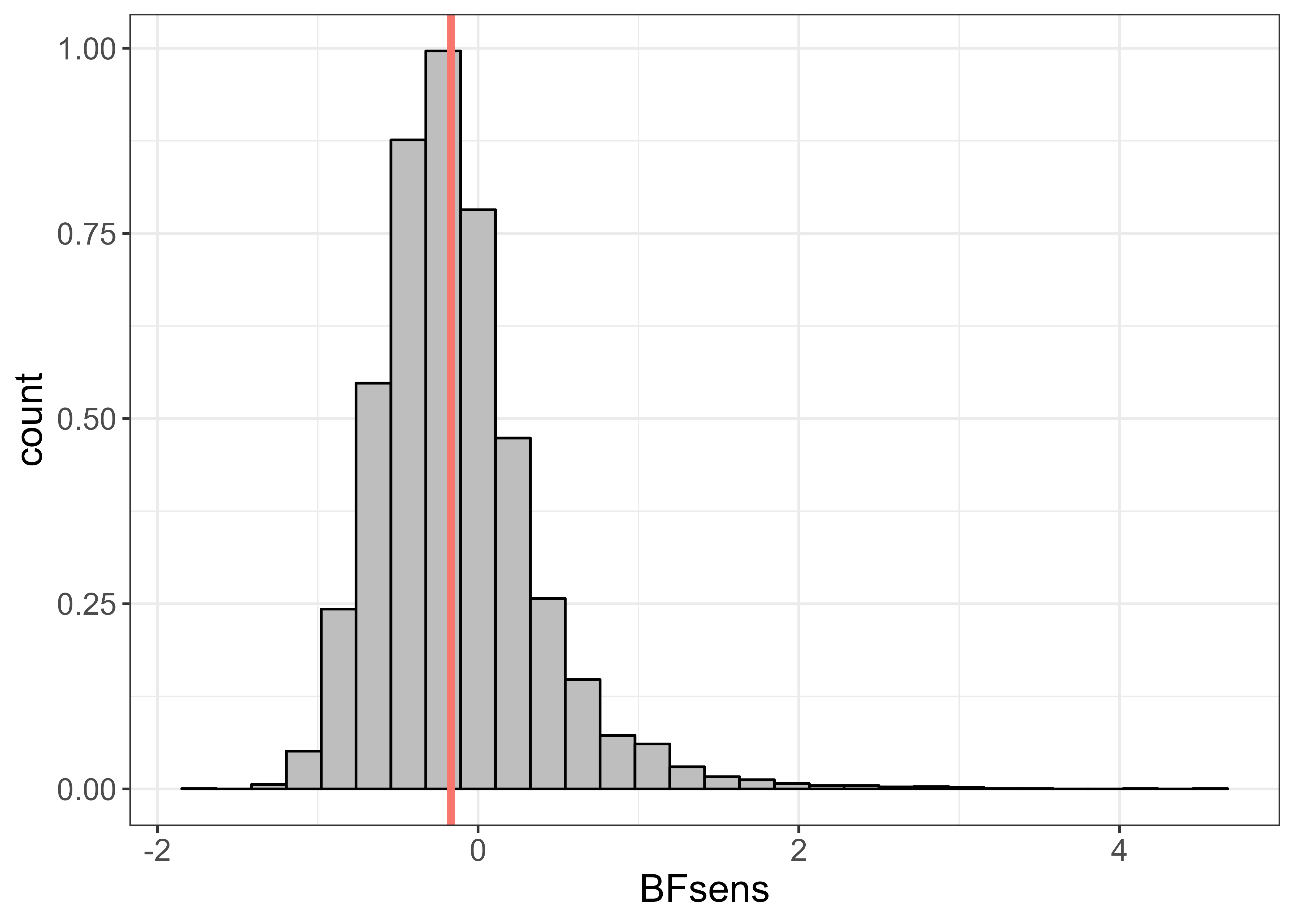}
      \includegraphics[width=0.49\linewidth]
      {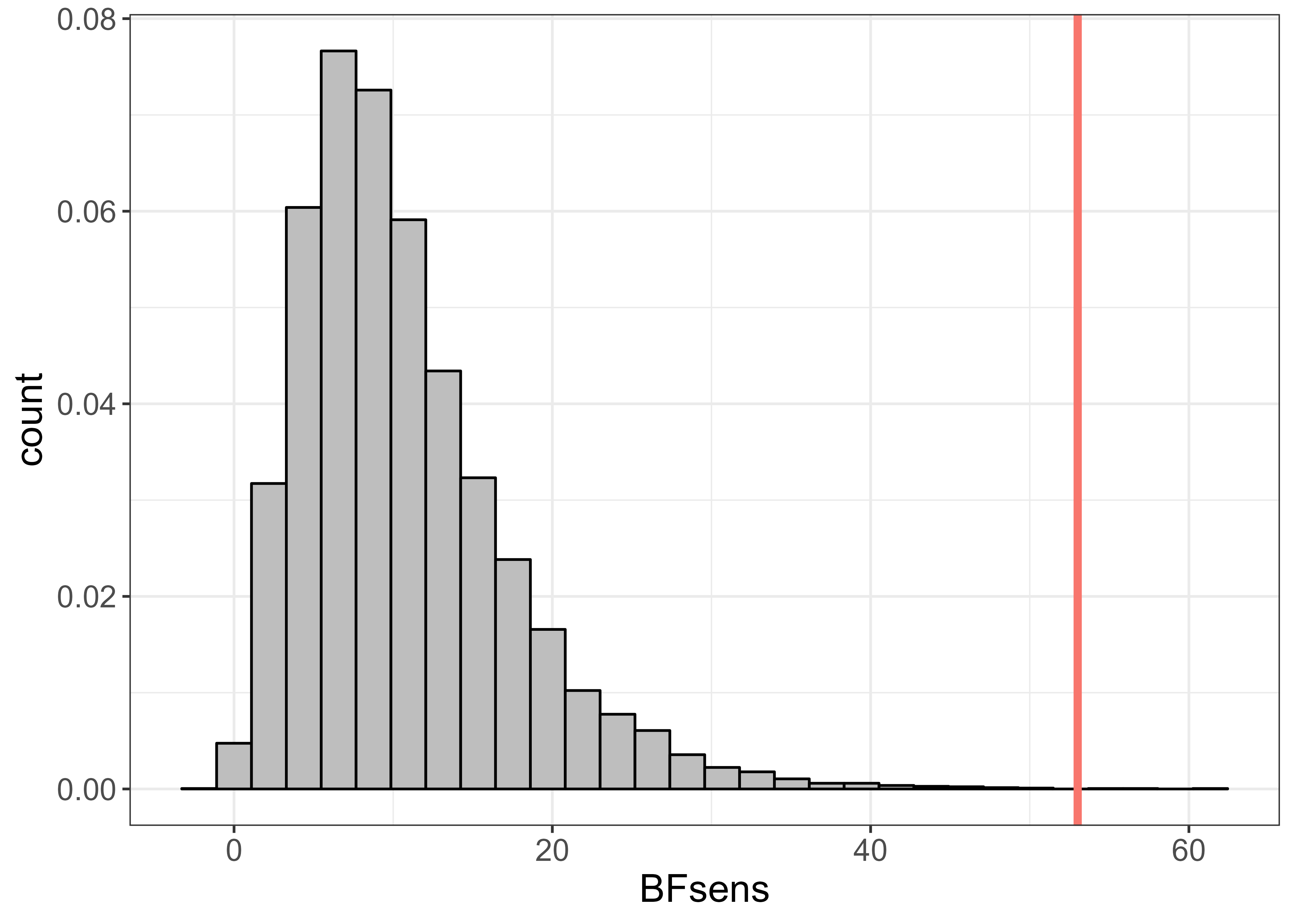}
   \caption{Reference distribution $s_0(\mathbf{y}_0^\text{post}, \hat{\boldsymbol{\gamma}})$ obtained for the original time series with no jumps (left) and the time series with the added jumps (right) using posterior predictive replicates. The red vertical line indicates the observed value.} 
  \label{fig:posteriorpred} 
\end{figure}

Posterior predictive replicates \citep{rubin1984bayesianly,gelman1996posterior} are drawn from
$$\pi(\mathbf{y}^{\text{post}}|\mathbf{y}^{\text{obs}},\eta=0) = \int \pi(\mathbf{y}^{\text{post}}|  \boldsymbol{\beta},\mathbf{w}, \boldsymbol{\theta}_1) \pi(\boldsymbol{\beta}, \mathbf{w}, \boldsymbol{\theta}|\mathbf{y}^{\text{obs}},\eta=0) d\boldsymbol{\beta}d\mathbf{w} d\boldsymbol{\theta}.$$
 \cite{gelman1995bayesian} rejected prior predictive checking by claiming that it considers the prior as true ``population distribution" while the posterior predictive distribution views the prior as an outdated initial estimate. Figure~\ref{fig:mixedrep} shows five posterior predictive replicates for the model and data of Section~\ref{sect:illust}.  Unlike prior predictive checking, if we use the posterior predictive distribution, the first time series with no jumps is not unusual along the $s_0$-dimension, as shown in Figure~\ref{fig:posteriorpred}, which compares  $s_0(\mathbf{y}_0^\text{post},\hat{\boldsymbol{\gamma}})$ with $s_0(\mathbf{y},\hat{\boldsymbol{\gamma}})$, where $\hat{\boldsymbol{\gamma}}$ is the posterior mode of the nuisance parameters.

\begin{figure}[htp]
   \centering
   \includegraphics[width=0.49\linewidth]{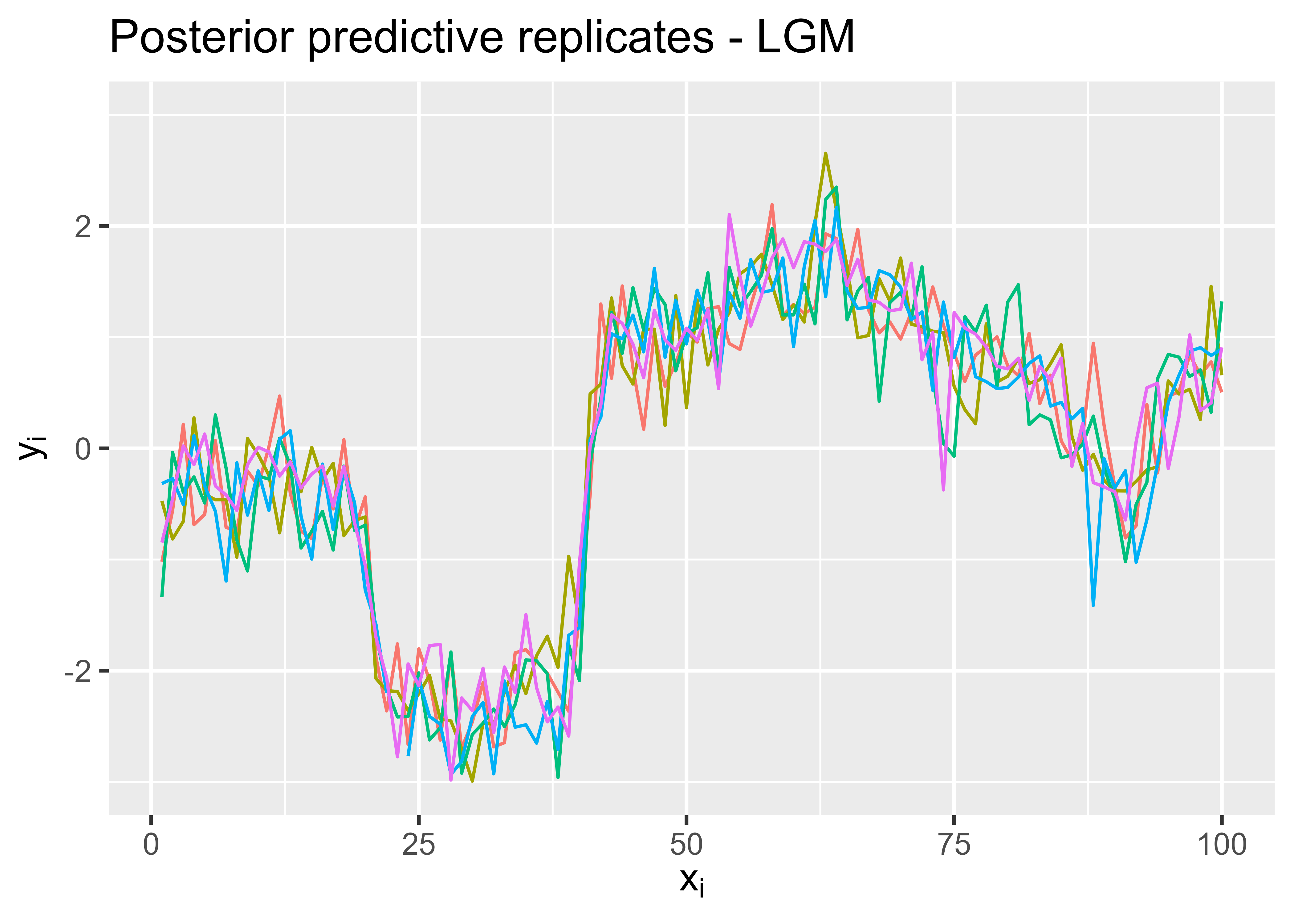}
      \includegraphics[width=0.49\linewidth]
      {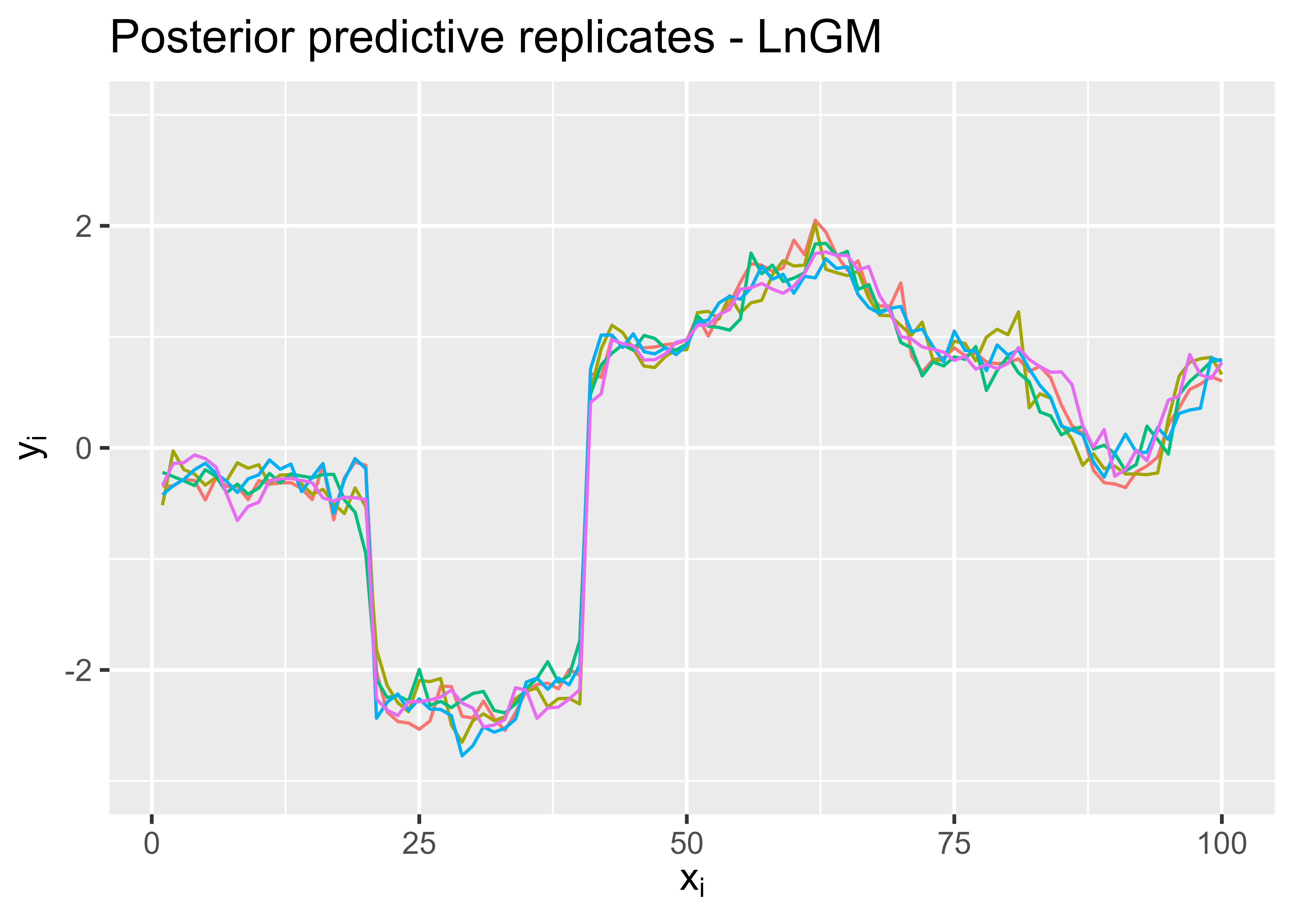}
   \caption{Posterior predictive replicates from the LGM (left) and LnGM (right) fit for the data with two sudden jumps.} 
  \label{fig:postrep} 
\end{figure}

\vspace{0.5cm}

\subsubsection{Mixed predictive replicates}

\sloppy More replicating schemes can be taken into account. To check if $\pi(\mathbf{w}|\boldsymbol{\theta}_2,\eta=0)$ is a good generative prior, we can think of \mbox{$\pi(\mathbf{y}^{\text{mixed}}|\boldsymbol{\beta},\boldsymbol{\theta},\eta=0) = \int \pi(\mathbf{y}^{\text{mixed}}|  \boldsymbol{\beta},\mathbf{w}, \boldsymbol{\theta}_1)\pi(\mathbf{w}|\boldsymbol{\theta}_2,\eta=0)d\mathbf{w}$} as the likelihood, obtained by marginalising out $\mathbf{w}$. By taking into account the prior distribution $\pi(\boldsymbol{\beta},\boldsymbol{\theta})$, we end up with a 2-stage hierarchical formulation for LGMs. Since the priors for the linear effects and hyperparameters are often not generative, it makes sense to sample from their posterior distribution. This leads to the mixed predictive distribution that we use in Section \ref{sect:ref_dist}: 
\begin{equation}\label{eq:mixed}
\pi(\mathbf{y}^{\text{mixed}}|\mathbf{y}^{\text{obs}},\eta=0) = \int \pi(\mathbf{y}^{\text{mixed}}|\boldsymbol{\beta},\boldsymbol{\theta},\eta=0) \pi(\boldsymbol{\beta}, \boldsymbol{\theta}|\mathbf{y}^{\text{obs}},\eta=0) d\boldsymbol{\beta}d\boldsymbol{\theta}.
\end{equation}

\begin{figure}[htp]
   \centering
   \includegraphics[width=0.49\linewidth]{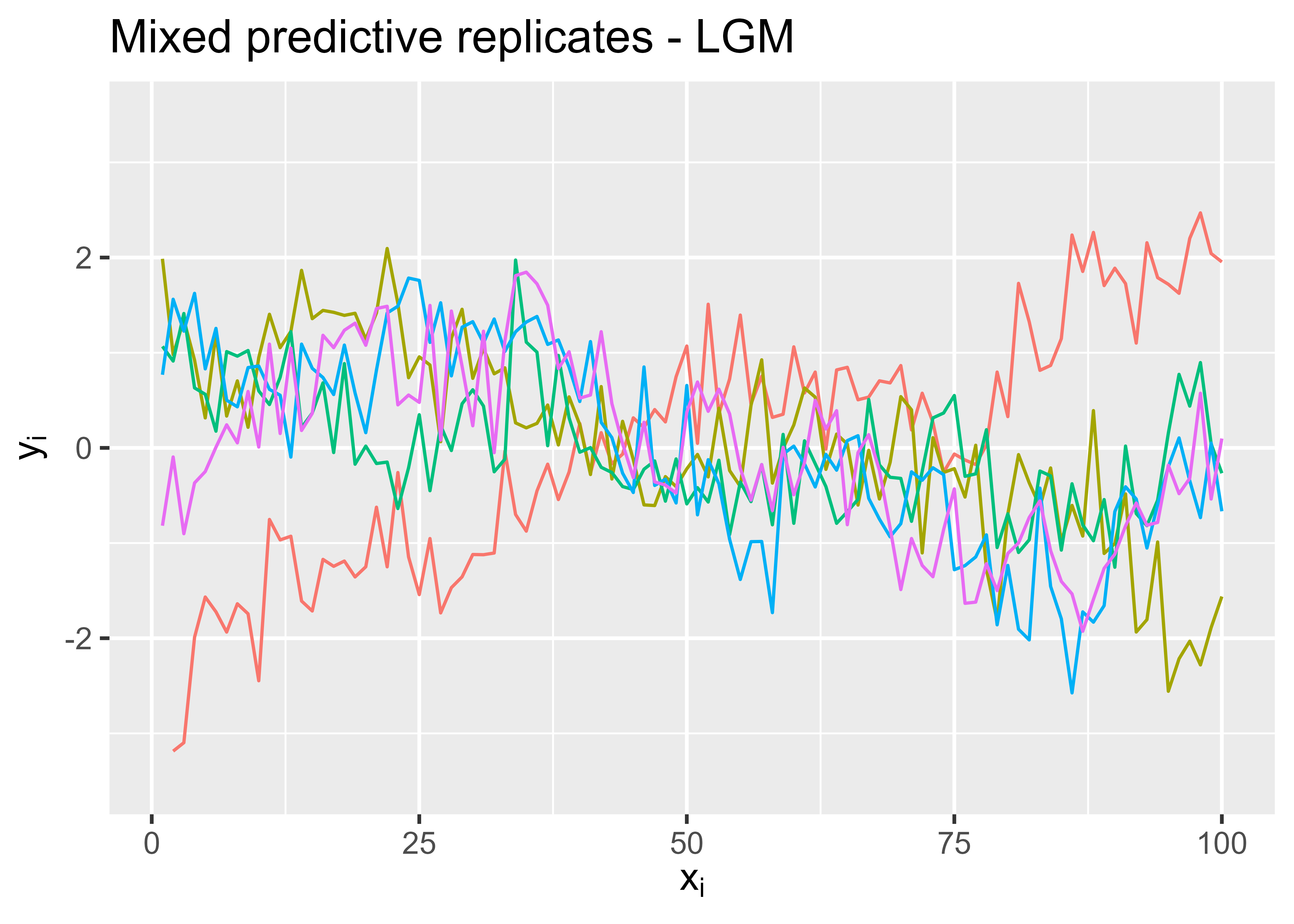}
      \includegraphics[width=0.49\linewidth]
      {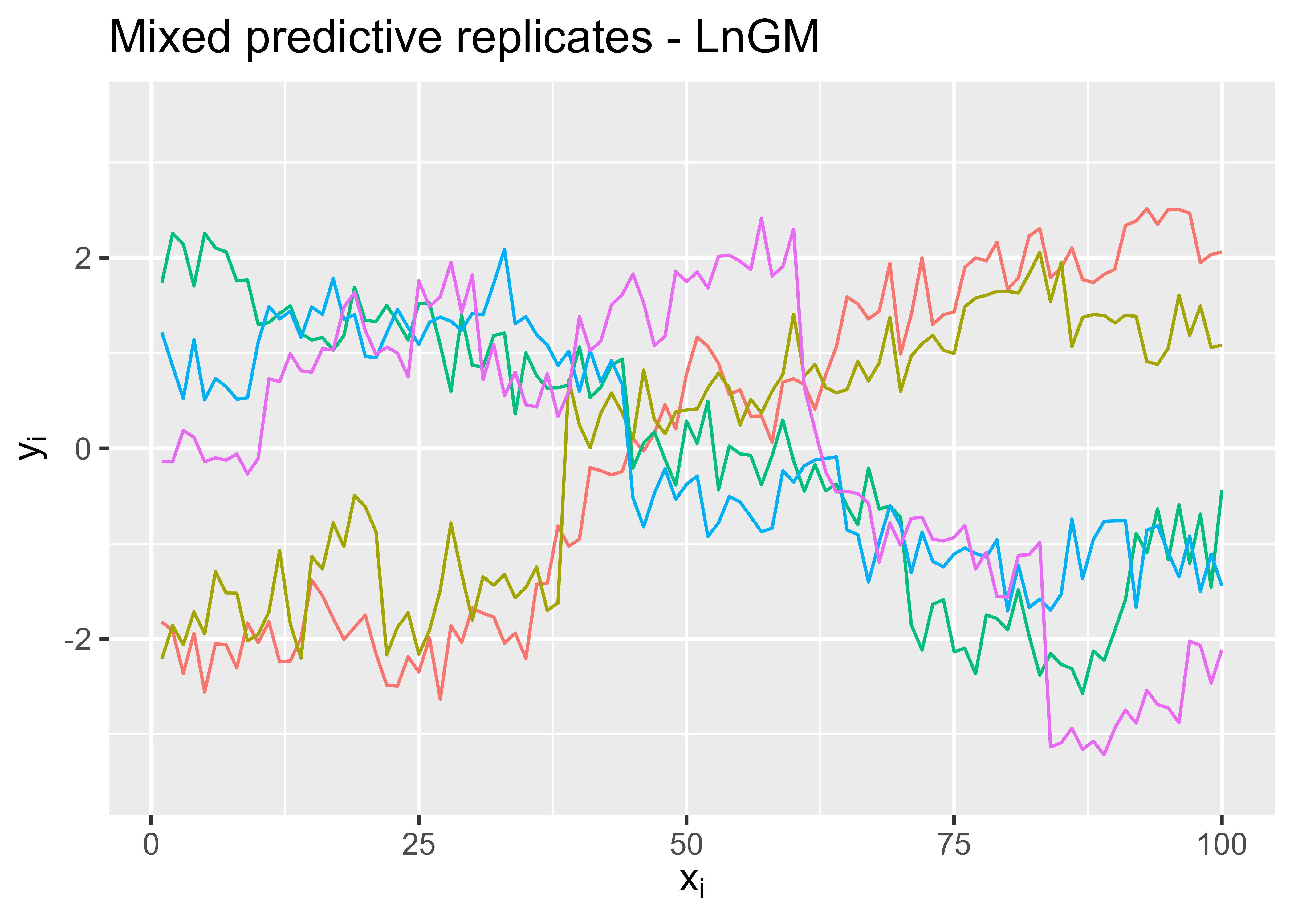}
   \caption{Mixed predictive replicates from the LGM (left) and LnGM (right) fit for the data with two sudden jumps.} 
  \label{fig:mixedrep} 
\end{figure}

\begin{figure}[htp]
   \centering
   \includegraphics[width=0.49\linewidth]{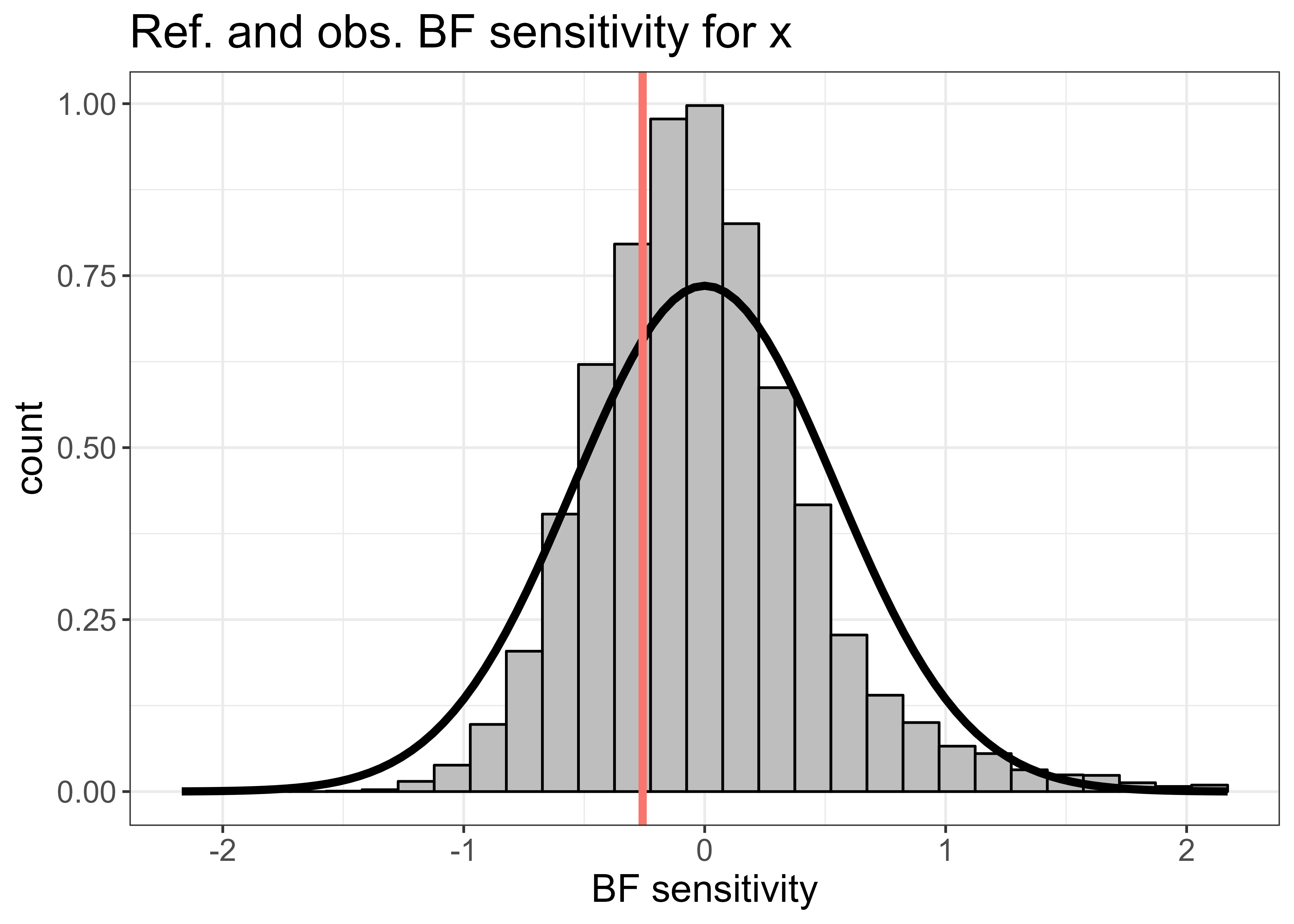}
      \includegraphics[width=0.49\linewidth]
      {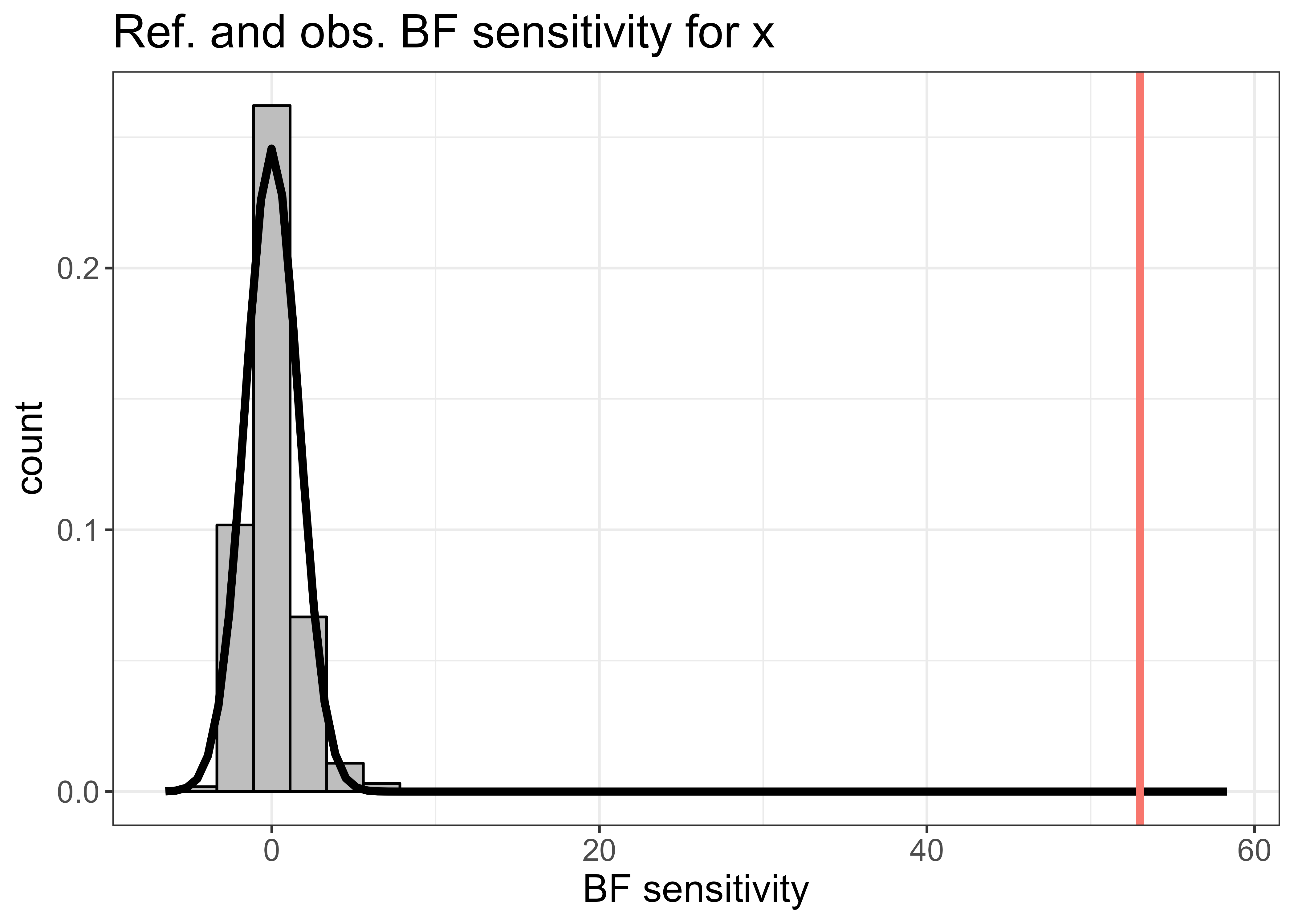}
   \caption{Reference distribution $s_0(\mathbf{y}_0^\text{pred}, \hat{\boldsymbol{\gamma}})$ obtained for the original time series with no jumps (left) and the time series with the added jumps (right). The red vertical line indicates the observed value, and the black curve approximates the reference distribution based on Proposition \ref{prop:variance_approx}.} 
  \label{fig:RW1score} 
\end{figure}

Predictive distributions of the previous kind that involve the prior distribution for the variables of one level of the hierarchical model and the posterior distribution of the variables at another level are referred to as ``mixed predictive" distributions in \cite{gelman1996posterior}.
Mixed predictive replicates for the LGM and LnGM fits are shown in Figure~\ref{fig:mixedrep} for the time series with two sudden jumps.
The sample paths of these replicates resemble the future predictions in Figure~\ref{fig:datafit}. Namely, we see a similar level of smoothness and jumps for the LnGM model (we explore this resemblance in the next paragraph). Thus, if the model checks reveal inadequacy using this mixed predictive distribution, it means there are features in the data that will not be replicated in future time points (in the example of this section, it is the sudden jumps). The reference distribution $s_0(\mathbf{y}^\text{mixed}, \hat{\boldsymbol{\gamma}})$ obtained from this predictive scheme is shown in Figure \ref{fig:RW1score}. For the modified time series with two jumps, we can see that this predictive check is less optimistic than the one in Figure \ref{fig:posteriorpred} since the samples are closer to 0.


The mixed predictive scheme density in \eqref{eq:mixed} is associated with several prediction tasks, including predicting the observations of an unobserved subject in a longitudinal study. In this case, the temporal random effects are often sampled from their prior distribution, while the linear effects and hyperparameters are sampled from the posterior distribution. Other predictive tasks are long-range forecasting in a time series model or predicting data on a territory far away from the location of the observations in a spatial model. The relationship between these prediction tasks and the mixed predictive scheme in \eqref{eq:mixed} is demonstrated next. Let us consider here predictions at unobserved locations $\mathbf{y}^{\text{unob}}$, and the latent vectors $\mathbf{w}^{\text{obs}}$ and $\mathbf{w}^{\text{unob}}$ at the observed and unobserved locations. Replicates draw from \eqref{eq:mixed} share similar sample path features as $\mathbf{y}^{\text{unob}}$, as suggested by the~following~relationship (we omit the conditioning on $\eta=0$):
\begin{align}\label{eq:ypred}
    &\pi(\mathbf{y}^{\text{unob}}|\mathbf{y}^{\text{obs}}) = \int \pi(\mathbf{y}^{\text{unob}}, \mathbf{w}^{\text{unob}}, \mathbf{w}^{\text{obs}}, \boldsymbol{\beta}, \boldsymbol{\theta}| \mathbf{y}^{\text{obs}}) d\mathbf{w}^{\text{unob}}d\mathbf{w}^{\text{obs}}d\boldsymbol{\beta}d\boldsymbol{\theta} \\ 
    &= \int \pi( \mathbf{y}^{\text{unob}}| \mathbf{w}^{\text{unob}},\boldsymbol{\beta}, \boldsymbol{\theta}_1) \pi(\mathbf{w}^{\text{unob}}|\mathbf{w}^{\text{obs}},\boldsymbol{\theta}_2) \pi(\mathbf{w}^{\text{obs}}|\mathbf{y}^{\text{obs}},\boldsymbol{\theta}_2) \times 
    \pi(\boldsymbol{\beta}, \boldsymbol{\theta}|\mathbf{y}^{\text{obs}}) 
    d\mathbf{w}^{\text{obs}} d\mathbf{w}^{\text{unob}}  d\boldsymbol{\beta}d\boldsymbol{\theta} \nonumber \\ 
    &\approx  \int \pi( \mathbf{y}^{\text{unob}}| \mathbf{w}^{\text{unob}}, \boldsymbol{\beta}, \boldsymbol{\theta}_1) \pi(\mathbf{w}^{\text{unob}}|\boldsymbol{\theta}_2) 
    \pi(\boldsymbol{\beta}, \boldsymbol{\theta}|\mathbf{y}^{\text{obs}})
     d\mathbf{w}^{\text{unob}}  d\boldsymbol{\beta} d\boldsymbol{\theta}. \nonumber
\end{align}
\sloppy For time series and spatial statistics models, the joint prior distribution $\pi(\mathbf{w}^{\text{obs}},\mathbf{w}^{\text{unob}}|\boldsymbol{\theta}_2)$ often considers decaying correlation with lag. Therefore, the unobserved latent process, which is sufficiently distant from the observed locations, will be practically independent of it, motivating the approximation   
\mbox{$\pi(\mathbf{w}^{\text{unob}}|\mathbf{w}^{\text{obs}},\boldsymbol{\theta}_2) \approx \pi(\mathbf{w}^{\text{unob}}|\boldsymbol{\theta}_2)$}, 
which serves as the basis for the approximation in \eqref{eq:ypred}. We note that $\pi(\mathbf{y}^{\text{unob}}|\mathbf{y}^{\text{obs}})$ in \eqref{eq:ypred} closely resembles the posterior predictive density of \eqref{eq:mixed}, where both involve the prior distribution for the latent field and the posterior distribution for the nuisance parameters and thus have similar sample path features.


\section{Asymptotic approximation for the reference distribution} \label{sect:refapprox}

For i.i.d. data and simple 1-level models with no nuisance parameters, the reference distribution of the Fisher's score $s^\text{RS}(\mathbf{y}_0^{\text{pred}}) = \lim_{\eta \to 0}\partial_\eta \pi(\mathbf{y}_0^{\text{pred}}|\eta)$ can be asymptotically approximated by a normal distribution with variance given by the Fisher's information $I^\text{RS} = -E\{\lim_{\eta \to 0}\partial^2_\eta \pi(\mathbf{y}_0^{\text{pred}}|\eta)\}$ \citep{rao2005score}, where $\mathbf{y}_0^{\text{pred}}$ stands for replicated data from the base model. The resulting test is Rao's score test, which considers the statistic $(s^\text{RS}(\mathbf{y}))^2/I^\text{RS}$, which asymptotically follows the $\chi^2$ distribution under the base model. In practice, the observed Fisher's information, $- \lim_{\eta \to 0}\partial^2_\eta \pi(\mathbf{y}|\eta)$, can be used as a consistent estimator of the Fisher's information, and it can be readily computed from the observed data. 

Likewise, we could approximate the variance of $s_0(\mathbf{y}_0^{\text{pred}}, {\boldsymbol{\gamma}})$ by the measure $I_0(\mathbf{y}, {\boldsymbol{\gamma}})$ given in Theorem \ref{theo:scorefisher}, for fixed nuisance parameters:
\begin{align*}
    I_0(\mathbf{y}, {\boldsymbol{\gamma}}) = -E\left.\left\{g(\mathbf{y},\mathbf{z}) \right\vert \mathbf{y}, {\boldsymbol{\gamma}}, \eta = 0 \right\} - V\left.\left\{ p(\mathbf{y},\mathbf{z}) \right\vert \mathbf{y},   {\boldsymbol{\gamma}}, \eta = 0 \right\}, \\
      p(\mathbf{y},\mathbf{z}) = \lim_{\eta\to0} \frac{d}{d\eta}  \log \pi(\mathbf{y}, \mathbf{z}| \eta), \ \ 
    g(\mathbf{y},\mathbf{z}) = \lim_{\eta\to0} \frac{d^2}{d\eta^2}  \log \pi(\mathbf{y}, \mathbf{z}| \eta).
\end{align*}
We can compute $I_0(\mathbf{y}, {\boldsymbol{\gamma}})$ by Monte Carlo from samples of the fitted model since it only involves the posterior expectation and variance of $g(\mathbf{y},\mathbf{z})$ and  $p(\mathbf{y},\mathbf{z})$, respectively. For the latent non-Gaussianity check of Section \ref{sect:sensmeasures}, if we consider random effects driven by NIG noise, we have:
$$
p(\mathbf{w},\boldsymbol{\theta}_2) = \lim_{\eta \to 0} \frac{d}{d\eta} \log \pi(\mathbf{w}|\boldsymbol{\theta}_2, \eta) = \sum_{i=1}^n \frac{(\{\mathbf{D}(\boldsymbol{\theta}_2)\mathbf{w}\}_i^2 - 3 h_i)^2 - 6 h_i^2}{8 h_i^3},$$
and,
\begin{align} \label{eq:ngpert2}
g(\mathbf{w},\boldsymbol{\theta}_2)& = \lim_{\eta \to 0} \frac{d^2}{d\eta^2} \log \pi(\mathbf{w}|\boldsymbol{\theta}_2, \eta)  \\ &= \sum_{i=1}^n \frac{-3 h_i^3 - 3 h_i^2 \{\mathbf{D}(\boldsymbol{\theta}_2)\mathbf{w}\}_i^2 + 6 h_i \{\mathbf{D}(\boldsymbol{\theta}_2)\mathbf{w}\}_i^4 - \{\mathbf{D}(\boldsymbol{\theta}_2)\mathbf{w}\}_i^6}{8 h_i^5}. \nonumber
\end{align}

We prove in Sections \ref{sect:meanproof} and \ref{sect:varproof} that \mbox{$E\{s_0(\mathbf{y}_0^{\text{pred}},{\boldsymbol{\gamma}})\} = 0$}, and \mbox{$I = V\{s_0(\mathbf{y}_0^{\text{pred}},{\boldsymbol{\gamma}})\} = E\{I_0(\mathbf{y}_0^{\text{pred}},{\boldsymbol{\gamma}})\}$}, for fixed values of $\boldsymbol{\gamma}$. However, establishing the conditions for the asymptotic normality of $s_0(\mathbf{y}_0^\text{pred}, {\boldsymbol{\gamma}})$ and if $I_0(\mathbf{y}, {\boldsymbol{\gamma}})$ is a consistent estimator of $I$ is not straightforward in general, because in our models the data is not i.i.d. and the dimension of the latent variables grows with the data in some LGMs. Nonetheless, simulations suggest that $s_0(\mathbf{y}_0^\text{pred},{\boldsymbol{\gamma}})/\sqrt{I_0(\mathbf{y},{\boldsymbol{\gamma}})}$ asymptotically follows a standard normal distribution, when the latent signal is detectable. These are shown in the next section.

The replicated data $\mathbf{y}_0^{\text{pred}}$ is draw from the base model:
$$
\pi(\mathbf{y}^\text{pred}_0| \boldsymbol{\gamma}, \eta = 0 ) 
 = \int \pi(\mathbf{y}|\mathbf{w},{\boldsymbol{\beta}},{\boldsymbol{\theta}}_1)\pi(\mathbf{w}|{\boldsymbol{\theta}}_2,\eta=0) d\mathbf{w}. 
$$
We consider that the nuisance parameters $\boldsymbol{\gamma} = (\boldsymbol{\beta}, \boldsymbol{\theta})$ are fixed, and for simplicity, from now on, we omit the conditioning on $\boldsymbol{\gamma}$.

\subsection{Simulations}

\begin{figure}[!htp]
   \centering
   \includegraphics[width=\linewidth]{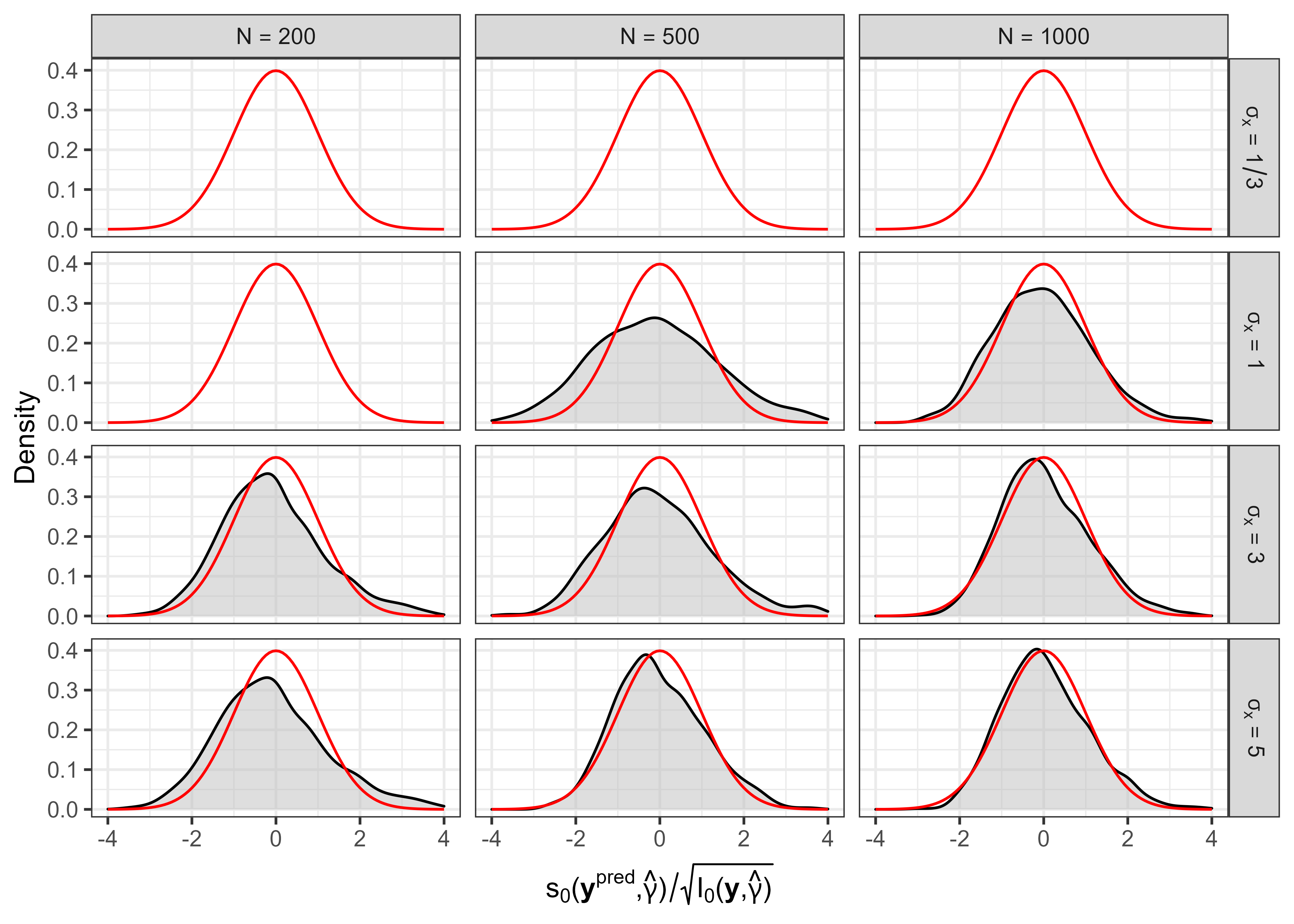}
   \caption{Density of a standard normal distribution (red) and kernel density estimate (black) of the distribution $s_0(\mathbf{y}_0^\text{pred},\hat{\boldsymbol{\gamma}})/\sqrt{I_0(\mathbf{y},\hat{\boldsymbol{\gamma}})}$ for simulated latent RW1 models with different dimensions $N$ and scales $\sigma_\mathbf{w}$.} 
  \label{fig:asympcomp} 
\end{figure}

Figure \ref{fig:asympcomp} shows samples of the distribution $s_0(\mathbf{y}_0^\text{pred},\hat{\boldsymbol{\gamma}})/\sqrt{I_0(\mathbf{y},\hat{\boldsymbol{\gamma}})}$ obtained for 12 simulated datasets $\mathbf{y}^{(i)}, i=1,\dotsc,12,$ from the latent Gaussian RW1 model in \eqref{eq:illusmodel}: $y_i = \sigma_{\mathbf{w}}w_i + \sigma_{\boldsymbol{\epsilon}} \epsilon_i$. The simulation parameters were $\sigma_{\boldsymbol{\epsilon}}=1$ and \mbox{$\sigma_\mathbf{w} \in \{1/3, 1, 3, 5\}$}, and the dimension was $N \in \{50,200,1000\}$. When $\sigma_\mathbf{w} = 1/3$, the signal is barely detectable, and $I_0$ is negative, and so $s_0(\mathbf{y}_0^\text{pred},\hat{\boldsymbol{\gamma}})/\sqrt{I_0(\mathbf{y},\hat{\boldsymbol{\gamma}})}$ cannot be computed, and the plots for these cases are empty. For the other cases, the asymptotic result can provide an adequate approximation.

\subsection{Mean of the reference distribution} \label{sect:meanproof}
The results rely again on conditions that allow derivative-integral and limit-integral interchange. The mean of the reference distribution under replicated data $\mathbf{y}_0^{\text{pred}}$ is
\begin{align*}
    E\{s_0(\mathbf{y}_0^{\text{pred}},\boldsymbol{\gamma})\} &= \int s_0(\mathbf{y}_0^{\text{pred}},\boldsymbol{\gamma})\pi(\mathbf{y}^\text{pred}_0|\eta = 0 ) d\mathbf{y}^\text{pred}_0 \\
&=  \int  \left(\lim_{\eta \to 0}\partial_\eta \log \pi(\mathbf{y}^\text{pred}_0|\eta)\right) \pi(\mathbf{y}^\text{pred}_0|\eta=0) d\mathbf{y}^\text{pred}_0 \\
 &=  \int  \frac{\lim_{\eta \to 0}\partial_\eta \pi(\mathbf{y}^\text{pred}_0|\eta)}{\pi(\mathbf{y}^\text{pred}_0|\eta=0)} \pi(\mathbf{y}^\text{pred}_0|\eta=0) d\mathbf{y}^\text{pred}_0 \\
 &= \lim_{\eta \to 0} \partial_\eta \int   \pi(\mathbf{y}^\text{pred}_0|\eta) d\mathbf{y}^\text{pred}_0 =  \lim_{\eta \to 0} \partial_\eta  1 = 0.
\end{align*} 

\subsection{Variance of the reference distribution}\label{sect:varproof}

The variance of the reference distribution under replicated data $\mathbf{y}_0^{\text{pred}}$ is
\begin{align*}
    V\left\{s_0(\mathbf{y}_0^{\text{pred}},\boldsymbol{\gamma})\right\} = E\left\{s_0^2(\mathbf{y}_0^{\text{pred}},\boldsymbol{\gamma})\right\} \overset{(a)}{=}  E\left\{I_0(\mathbf{y}_0^{\text{pred}}, \boldsymbol{\gamma})\right\}.
\end{align*}
Step (a) results from the relationship
\begin{equation}\label{eq:r1}
E\left[\left\{\lim_{\eta \to 0} \partial_\eta \log \pi(\mathbf{y}_0^{\text{pred}}|\eta)\right\}^2 \right] = - E\left\{ \lim_{\eta \to 0} \partial_\eta^2 \log \pi(\mathbf{y}_0^{\text{pred}}|\eta) \right\}.
\end{equation}
The previous equation can be derived from the expansion
\begin{align} \label{eq:r2}
 \partial_\eta^2 \log \pi(\mathbf{y}_0^{\text{pred}}|\eta) = \frac{\partial_\eta^2 \pi(\mathbf{y}_0^{\text{pred}}|\eta)}{\pi(\mathbf{y}_0^{\text{pred}}|\eta)}-(\partial_\eta \log \pi(\mathbf{y}_0^{\text{pred}}|\eta))^2, 
\end{align}
and $$E\left\{\frac{\partial_\eta^2 \pi(\mathbf{y}_0^{\text{pred}}|\eta)}{\pi(\mathbf{y}_0^{\text{pred}}|\eta)}\right\} = \int \frac{\partial_\eta^2 \pi(\mathbf{y}_0^{\text{pred}}|\eta)}{\pi(\mathbf{y}_0^{\text{pred}}|\eta)} \pi(\mathbf{y}_0^{\text{pred}}|\eta) d\mathbf{y}_0^{\text{pred}} = \partial_\eta^2 \int \pi(\mathbf{y}_0^{\text{pred}}|\eta) d\mathbf{y}_0^{\text{pred}} =0.$$
Thus, taking the limit of both sides of \eqref{eq:r2} and then taking the expectation leads to \eqref{eq:r1}.

\section{Implementation in R-INLA} \label{app:INLA}

The R-INLA software computes the posterior distributions of the parameters of LGMs, which, in Section \ref{sect:sensmeasures} corresponds to setting $\eta=0$ (we omit the conditioning on $\eta=0$ here). We leverage on the Gaussian mixture approximation in \eqref{eq:Gaussian_mixture} 
where \mbox{$\mathbf{x} \mid \mathbf{y}, \boldsymbol{\theta}_k \sim N(\boldsymbol{\mu}_k, \mathbf{\Sigma}_k = \mathbf{Q}^{-1}_k)$}, and  $\mathbf{x} = (\boldsymbol{\ell}, \boldsymbol{\beta}, \mathbf{w})$. The quantities
\mbox{$\pi(\boldsymbol{\theta}_k \mid \mathbf{y})$}, $\boldsymbol{\mu}_k$, $\mathbf{Q}_k$, and $\mathbf{\Sigma}_k$ are returned by R-INLA when the argument \verb|control.compute = list(config = TRUE)| is added to the \verb|inla| function. With the previous approximation, posterior expectations and covariances can be computed by conditioning on each hyperparameter configuration $\boldsymbol{\theta}_k$, and then computing the weighted average:
\begin{equation}
    E\{f(\mathbf{x}, \boldsymbol{\theta})|\mathbf{y}\} = \sum_{k=1}^K E\{f(\mathbf{x}, \boldsymbol{\theta}_k)|\mathbf{y}, \boldsymbol{\theta}_k\}  \pi(\boldsymbol{\theta}_k \mid \mathbf{y})  \Delta_k.
\end{equation}

All sensitivity measures can be expressed as posterior expectations or covariances, so below, we will derive the results conditioned on $\boldsymbol{\theta}_k$. We consider the posterior distribution of the random effects $\mathbf{w}|\mathbf{y}, \boldsymbol{\theta}_k \sim N(\boldsymbol{\mu}^{\mathbf{w}}, \mathbf{\Sigma}^{\mathbf{w}})$, where $\boldsymbol{\mu}^{\mathbf{w}}$ and $\mathbf{\Sigma}^{\mathbf{w}}$ can be obtained from R-INLA. Another useful random vector is the latent residual vector: $$\boldsymbol{r} | \mathbf{y}, \boldsymbol{\theta}_k= \mathbf{D}(\boldsymbol{\theta}_k){\mathbf{w}} | \mathbf{y}, \boldsymbol{\theta}_k \sim N\left\{\mathbf{D}(\boldsymbol{\theta}_k)\boldsymbol{\mu}^{\mathbf{w}}, \ \mathbf{D}(\boldsymbol{\theta}_k)\mathbf{\Sigma}^{\mathbf{w}} \mathbf{D}(\boldsymbol{\theta}_k)^T \right\}.$$ We also define 
$\mathbf{b} = \mathbf{D}(\boldsymbol{\theta}_k)\boldsymbol{\mu}^{\mathbf{w}}$ and $\boldsymbol{\Gamma} = -\mathbf{D}(\boldsymbol{\theta}_k)\mathbf{\Sigma}^{\mathbf{w}} \mathbf{D}(\boldsymbol{\theta}_k)^T + \text{diag}(\mathbf{h})$. We show in the next sections how various quantities related to sensitivity analysis and model checking are computed in R-INLA. The results in Sections \ref{sect:E1}, \ref{sect:E2} and \ref{sect:E3} apply to any LGM (with Gaussian or non-Gaussian response), while Section \ref{sect:E4} only applies for LGMs with a Gaussian response.


\subsection{BF sensitivity} \label{sect:E1}
The BF sensitivity $s_0$ is $s_0(\boldsymbol{\theta}_k) = \sum_i^n d_i(\boldsymbol{\theta}_k)$, and as already seen in Section \ref{sect:LGMnormal}: 
\begin{align*}
d_i(\mathbf{y},\boldsymbol{\theta}_k) = E\left.\left\{\frac{(r_i^2-3h_i)^2-6h_i^2}{8h_i^3} \right\vert \mathbf{y}, \boldsymbol{\theta}_k \right\} = \frac{ b_i^4 + 3\Gamma_{ii}^2 - 6 b_i^2\Gamma_{ii}}{8 h_i^3}.
\end{align*} 

\subsection{Sensitivity of posterior expectations $s_l$} \label{sect:E2}
We consider the sensitivity of posterior expectations of $\mathbf{w}$, $\boldsymbol{\beta}$ and $\boldsymbol{\ell}$, for non-Gaussian prior perturbations on $\mathbf{w}$. We compute here the sensitivity of the posterior mean of the first fixed effect $\beta_1$, and similar calculations apply for other posterior means. From Theorem \ref{theo:sens} and equation \eqref{eq:ngpert}, the sensitivity is given by:
\begin{align*}
s_{\beta_1}(\boldsymbol{\mathbf{y},\theta}_k) = \sum_{i=1}^n Cov\left.\left\{ \beta_1, \frac{(r_i^2-3h_i)^2-6h_i^2}{8h_i^3} \right\vert \mathbf{y}, \boldsymbol{\theta}_k\right\}.
\end{align*} 
Under the Gaussian mixture approximation:
$$
(\beta_1, \mathbf{r})^T | \mathbf{y}, \boldsymbol{\theta}_k \sim N\left(\begin{bmatrix}\mu^{\beta_1} \\ \mathbf{b} \end{bmatrix}, \begin{bmatrix}\mathbf{\Sigma}^{\beta_1} & \mathbf{\Sigma}^{\beta_1, \mathbf{r}} \\ \mathbf{\Sigma}^{\mathbf{r}, \beta_1} &  -\mathbf{\Gamma} + \text{diag}(\mathbf{h})  \end{bmatrix} \right),
$$
where $\mathbf{\Sigma}^{\beta_1, \mathbf{r}} = Cov(\beta_1, \mathbf{w} \ | \ \mathbf{y}, \boldsymbol{\theta}_k) \mathbf{D}^T(\boldsymbol{\theta}_k)$. Then, it can be shown that:
$$
s_{\beta_1}(\mathbf{y},\boldsymbol{\theta}_k) = \sum_{i=1}^n \frac{b_i \mathbf{\Sigma}^{\beta_1, \mathbf{r}}_{i} (-3 \mathbf{\Gamma}_{ii} +  b_i^2) }{2h_i^3}.
$$

\subsection{Measure $I_0$} \label{sect:E3}

From Theorem \ref{theo:scorefisher}, and equations \eqref{eq:ngpert} and \eqref{eq:ngpert2}, the measure $I_0$ is given by:
\begin{align*} \label{eq:ofibayes}
I_0(\mathbf{y},\boldsymbol{\theta}_k) &=  - \sum_{i=1}^n E\left.\left\{g_i(r_i) \right\vert \mathbf{y}, \boldsymbol{\theta}_k \right\} - V\left\{ \sum_{i=1}^n p_i(r_i) \ | \ \mathbf{y}, \boldsymbol{\theta}_k \right\}. 
\end{align*}
where $g_i(r_i)$ is given in \eqref{eq:ngpert2}. Then, we have:
\begin{align*}
     E\left.\left\{g_i(r_i) \right\vert \mathbf{y}, \boldsymbol{\theta}_k \right\} &= E\left.\left\{(-3 h_i^3 - 3 h_i^2 r_i^2 + 6 h_i r_i^4 - r_i^6)/(8 h_i^5) \right\vert \mathbf{y}, \boldsymbol{\theta}_k  \right\}  \\
     &= -\frac{(b_i^6 + b_i^4 (-15 \Gamma_{ii} + 9 h_i) + 3 b_i^2 (15 \Gamma_{ii}^2 - 18 \Gamma_{ii} h_i + 4 h_i^2)}{8 h^5} -    \\
     &- \frac{3 (-5 \Gamma_{ii}^3 + 9 \Gamma_{ii}^2 h_i - 4 \Gamma_{ii} h_i^2 + h_i^3)}{8 h^5}. 
\end{align*}
Also, $V\left\{ \sum_{i=1}^n p_i(r_i) \vert \mathbf{y}, \boldsymbol{\theta}_k \right\} = \sum_{i,j} Cov\{ p_i(r_i),  p_j(r_j)  \vert \mathbf{y}, \boldsymbol{\theta}_k \}$, and
\begin{align*}
&Cov\{ p_i(r_i),  p_j(r_j)  \vert \mathbf{y}, \boldsymbol{\theta}_k \}= \\ &= \frac{\Gamma_{ij} (3 \Gamma_{ij}^3 - 12 \Gamma_{ij}^2 b_i b_j + 9 \Gamma_{ij} (\Gamma_{ii} - b_i^2) (\Gamma_{jj} - b_j^2) - 2 b_1 (-3 \Gamma_{ii} + b_i^2) b_j (-3 \Gamma_{jj} + b_j^2))}{8 h_i^3 h_j^3}.
\end{align*}

\subsection{Model checking for LGMs with Gaussian response} \label{sect:E4}

The \verb|ng.check| function produces a diagnostic plot, which can be seen in Figure \ref{fig:press_score} (except the histogram) comparing the Gaussian approximation of the reference distribution \mbox{$N[0, V\{s_0(\mathbf{y}^{\text{pred}},\hat{\boldsymbol{\theta}})|\eta=0\}]$} with the observed BF sensitivity value, where
the variance of the reference distribution is given in Proposition \eqref{prop:variance_approx}, and $\hat{\boldsymbol{\theta}}$ is fixed at the posterior mode. The upper-tailed probability $p$ in \eqref{eq:pv} is also computed, and it takes into account the uncertainty associated with the hyperparameters,  \mbox{$p = \sum_{k=1}^K p(\boldsymbol{\theta}_k) \pi(\boldsymbol{\theta}_k \mid \mathbf{y})  \Delta_k$}, where:
\begin{align*}
    p(\boldsymbol{\theta}_k) &= P\left\{s_0(\mathbf{y}^{\text{pred}},\boldsymbol{\theta}_k) > s_0(\mathbf{y},\boldsymbol{\theta}_k) |  \boldsymbol{\theta}_k, \eta = 0\right\} \approx \Phi\left[-s_0(\mathbf{y}, \boldsymbol{\theta}_k)/SD\{s_0(\mathbf{y}^{\text{pred}},\boldsymbol{\theta}_k)|\eta=0\}\right].
\end{align*}

\end{appendices}

\end{document}